\documentclass[12pt]{article}
\pdfoutput=1

\usepackage{amsmath}
\usepackage{amssymb}
\usepackage{epsfig}
\usepackage{epstopdf}
\usepackage{latexsym}
\usepackage{color}
\numberwithin{equation}{section}
\usepackage[
      colorlinks=false,
      linkcolor=darkblue,  
      urlcolor=blue,    
      filecolor=blue,     
      citecolor=red,
linktocpage=true,
      pdfstartview=FitV,
      bookmarksopen=true    
      ]{hyperref}

\begin{document}

\begin{titlepage}

\centerline
\centerline
\centerline
\bigskip
\bigskip
\bigskip
\centerline{\Huge \rm Non-conformal branes wrapped on a disk}
\bigskip
\bigskip
\bigskip
\bigskip
\bigskip
\bigskip
\bigskip
\bigskip
\centerline{\rm Minwoo Suh}
\bigskip
\centerline{\it School of General Education, Kumoh National Institute of Technology,}
\centerline{\it Gumi, 39177, Korea}
\bigskip
\centerline{\tt minwoosuh1@gmail.com} 
\bigskip
\bigskip
\bigskip
\bigskip
\bigskip
\bigskip
\bigskip
\bigskip

\begin{abstract}
\noindent We classify the disk solutions which are obtained from the spindle solutions from non-conformal D$p$-branes recently constructed by Boisvert and Ferrero. Then we study the global geometry of the disk solutions. We discover several common features of disk solutions, $e.g.$, monopoles and smeared branes, in most cases, but there are also exceptions.
\end{abstract}

\vskip 6cm

\flushleft {August, 2025}

\end{titlepage}

\tableofcontents

\vspace{4.5cm}

\section{Introduction and conclusions}

String and M-theory has extended objects of branes. The brane solutions were constructed in ten- and eleven-dimensional supergravity and they preserve half maximal supersymmetry, \cite{Horowitz:1991cd}. It was understood that they naturally fit in string and M-theory as the extended objects, D-branes and M-branes, \cite{Dai:1989ua, Leigh:1989jq, Polchinski:1995mt}. For D$p$-branes the worldvolume is ($p$+1)-dimensional flat space, $\mathbb{R}^{1,p}$. For $p\ne3$, the near-horizon geometry of D$p$-brane develops a singularity where the running dilaton diverges. Via the extension of AdS/CFT correspondence, \cite{Maldacena:1997re}, the near-horizon geometry is dual to ($p$+1)-dimensional maximally supersymmetric Yang-Mills theory living on $\mathbb{R}^{1,p}$ which is non-conformal for $p\ne3$, \cite{Itzhaki:1998dd}.

In the study of superconformal field theories, \cite{Witten:1988ze}, and supergravity, \cite{Maldacena:2000mw}, supersymmetry preserved by topological twist has provided important examples and lessons. Recently, new examples preserving supersymmetry in more general ways were discovered: twisted compactification on non-constant curvature manifolds with orbifold singularities. Largely, two classes of orbifolds were discovered. The first class involves a sphere with two orbifold singularities at the north and south poles and they are known as spindles, \cite{Ferrero:2020laf, Ferrero:2020twa}. The second class uses a disk which has a single orbifold singularity at the center, \cite{Bah:2021mzw, Bah:2021hei}. Although the spindles and the disks have distinct topologies, they originate from an identical local solution.

So far the orbifold solutions were mainly constructed by wrapping D3-, M2-, M5-branes and D4-D8-brane system on an orbifold. Thus the spindle and disk solutions constructed so far asymptote to an $AdS$ vacuum dual to superconformal field theories. Recently, a general class of solutions were constructed by wrapping so-called ``non-conformal" branes on various manifolds including orbifolds by Boisvert and Ferrero, \cite{Ferrero:2024vmz, Boisvert:2024jrl}. In particular, they studied the near-horizon limit of D$p$-branes for $p=2,4,5,6$ and NS5-branes. They are solutions of ($p$+2)-dimensional gauged supergravity theories from dimensional reductions of type II supergravity on a (8-$p$)-sphere.

In this paper, we present the classification of disk solutions which are obtained from the spindle solutions from non-conformal D$p$-branes constructed in \cite{Ferrero:2024vmz, Boisvert:2024jrl}. By the choice of non-trivial gauge fields, we classify the disk solutions from non-conformal branes. Largely, there are four classes of disk solutions.
\begin{itemize}

\item No charge solutions: $A^1=\ldots=0$,

\item Single charge solutions: $e.g.$, $A^1\ne0$, $A^2=\ldots=0$,

\item Equal charge solutions: $e.g.$, $A^1=A^2\ne0$, $A^3=\ldots=0$,

\item Multi charge solutions: $e.g.$, $A^1\ne{A}^2$, $A^3=\ldots=0$.

\end{itemize}
In \cite{Boisvert:2024jrl} disk solutions with no gauge fields were first considered for the case of D6-branes. In this work, we show that no charge solutions are ubiquitous and further find them from D4- and D2-branes. On the other hand, we could not find one from D5/NS5-branes. It would be interesting to study how these solutions preserve supersymmetry. 

We can also classify solutions by the boundary of the disk. For a disk, $\Sigma(y,z)$, where $y$ and $z$ are radial and angular coordinates, respectively, there is an orbifold singularty at the center of the disk, $e.g.$, $y=y_1$. Then there can be two classes of solutions.
\begin{itemize}

\item Compact solutions with $y\in[0,y_1]$: the boundary at $y=0$,

\item Non-compact solutions with $y\in[y_1,\infty)$: the boundary at $y=\infty$.

\end{itemize}
The second class of solutions are non-compact and are in relevance with the defect or domain wall solutions, \cite{Gutperle:2022pgw, Gutperle:2023yrd, Capuozzo:2023fll}.{\footnote {See also the discussion in appendix A.5 of \cite{Bah:2021hei}.}}

As an aside, for the conformal branes which asymptote to an $AdS$ vacuum, we briefly study no charge solutions. We find that the no charge solutions from D3-, M2-, M5-branes and D4-D8-brane system are in the non-compact class, $y\in[y_1,\infty)$. Even from the non-conformal branes, solution from D6-branes is the only compact solution, $y\in[0,y_1]$, and solutions from D4- and D2-branes are in the non-compact class, $y\in[y_1,\infty)$. As a representative case, we present no charge solutions from M5-branes in appendix \ref{appD}.

Then we study the global geometry of disk solutions from non-conformal branes. The global geometry of disk solutions from conformal branes has several common structures. (i) The Euler characteristic of the disk is characterized by the orbifold number of the disk. (ii) The solutions feature a monopole source at the center of the disk. (iii) The solutions asymptote to smeared branes when approaching the boundary of the disk. These are some of the common characters of disk solutions from D3-branes, \cite{Couzens:2021tnv, Suh:2021ifj}, M2-branes, \cite{Suh:2021hef, Couzens:2021rlk}, M5-branes, \cite{Bah:2021mzw, Bah:2021hei, Karndumri:2022wpu, Couzens:2022yjl, Bah:2022yjf, Couzens:2023kyf}, D4-D8-branes, \cite{Suh:2021aik, Couzens:2022lvg, Suh:2024fru} and also of disk$\times$disk and spindle$\ltimes$disk solutions from M5-branes and D4-D8-branes, \cite{Bomans:2023ouw, Suh:2024fru}.

Interestingly, we find smeared branes at the boundary of the solutions with $y\in[0,y_1]$. On the other hand, the branes do not get smeared at the boundary of the solutions with $y\in[y_1,\infty)$.

The most interesting open question would be to calculate the gravitational partition function of the disk solutions from non-conformal branes. Calculation of gravitational partition function is less understood compared to the case of conformal branes.{\footnote{Recently, off-shell central charge in the gravitational block form for disk solutions has been proposed in \cite{Kim:2025ziz}.}} One could perform the holographic renormalization of the on-shell action on non-conformal branes, \cite{Peet:1998wn, Boonstra:1998mp, Kanitscheider:2008kd}, as in \cite{Bobev:2018ugk, Bobev:2019bvq, Bobev:2024gqg}. Holographic entanglement entropy could provide an alternative method, \cite{Ryu:2006ef, vanNiekerk:2011yi}. 

In section \ref{d6}, \ref{d5}, \ref{d4}, \ref{d2}, we classify the disk solutions and study the global geometries from D6-, D5/NS5-, D4-, and D2-branes, respectively. A review of smeared brane solutions are given in appendix \ref{appA}. The uplift formula for $D=6$ gauged supergravity to type IIA supergravity is reviewed in appendix \ref{appB}. We construct the uplift formula for the $U(1)^3$ subsector of $D=4$ $ISO(7)$-gauged maximal supergravity to type IIA supergravity in appendix \ref{appC}. In appendix \ref{appD}, we present no charge solutions from M5-branes.

\section{D6-branes wrapped on a disk} \label{d6}

\subsection{No charge solution: $A=0$: $y\in[0,y_1]$}

\subsubsection{$D=8$ gauged supergravity}

We consider $D=8$ $SU(2)$-gauged maximal supergravity, \cite{Salam:1984ft}, which is obtained from the reduction of type IIA supergravity on a two-sphere. In particular, a $U(1)$ subsector of the theory, \cite{Edelstein:2001pu}, is of our interest. The field content is the metric, a $U(1)$ gauge field, $A$, and two real scalar fields, $\phi$, $\lambda$. For details of $D=8$ $U(1)$-gauged supergravity, we refer section 3 of \cite{Boisvert:2024jrl}.

\begin{figure}[t] 
\begin{center}
\includegraphics[width=2.0in]{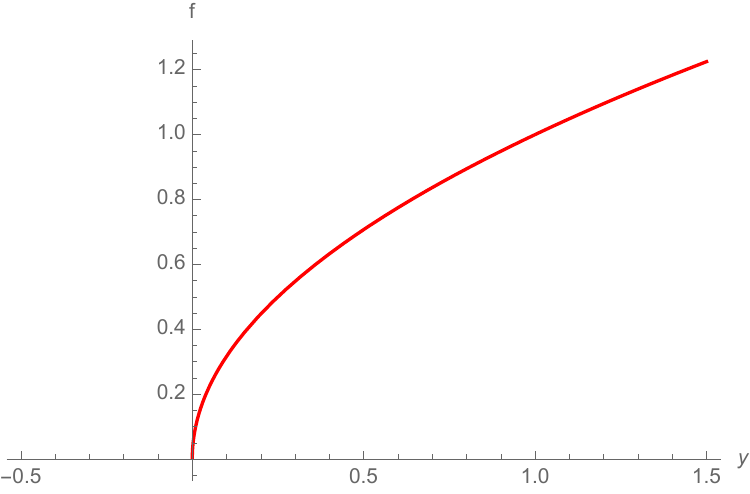} \qquad \includegraphics[width=2.0in]{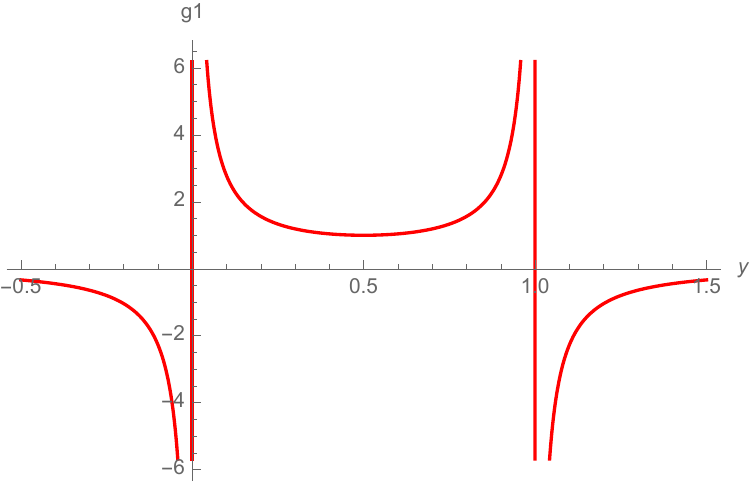} \qquad \includegraphics[width=2.0in]{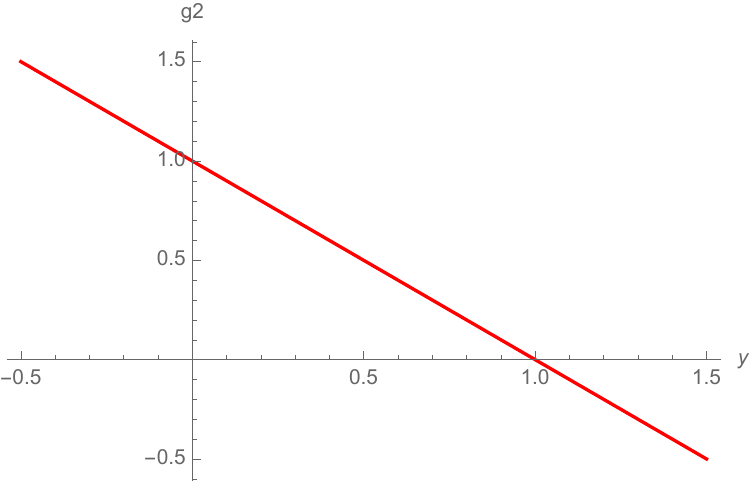}
\caption{{\it A representative solution with $g=2$. The solution is regular in the range of $0\,<\,y\,<\,y_1=1$.}} \label{d61}
\end{center}
\end{figure}

In this section, we consider solution without a gauge field by setting $A=0$ or, equivalently, $q=0$ in the the spindle solution in (3.15) of \cite{Boisvert:2024jrl}. Then the solution is given by
\begin{align}
ds_8^2\,=&\,r^3y^{1/2}\left[\frac{ds_{1,4}^2+dr^2}{r^2}+\frac{1}{4yh(y)}dy^2+h(y)dz^2\right]\,,  \notag \\
e^{2\phi}\,=&\,r^6y^3\,,
\end{align}
where $ds_{1,4}^2$ is the metric on $\mathbb{R}^{1,4}$ and 
\begin{equation}
h(y)\,=\,\frac{g^2}{4}-y\,.
\end{equation}
The scalar field, $\lambda$, and the gauge field, $A$, are trivial. For $h(y)=0$, there is a single root, $y_1\,\equiv\,g^2/4$. We consider solutions with
\begin{equation}
0<y<y_1\,\equiv\,\frac{g^2}{4}\,.
\end{equation}
We plot a representative solution with $g=2$ in figure \ref{d61}. The metric functions, $f(y)$, $g_1(y)$, and $g_2(y)$, are defined by
\begin{equation}
ds_8^2\,=\,r^3f(y)\left[\frac{ds_{1,2}^2+dr^2}{r^2}+g_1(y)dy^2+g_2(y)dz^2\right]\,.
\end{equation}

Near $y\rightarrow0$ the warp factor vanishes and it is a curvature singularity of the metric, 
\begin{equation}
ds_8^2\,\approx\,r^3y^{1/2}\left[\frac{ds_{1,4}^2+dr^2}{r^2}+\frac{1}{g^2y}dy^2+\frac{g^2}{4}dz^2\right]\,.
\end{equation}
Approaching $y\rightarrow{y}_1$, the metric becomes to be
\begin{equation}
ds_8^2\,\approx\,r^3y_1^{1/2}\left[\frac{ds_{1,4}^2+dr^2}{r^2}+\frac{d\rho^2+\frac{g^2}{4}\rho^2dz^2}{-y_1h'(y_1)}\right]\,,
\end{equation}
where we introduced a new parametrization of coordinate, $\rho^2\,=\,y_1-y$. Then,  the $\rho-z$ surface is locally an $\mathbb{R}^2/\mathbb{Z}_\ell$ orbifold if we set
\begin{equation}
\frac{\ell\Delta{z}}{2\pi}\,=\,\frac{2}{g}\,,
\end{equation}
where $\Delta{z}$ is the period of coordinate, $z$, and $\ell\,=\,1,2,3,\ldots$. The metric spanned by $(y,z)$ has a topology of disk, $\Sigma$, with the center at $y=y_1$ and the boundary at $y=0$.

We calculate the Euler characteristic of the $y-z$ surface, $\Sigma$, 
\begin{equation}
\chi\,=\,\frac{1}{4\pi}\int_\Sigma{R}_\Sigma\text{vol}_\Sigma\,=\,\frac{1}{4\pi}2y_1^{1/2}\Delta{z}\,=\,\frac{1}{\ell}\,.
\end{equation}
This is a natural result for a disk in an $\mathbb{R}^2/\mathbb{Z}_\ell$ orbifold.

\subsubsection{Uplift to type IIA supergravity}

By employing the uplift formula to type IIA supergravity, $e.g.$, in (3.5) of \cite{Boisvert:2024jrl}, we obtain the metric in the string frame, dilaton, and RR one-form potential, respectively,
\begin{align}
ds_{10}^2\,=&\,r^3\left[y^{1/2}\left(\frac{ds_{1,4}^2+dr^2}{r^2}+\frac{1}{4yh(y)}dy^2+h(y)dz^2\right)+\frac{1}{g^2}y^{3/2}\left(d\theta^2+\sin^2\theta{d}\alpha^2\right)\right]\,, \notag \\
e^{4\Phi}\,=&\,\frac{r^6y^3}{g^2}\,, \notag \\
C_{(1)}\,=&\,-\cos\theta{d}\alpha\,,
\end{align}
where $0\le\theta\le\pi$ and $0\le\alpha\le2\pi$.

We consider the singularity at $y\rightarrow0$ of the uplifted metric. As $y\rightarrow0$, the uplifted metric becomes
\begin{equation}
ds_{10}^2\,\approx\,r^3\left[y^{1/2}\left(\frac{ds_{1,4}^2+dr^2}{r^2}+\frac{g^2}{4}dz^2\right)+\frac{1}{g^2y^{1/2}}\Big(dy^2+y^2\left(d\theta^2+\sin^2\theta{d}\alpha^2\right)\Big)\right]\,,
\end{equation}
and the dilaton is
\begin{equation}
e^\Phi\,\approx\,\frac{r^{3/2}y^{3/4}}{g^2}\,.
\end{equation}
The branes are extended along the $\mathbb{R}^{1,4}$, $r$, $z$ directions and localized at the center of $y$, $\theta$, $\alpha$ directions. We observe that there is no smeared direction. The $y^{1/2}$ and $1/y^{1/2}$ factors of the metric and the dilaton match the un-smeared branes reviewed in appendix \ref{appA}.

\subsection{Single charge solution: $A\ne0$: $y\in[0,y_1]$}

\subsubsection{$D=8$ gauged supergravity}

\begin{figure}[t] 
\begin{center}
\includegraphics[width=2.0in]{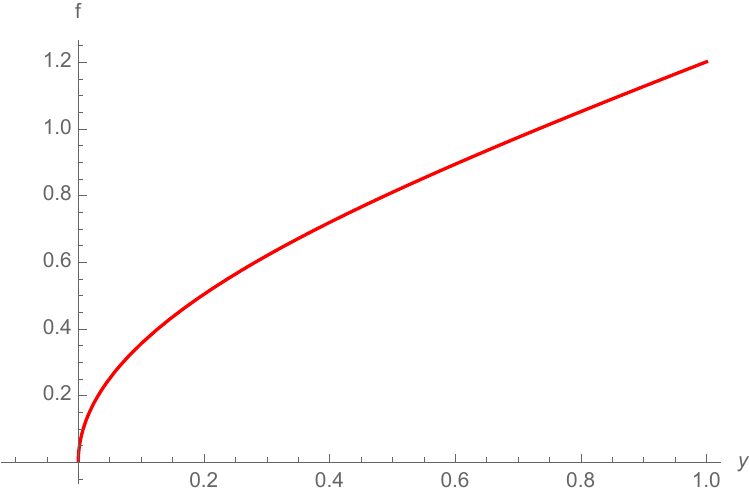} \qquad \includegraphics[width=2.0in]{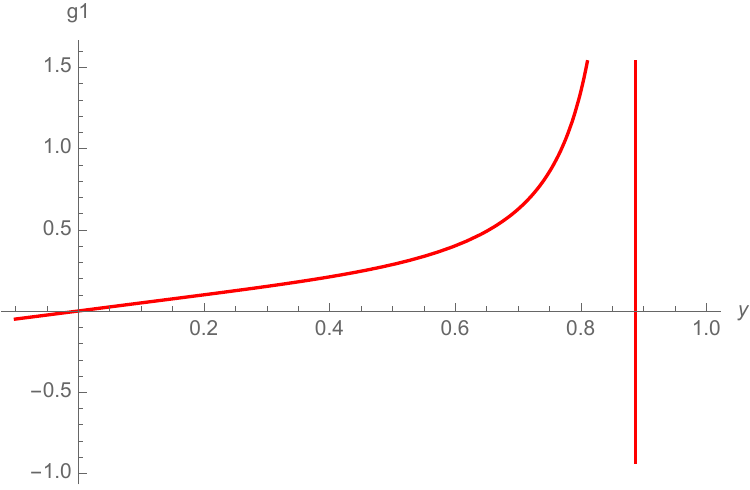} \qquad \includegraphics[width=2.0in]{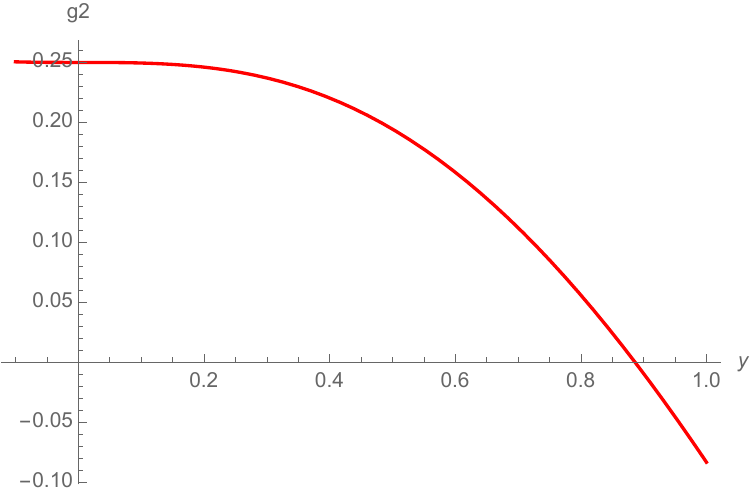}
\caption{{\it A representative solution with $g=1$ and $q=2$. The solution is regular in the range of $0\,<\,y\,<\,y_1=0.886$.}} \label{d62}
\end{center}
\end{figure}

In this section, we consider single charge solutions from the the spindle solution in (3.15) of \cite{Boisvert:2024jrl}. Then the solution is given by
\begin{align}
ds_8^2\,=&\,r^3y^{1/6}\left(y^2+q\right)^{1/6}\left[\frac{ds_{1,4}^2+dr^2}{r^2}+\frac{y}{4\left(y^2+q\right)h(y)}dy^2+h(y)dz^2\right]\,,  \notag \\
e^{2\phi}\,=&\,r^6y\left(y^2+q\right)\,, \qquad e^{6\lambda}\,=\,\frac{y^2+q}{y^2}\,, \notag \\
A\,=&\,\frac{q}{y^2+q}dz\,,
\end{align}
where $ds_{1,4}^2$ is the metric on $\mathbb{R}^{1,4}$ and 
\begin{equation}
h(y)\,=\,\frac{g^2\left(y^2+q-\frac{4}{g^2}y^3\right)}{4\left(y^2+q\right)}\,.
\end{equation}
For $h(y)=0$, we consider the case with a single real root, $y_1$, and two complex roots. We find solutions with
\begin{equation}
0<y<y_1\,,
\end{equation}
where we do not present the explicit expression of $y_1$ for its complexity. We plot a representative solution with $g=1$ and $q=2$ in figure \ref{d62}. The metric functions, $f(y)$, $g_1(y)$, and $g_2(y)$, are defined by
\begin{equation}
ds_8^2\,=\,r^3f(y)\left[\frac{ds_{1,2}^2+dr^2}{r^2}+g_1(y)dy^2+g_2(y)dz^2\right]\,.
\end{equation}

Near $y\rightarrow0$ the warp factor vanishes and it is a curvature singularity of the metric, 
\begin{equation}
ds_8^2\,\approx\,r^3q^{1/6}y^{1/6}\left[\frac{ds_{1,4}^2+dr^2}{r^2}+\frac{y}{g^2q}dy^2+\frac{g^2}{4}dz^2\right]\,.
\end{equation}
Approaching $y\rightarrow{y}_1$, the metric becomes to be
\begin{equation}
ds_8^2\,\approx\,r^3y_1^{1/2}\left[\frac{ds_{1,4}^2+dr^2}{r^2}+\frac{d\rho^2+\mathcal{E}(q)^2\rho^2dz^2}{-y_1^{-1}\left(y_1^2+q\right)h'(y_1)}\right]\,,
\end{equation}
where we have
\begin{equation}
\mathcal{E}(q)\,\equiv\,y_1^{-1}\left(y_1^2+q\right)h'(y_1)^2\,=\,\left(\frac{y_1^{3/2}\left(y_1^2+3q\right)}{\left(y_1^2+q\right)^{3/2}}\right)^2\,,
\end{equation}
and we introduced a new parametrization of coordinate, $\rho^2\,=\,y_1-y$. Then,  the $\rho-z$ surface is locally an $\mathbb{R}^2/\mathbb{Z}_\ell$ orbifold if we set
\begin{equation}
\frac{\ell\Delta{z}}{2\pi}\,=\,\frac{1}{\mathcal{E}(q)}\,,
\end{equation}
where $\Delta{z}$ is the period of coordinate, $z$, and $\ell\,=\,1,2,3,\ldots$. The metric spanned by $(y,z)$ has a topology of disk, $\Sigma$, with the center at $y=y_1$ and the boundary at $y=0$.

We calculate the Euler characteristic of the $y-z$ surface, $\Sigma$, 
\begin{equation}
\chi\,=\,\frac{1}{4\pi}\int_\Sigma{R}_\Sigma\text{vol}_\Sigma\,=\,\frac{1}{4\pi}2\mathcal{E}(q)\Delta{z}\,=\,\frac{1}{\ell}\,.
\end{equation}
This is a natural result for a disk in an $\mathbb{R}^2/\mathbb{Z}_\ell$ orbifold.

\subsubsection{Uplift to type IIA supergravity}

By employing the uplift formula to type IIA supergravity, $e.g.$, in (3.5) of \cite{Boisvert:2024jrl}, we obtain the metric in the string frame, dilaton, and RR one-form potential, respectively,
\begin{align}
ds_{10}^2\,=&\,r^3\left[\frac{\left(y^2+q\right)^{1/2}}{y^{1/2}}\left(\frac{ds_{1,4}^2+dr^2}{r^2}+\frac{y}{4\left(y^2+q\right)h(y)}dy^2+h(y)dz^2\right)\right. \notag \\
& \,\,\,\,\,\,\,\,\,\, \left.+\frac{1}{g^2}y^{1/2}\left(y^2+q\right)^{1/2}\left(d\theta^2+\sin^2\theta{d}\alpha^2\right)\right]\,, \notag \\
e^{\Phi}\,=&\,\frac{r^{3/2}\left(y^2+q\right)^{3/4}}{g^{1/2}y^{3/4}}\,, \notag \\
C_{(1)}\,=&\,-\cos\theta{d}\alpha+\frac{gq}{y^2+q}dz\,,
\end{align}
where $0\le\theta\le\pi$ and $0\le\alpha\le2\pi$.

We consider the singularity at $y\rightarrow0$ of the uplifted metric. As $y\rightarrow0$, the uplifted metric becomes
\begin{equation}
ds_{10}^2\,\approx\,r^3\left[\frac{q^{1/2}}{y^{1/2}}\left(\frac{ds_{1,4}^2+dr^2}{r^2}+\frac{g^2}{4}dz^2\right)+\frac{y^{1/2}}{g^2q^{1/2}}\Big(dy^2+q\left(d\theta^2+\sin^2\theta{d}\alpha^2\right)\Big)\right]\,,
\end{equation}
and the dilaton is
\begin{equation}
e^\Phi\,\approx\,\frac{r^{3/2}q^{3/4}}{g^{1/2}y^{3/4}}\,.
\end{equation}
The branes are extended along the $\mathbb{R}^{1,4}$, $r$, $z$ directions, localized at the center of $y$ directions, and smeared along the $\theta$ and $\alpha$ directions. The $1/y^{1/2}$ and $y^{1/2}$ factors of the metric and the dilaton match the smeared branes reviewed in appendix \ref{appA}.

\vspace{4.2cm}

\section{D5/NS5-branes wrapped on a disk} \label{d5}

\subsection{Single charge solution: $A^1\ne0$, $A^2=0$: $y\in[0,y_1]$}

\subsubsection{$D=7$ gauged supergravity}

We consider $D=7$ $ISO(4)$-, in \cite{Cvetic:2000ah}, and $SO(4)$-gauged maximal supergravity which uplift to the NSNS sectors of type IIA, \cite{Giani:1984wc, Campbell:1984zc, Huq:1983im}, and IIB, \cite{Schwarz:1983qr, Howe:1983sra}, supergravity, respectively. In particular, the $U(1)^2$ subsector which is common to both theories, \cite{Gauntlett:2001ps, Bigazzi:2001aj}, is of our interest. The field content is the metric, two $U(1)$ gauge fields, $A^1$, $A^2$, and two real scalar fields, $\lambda_1$, $\lambda_2$. For details of $D=7$ $U(1)^2$-gauged supergravity, we refer section 4 of \cite{Boisvert:2024jrl}.

Let us briefly discuss the possiblity of no charge solutions, $A^1=A^2=0$. We set the parameters, $q_1=q_2=0$, in the the spindle solution in (4.26) of \cite{Boisvert:2024jrl}. Then the solution is given by
\begin{align}
ds_7^2\,=&\,e^{\frac{4g}{5}\rho}y^{2/5}\left[ds_{1,3}^2+d\rho^2+\frac{1}{4g^2\left(p^2-1\right)y^2}dy^2+\left(p^2-1\right)dz^2\right]\,, \notag \\
e^{\lambda_1+\frac{g}{5}\rho}\,=&\,e^{\lambda_2+\frac{g}{5}\rho}\,=\,\frac{p^{1/2}}{y^{1/10}}\,, \notag \\
A^1\,=&\,A^2\,=\,0\,,
\end{align}
where $ds_{1,4}^2$ is the metric on $\mathbb{R}^{1,4}$. This solution cannot have the disk structure.

\begin{figure}[t] 
\begin{center}
\includegraphics[width=2.0in]{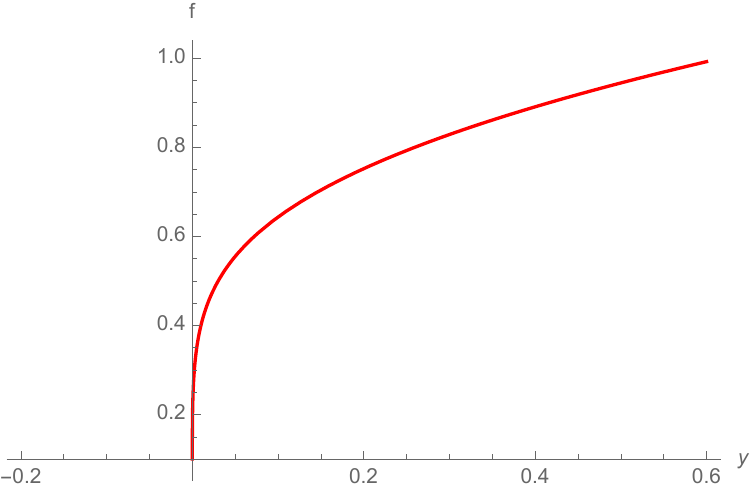} \qquad \includegraphics[width=2.0in]{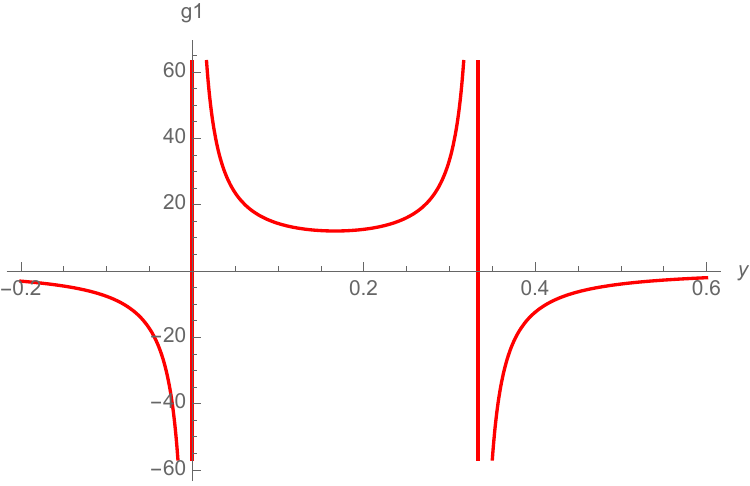} \qquad \includegraphics[width=2.0in]{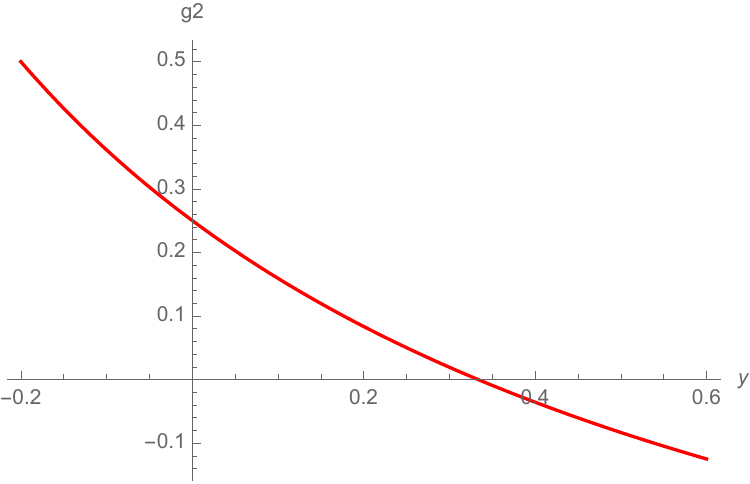}
\caption{{\it A representative solution with $g=1$ and $p=0.5$. The solution is regular in the range of $0<\,y<\,y_1=1/3$.}} \label{d51}
\end{center}
\end{figure}

In this section, we study single charge disk solutions: $A^1\ne0$, $A^2=0$. We set the parameters, $q_1=1$ and $q_2=0$, for the spindle solution given in (4.26) of \cite{Boisvert:2024jrl}. Then the solution is given by
\begin{align} \label{d5sol}
ds_7^2\,=&\,e^{\frac{4g}{5}\rho}y^{1/5}(y+1)^{1/5}\left[ds_{1,3}^2+d\rho^2+\frac{1}{4g^2y\left(y+1\right)h(y)}dy^2+h(y)dz^2\right]\,, \notag \\
A\,\equiv&\,A^1\,=\,\frac{p}{y+1}dz\,, \qquad A^2\,=\,0\,, \notag \\
e^{\lambda_1+\frac{g}{5}\rho}\,=&\,\frac{p^{1/2}y^{1/5}}{(y+1)^{3/10}}\,, \qquad e^{\lambda_2+\frac{g}{5}\rho}\,=\,\frac{p^{1/2}(y+1)^{1/5}}{y^{3/10}}\,,
\end{align}
where $ds_{1,3}^2$ is the metric on $\mathbb{R}^{1,3}$ and 
\begin{equation}
h(y)\,=\,\frac{p^2(y+1)-y}{y+1}\,.
\end{equation}
For $h(y)=0$, there is a single root, $y_1\,\equiv\,\frac{p^2}{1-p^2}$. We consider solutions with
\begin{equation}
-1<p<1\,, \qquad 0<y<y_1\,\equiv\,\frac{p^2}{1-p^2}\,.
\end{equation}
We plot a representative solution with $g=1$ and $p=0.5$ in figure \ref{d51}. The metric functions, $f(y)$, $g_1(y)$, and $g_2(y)$, are defined by
\begin{equation}
ds_7^2\,=\,e^{\frac{4g}{5}\rho}f(y)\left[ds_{1,3}^2+d\rho^2+g_1(y)dy^2+g_2(y)dz^2\right]\,.
\end{equation}

Near $y\rightarrow0$ the warp factor vanishes and it is a curvature singularity of the metric, 
\begin{equation}
ds_7^2\,\approx\,e^{\frac{4g}{5}\rho}y^{1/5}\left[ds_{1,3}^2+d\rho^2+\frac{1}{4g^2p^2y}dy^2+p^2dz^2\right]\,.
\end{equation}
Approaching $y\rightarrow{y}_1$, the metric becomes to be
\begin{equation}
ds_7^2\,\approx\,e^{\frac{4g}{5}\rho}y_1^{1/5}\left[ds_{1,3}^2+d\rho^2+\frac{d\tilde{\rho}^2+\mathcal{E}(q)^2\tilde{\rho}^2dz^2}{-g^2y_1(y_1+1)h'(y_1)}\right]\,,
\end{equation}
where we have
\begin{equation}
\mathcal{E}(p)^2\,\equiv\,g^2y_1(y_1+1)h'(y_1)^2\,=\,\Big(gp\left(1-p^2\right)\Big)^2\,,
\end{equation}
and we introduced a new parametrization of coordinate, $\tilde{\rho}^2\,=\,y_1-y$. Then,  the $\tilde{\rho}-z$ surface is locally an $\mathbb{R}^2/\mathbb{Z}_\ell$ orbifold if we set
\begin{equation} \label{ldz5}
\frac{\ell\Delta{z}}{2\pi}\,=\,\frac{1}{gp\left(1-p^2\right)}\,,
\end{equation}
where $\Delta{z}$ is the period of coordinate, $z$, and $\ell\,=\,1,2,3,\ldots$. The metric spanned by $(y,z)$ has a topology of disk, $\Sigma$, with the center at $y=y_1$ and the boundary at $y=0$.

We calculate the Euler characteristic of the $y-z$ surface, $\Sigma$, 
\begin{equation}
\chi\,=\,\frac{1}{4\pi}\int_\Sigma{R}_\Sigma\text{vol}_\Sigma\,=\,\frac{1}{4\pi}2gp\left(1-p^2\right)\Delta{z}\,=\,\frac{1}{\ell}\,.
\end{equation}
This is a natural result for a disk in an $\mathbb{R}^2/\mathbb{Z}_\ell$ orbifold.

We perform the flux quantization of the $U(1)$ gauge field, $A$,
\begin{equation} \label{gf5}
\mathfrak{p}\,\equiv\,\text{hol}_{\partial\Sigma}\left(A\right)\,=\,\frac{g}{2\pi}\oint_{y=0}A\,=\,-\frac{g}{2\pi}\int_\Sigma{F}\,=\,\frac{g\Delta{z}}{2\pi}p^3\,.
\end{equation}
By solving \eqref{ldz5} and \eqref{gf5}, $p$ and $\Delta{z}$ can be expressed in terms of $\ell$ and $\mathfrak{p}$, \cite{Boisvert:2024jrl},
\begin{equation}
p\,=\,\frac{\sqrt{\mathfrak{p}\ell}}{\sqrt{1+\mathfrak{p}\ell}}\,, \qquad \Delta{z}\,=\,\frac{2\pi}{g\ell}\frac{\left(1+\mathfrak{p}\ell\right)^{3/2}}{\sqrt{\mathfrak{p}\ell}}\,.
\end{equation}

\subsubsection{Uplift to type II supergravity}

By employing the uplift formula to the NSNS sector of type IIA and IIB supergravity, $e.g.$, in \cite{Cvetic:2000dm} and in (4.8) and (4.9) of \cite{Boisvert:2024jrl}, we obtain the uplifted metric in the string frame and the dilaton, respectively,{\footnote{We suspect a typographical error in (4.9) of \cite{Boisvert:2024jrl}: $e^{2\Phi}$ should be $e^\Phi$.}}
\begin{align}
ds_\text{NS5}^2\,=&\,p^2\left[ds_{1,3}^2+d\rho^2+\frac{1}{4g^2y(y+1)h(y)}dy^2+h(y)dz^2\right] \notag \\
& +\frac{1}{g^2}\left[d\eta^2+\frac{(y+1)\cos^2\eta{D}\xi_1^2+y\sin^2\eta{d}\xi_2^2}{y\cos^2\eta+(y+1)\sin^2\eta}\right]\,, \notag \\
e^{\Phi_\text{NS5}}\,=&\,\frac{p^5e^{-2g\rho}}{y\cos^2\eta+(y+1)\sin^2\eta}\,,
\end{align}
where we have
\begin{equation}
D\xi_1\,=\,d\xi_1-gA\,.
\end{equation}
The NSNS three-form flux is given by
\begin{align}
H_{(3)}\,=&\,\left[-\frac{y(y+1)\sin(2\eta)}{g^2\left(y+\sin^2\eta\right)^2}d\eta+\frac{\sin^2(2\eta)}{4g^2\left(y+\sin^2\eta\right)^2}dy\right]\wedge{D}\xi_1\wedge{d}\xi_2 \notag \\
&-\frac{p\sin^2\eta}{g(y+1)\left(y+\sin^2\eta\right)}dy\wedge{d}z\wedge{d}\xi_2\,,
\end{align}
and it can also be obtained from the two-form potential by $H_{(3)}=dB_{(2)}$,
\begin{equation}
B_{(2)}\,=\,\frac{1}{2g^2\left(y+\sin^2\eta\right)}\Big[y\cos^2\eta\,d\xi_1\wedge{d}\xi_2-(y+1)\sin^2\eta\left(d\xi_1-2gA\right)\wedge{d}\xi_2\Big]\,.
\end{equation}

When the NS5-brane disk solutions are uplifted to type IIB supergravity, D5-brane solutions can be obtained by S-dualization,
\begin{align}
ds^2_{\text{D5}}\,=&\,e^{-\Phi_{\text{NS5}}}ds^2_{\text{NS5}}\,, \notag \\
\Phi_{\text{D5}}\,=&\,-\Phi_{\text{NS5}}\,, \notag \\
*_{\text{D5}}F_{(3)}\,=&\,e^{-2\Phi_{\text{NS5}}}*_{\text{NS5}}H_{(3)}\,,
\end{align} 
where $ds^2_{\text{D5}}$, $\Phi_{\text{D5}}$, and $F_{(3)}$ are the metric, dilaton, and RR three-form flux of D5-brane solutions, respectively.

We present the metric in the string frame and the dilaton of D5-brane solutions, respectively,
\begin{align}
ds_\text{D5}^2\,=\,\frac{y\cos^2\eta+\left(y+1\right)\sin^2\eta}{p^5e^{-2g\rho}}&\left\{p^2\left[ds_{1,3}^2+d\rho^2+\frac{1}{4g^2y(y+1)h(y)}dy^2+h(y)dz^2\right]\right. \notag \\
&\left.+\frac{1}{g^2}\left[d\eta^2+\frac{(y+1)\cos^2\eta{D}\xi_1^2+y\sin^2\eta{d}\xi_2^2}{y\cos^2\eta+(y+1)\sin^2\eta}\right]\right\}\,, \notag \\
e^{\Phi_\text{D5}}\,=\,\frac{y\cos^2\eta+(y+1)\sin^2\eta}{p^5e^{-2g\rho}}\,.&
\end{align}
We perform the global analysis of the D5-brane solutions in the following.

\begin{figure}[t] 
\begin{center}
\includegraphics[width=4.5in]{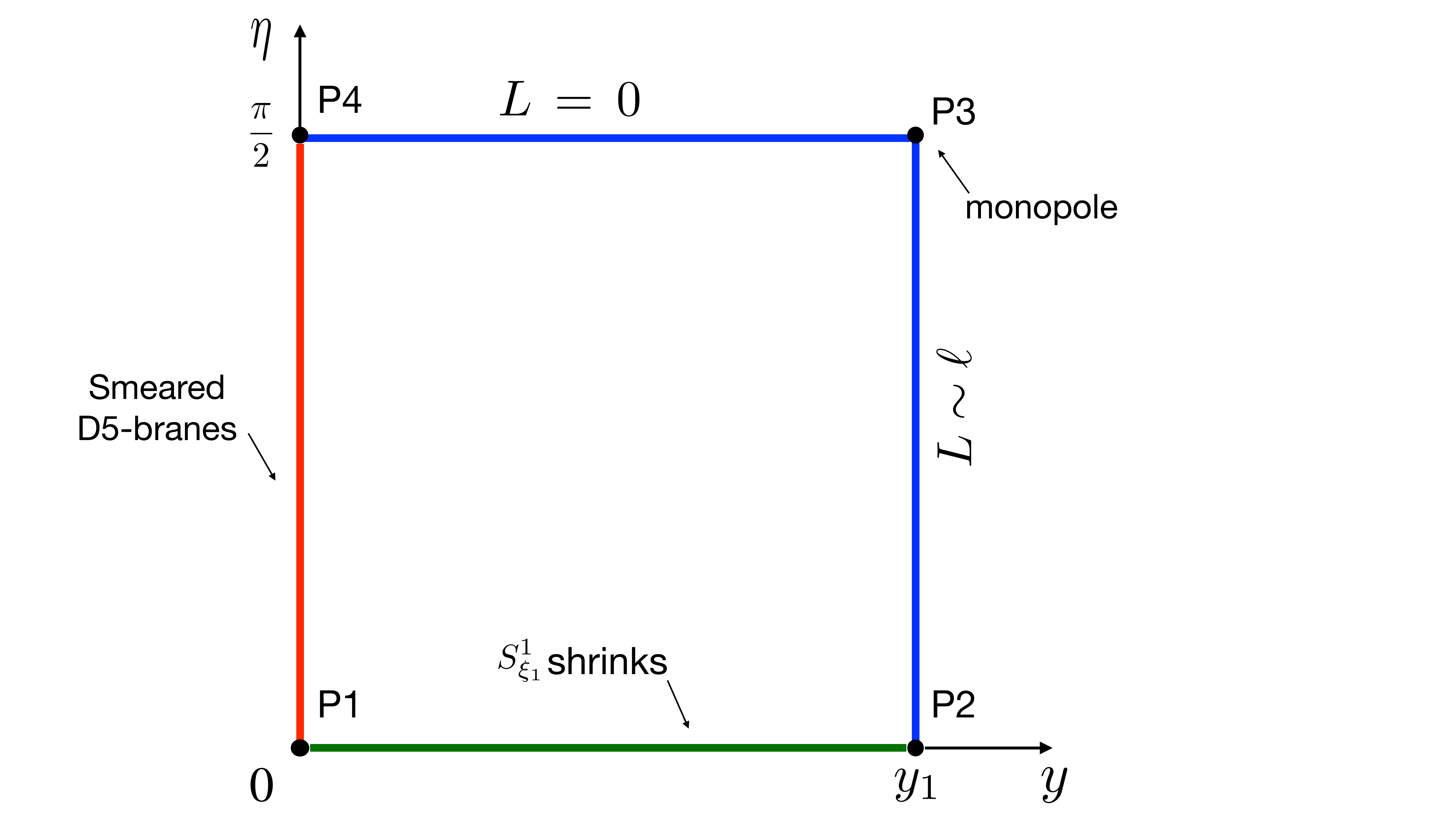}
\caption{{\it The two-dimensional base space, $B_2$, spanned by $y$ and $\xi$.}} \label{figd52}
\end{center}
\end{figure}

The five-dimensional internal space of the uplifted metric is an $S_z^1\times{S}_{\xi_1}^1\times{S}_{\xi_2}^1$ fibration over the 2d base space, $B_2$, of $(y,\eta)$. The 2d base space is a rectangle of $(y,\eta)$ over $[0,y_1]\,\times\left[0,\frac{\pi}{2}\right]$ and $\xi_i\in[0,2\pi)$. See figure \ref{figd52}. We explain the geometry of the internal space by three regions of the 2d base space, $B_2$.

\begin{itemize}
\item Region I: The side of $\mathsf{P}_1\mathsf{P}_2$.
\item Region II: The sides of $\mathsf{P}_2\mathsf{P}_3$ and $\mathsf{P}_3\mathsf{P}_4$.
\item Region III: The side of $\mathsf{P}_1\mathsf{P}_4$.
\end{itemize}

\noindent {\bf Region I:} On the side of $\eta\,=\,0$, the circle, $S_{\xi_2}^1$, shrinks and the internal space caps off.

\bigskip

\noindent {\bf Region II: Monopole} We break $D\xi_1$ and complete the square of $dz$ to obtain the metric of
\begin{align}
ds_\text{D5}^2\,=\,\frac{y\cos^2\eta+\left(y+1\right)\sin^2\eta}{p^5e^{-2g\rho}}&\left\{p^2\left[ds_{1,3}^2+d\rho^2+\frac{1}{4g^2y(y+1)h(y)}dy^2\right]\right. \notag \\
& +\frac{1}{g^2}\left[d\eta^2+\frac{y\sin^2\eta{d}\xi_2^2}{y\cos^2\eta+(y+1)\sin^2\eta}\right] \notag \\
&+R_z^2\left(dz+Ld\xi_1\right)^2+R_{\xi_1}^2d\xi_1^2\Big\}\,.
\end{align}
The metric functions are defined to be
\begin{align}
R_z^2\,=&\,\frac{p^2\left[\left(1+p^2\right)+\left(1-p^2\right)\left(\cos\left(2\eta\right)-2y\right)\right]}{2\left(y+\sin^2\eta\right)}\,, \notag \\
R_{\xi_1}^2\,=&\,\frac{2\left(p^2-\left(1-p^2\right)y\right)\cos^2\eta}{g^2\left[\left(1+p^2\right)+\left(1-p^2\right)\left(\cos\left(2\eta\right)-2y\right)\right]}\,, \notag \\
L\,=&\,\frac{2\cos^2\eta}{gp\left[\left(1+p^2\right)+\left(1-p^2\right)\left(\cos\left(2\eta\right)-2y\right)\right]}\,.
\end{align}

The function, $L(y,\eta)$, is piecewise constant along the sides of $y\,=\,y_1$ and $\eta\,=\,\frac{\pi}{2}$ of the 2d base, $B_2$,
\begin{equation}
L\left(y,\frac{\pi}{2}\right)\,=\,0\,, \qquad L\left(y_1,\eta\right)\,=\,\frac{1}{gp\left(1-p^2\right)}\,=\,\frac{\ell\Delta{z}}{2\pi}\,,
\end{equation}
The jump in $L$ at the corner, $(y,\eta)\,=\,\left(y_1,\frac{\pi}{2}\right)$, indicates the existence of a monopole source for the $Dz$ fibration. Note that, to observe the monopole source, the gauge choice  for the gauge field is 0 in \eqref{d5sol}.

\bigskip

\noindent {\bf Region III: Smeared D5-branes} We consider the singularity at $y\rightarrow0$ of the uplifted metric. As $y\rightarrow0$, the uplifted metric becomes
\begin{equation}
ds_\text{D5}^2\,\approx\,\frac{\sin^2\eta}{p^5e^{-2g\rho}}\left\{p^2\Big[ds_{1,3}^2+d\rho^2+p^2dz^2\Big]+\frac{1}{g^2}\left[\frac{1}{p^2}dr^2+r^2d\xi_2^2+d\eta^2+\cot^2\eta{D}\xi_1^2\right]\right\}\,,
\end{equation}
and the dilaton is
\begin{equation}
e^{\Phi_\text{D5}}\,\approx\,\frac{\sin^2\eta}{p^5e^{-2g\rho}}\,.
\end{equation}
where we introduced $r\,\equiv\,y^{1/2}$. The metric implies the smeared D5-brane sources. The D5-branes are
\begin{itemize}
\item extended along the $\mathbb{R}^{1,3}$, $\rho$, and $z$ directions;
\item localized at the origin of $r$ and $\xi_2$ directions;
\item smeared along the $\eta$ and $\xi_1$ directions. 
\end{itemize}
The $r^0$ factor of the metric and the dilaton match the smeared branes reviewed in appendix \ref{appA}.

\subsection{Equal charge solution:  $A^1=A^2\ne0$: $y\in[y_1,\infty)$}

\subsubsection{$D=7$ gauged supergravity}

\begin{figure}[t] 
\begin{center}
\includegraphics[width=2.0in]{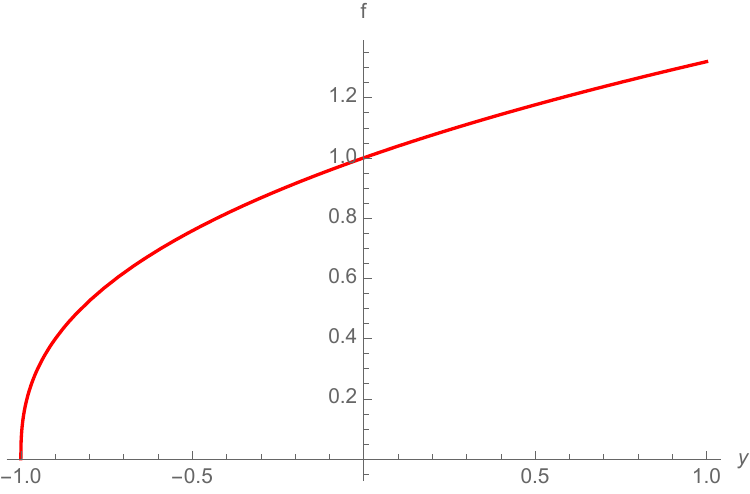} \qquad \includegraphics[width=2.0in]{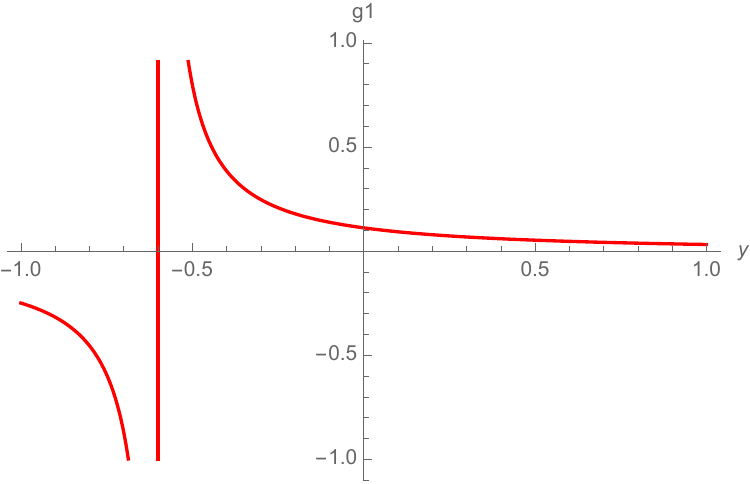} \qquad \includegraphics[width=2.0in]{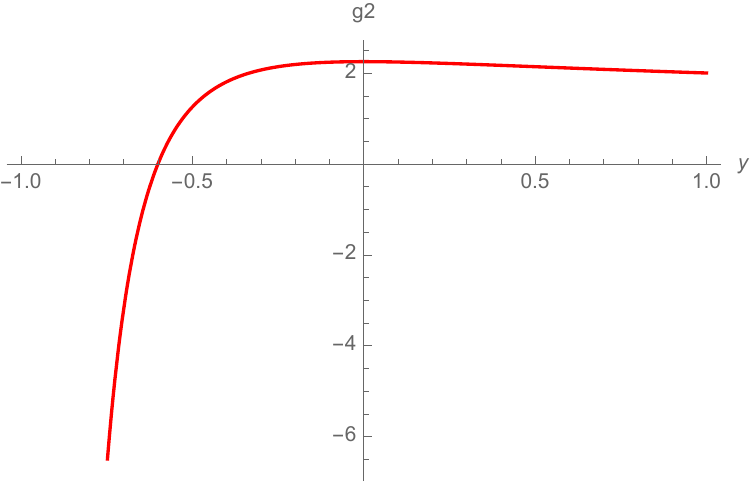}
\caption{{\it A representative solution with $g=1$ and $p=1.5$. The solution is regular in the range of $y_1=-0.6<\,y<\,\infty$.}} \label{d51two}
\end{center}
\end{figure}

In this section, we study equal charge disk solutions: $A^1=A^2\ne0$. We set the parameters, $q_1=q_2=1$, for the spindle solution given in (4.26) of \cite{Boisvert:2024jrl}. Then the solution is given by
\begin{align} \label{d5soltwo}
ds_7^2\,=&\,e^{\frac{4g}{5}\rho}(y+1)^{2/5}\left[ds_{1,3}^2+d\rho^2+\frac{1}{4g^2\left(y+1\right)^2h(y)}dy^2+h(y)dz^2\right]\,, \notag \\
A\,\equiv&\,A^1\,=\,A^2\,=\,\frac{p}{y+1}dz\,, \notag \\
e^{\lambda_1+\frac{g}{5}\rho}\,=&\,e^{\lambda_1+\frac{g}{5}\rho}\,=\,\frac{p^{1/2}}{(y+1)^{1/10}}\,,
\end{align}
where $ds_{1,3}^2$ is the metric on $\mathbb{R}^{1,3}$ and 
\begin{equation}
h(y)\,=\,\frac{p^2\left(y+1\right)^2-y^2}{\left(y+1\right)^2}\,.
\end{equation}
For $h(y)=0$, there are two roots, $y_1\,\equiv\,-\frac{p}{p+1}$ and $y_2\,\equiv\,-\frac{p}{p-1}$. It appears that, for $p<g$, we find spindle solutions and, for $p>g$, disk solutions with
\begin{equation}
y_a<y<\infty\,,
\end{equation}
where $a=1,2$. We plot a representative solution with $g=1$ and $p=1.5$ in figure \ref{d51two}. The metric functions, $f(y)$, $g_1(y)$, and $g_2(y)$, are defined by
\begin{equation}
ds_7^2\,=\,e^{\frac{4g}{5}\rho}f(y)\left[ds_{1,3}^2+d\rho^2+g_1(y)dy^2+g_2(y)dz^2\right]\,.
\end{equation}

Near $y\rightarrow\infty$ the warp factor vanishes and it is a curvature singularity of the metric, 
\begin{equation}
ds_7^2\,\approx\,e^{\frac{4g}{5}\rho}y^{2/5}\left[ds_{1,3}^2+d\rho^2+\frac{1}{4g^2\left(p^2-1\right)y^2}dy^2+\left(p^2-1\right)dz^2\right]\,.
\end{equation}
Approaching $y\rightarrow{y}_a$, the metric becomes to be
\begin{equation}
ds_7^2\,\approx\,e^{\frac{4g}{5}\rho}y_1^{1/5}\left[ds_{1,3}^2+d\rho^2+\frac{d\tilde{\rho}^2+\mathcal{E}(p)^2\tilde{\rho}^2dz^2}{-g^2(y_a+1)^2h'(y_a)}\right]\,,
\end{equation}
where we have
\begin{equation}
\mathcal{E}(p)^2\,\equiv\,g^2(y_a+1)^2h'(y_a)^2\,=\,\Big(2gp\left(1\pm{p}\right)\Big)^2\,,
\end{equation}
and we introduced a new parametrization of coordinate, $\tilde{\rho}^2\,=\,y_a-y$. Then,  the $\tilde{\rho}-z$ surface is locally an $\mathbb{R}^2/\mathbb{Z}_\ell$ orbifold if we set
\begin{equation} \label{ldz5two}
\frac{\ell\Delta{z}}{2\pi}\,=\,\frac{1}{2gp\left(1\pm{p}\right)}\,,
\end{equation}
where $\Delta{z}$ is the period of coordinate, $z$, and $\ell\,=\,1,2,3,\ldots$. The metric spanned by $(y,z)$ has a topology of disk, $\Sigma$, with the center at $y=y_a$ and the boundary at $y=0$.

We calculate the Euler characteristic of the $y-z$ surface, $\Sigma$, 
\begin{equation}
\chi\,=\,\frac{1}{4\pi}\int_\Sigma{R}_\Sigma\text{vol}_\Sigma\,=\,\frac{1}{4\pi}4gp\left(1\pm{p}\right)\Delta{z}\,=\,\frac{1}{\ell}\,.
\end{equation}
This is a natural result for a disk in an $\mathbb{R}^2/\mathbb{Z}_\ell$ orbifold.

We perform the flux quantization of the $U(1)$ gauge field, $A$,
\begin{equation} \label{gf5two}
\mathfrak{p}\,\equiv\,\text{hol}_{\partial\Sigma}\left(A\right)\,=\,\frac{g}{2\pi}\oint_{y=\infty}A\,=\,-\frac{g}{2\pi}\int_\Sigma{F}\,=\,-\frac{g\Delta{z}}{2\pi}p(1\pm{p})\,.
\end{equation}
Unlike any other solutions we study in this work, $\Delta{z}\,p(1\pm{p})$ is common in both of \eqref{ldz5two} and \eqref{gf5two}, and one cannot solve for $p$ and $\Delta{z}$ in terms of $\ell$ and $\mathfrak{p}$.

\subsubsection{Uplift to type II supergravity}

By employing the uplift formula to the NSNS sector of type IIA and IIB supergravity, $e.g.$, in \cite{Cvetic:2000dm} and in (4.8) and (4.9) of \cite{Boisvert:2024jrl}, we obtain the uplifted metric in the string frame and the dilaton, respectively,
\begin{align}
ds_\text{NS5}^2\,=&\,p^2\left[ds_{1,3}^2+d\rho^2+\frac{1}{4g^2(y+1)^2h(y)}dy^2+h(y)dz^2\right] \notag \\
& +\frac{1}{g^2}\left[d\eta^2+\cos^2\eta\,D\xi_1^2+\sin^2\eta\,D\xi_2^2\right]\,, \notag \\
e^{\Phi_\text{NS5}}\,=&\,\frac{p^5e^{-2g\rho}}{y+1}\,,
\end{align}
where we have
\begin{equation}
D\xi_1\,=\,d\xi_1-gA\,, \qquad D\xi_2\,=\,d\xi_2-gA\,.
\end{equation}
We do not present the NSNS three-form flux.

When the NS5-brane disk solutions are uplifted to type IIB supergravity, D5-brane solutions can be obtained by S-dualization,
\begin{align}
ds^2_\text{D5}\,=&\,e^{-\Phi_\text{NS5}}ds^2_\text{NS5}\,, \notag \\
\Phi_\text{D5}\,=&\,-\Phi_\text{NS5}\,, \notag \\
*_\text{D5}F_{(3)}\,=&\,e^{-2\Phi_\text{NS5}}*_\text{NS5}H_{(3)}\,,
\end{align} 
where $ds^2_{\text{D5}}$, $\Phi_{\text{D5}}$, and $F_{(3)}$ are the metric, dilaton, and RR three-form flux of D5-brane solutions, respectively.

We present the metric in the string frame and the dilaton of D5-brane solutions, respectively,
\begin{align}
ds_\text{D5}^2\,=&\,\frac{y+1}{p^5e^{-2g\rho}}\left\{p^2\left[ds_{1,3}^2+d\rho^2+\frac{1}{4g^2(y+1)^2h(y)}dy^2+h(y)dz^2\right]\right. \notag \\
& \,\,\,\,\,\,\,\,\,\,\,\,\,\,\,\,\,\,\,\,\,\,\,\,\,\,\,\,\,\,\,\,\, \left.+\frac{1}{g^2}\left[d\eta^2+\cos^2\eta\,D\xi_1^2+\sin^2\eta\,D\xi_2^2\right]\right\}\,, \notag \\
e^{\Phi_\text{D5}}\,=&\,\frac{y+1}{p^5e^{-2g\rho}}\,,
\end{align}
We perform the global analysis of the D5-brane solutions in the following.

The five-dimensional internal space of the uplifted metric is an $S_z^1\times{S}_{\xi_1}^1\times{S}_{\xi_2}^1$ fibration over the 2d base space, $B_2$, of $(y,\eta)$. The 2d base space is a rectangle of $(y,\eta)$ over $[y_a,\infty)\,\times\left[0,\frac{\pi}{2}\right]$ and $\xi_i\in[0,2\pi)$. However, unlike any other solutions we study in this work, there is no monopole structure in the solution. Thus we only present the geometry near the boundary, $y\rightarrow\infty$.

\bigskip

\noindent {\bf Un-smeared D5-branes} We consider the singularity at $y\rightarrow\infty$ of the uplifted metric. As $y\rightarrow\infty$, the uplifted metric becomes
\begin{align} \label{usd5}
ds_\text{D5}^2\,\approx\,&\left.\frac{1}{p^5e^{-2g\rho}}\right\{p^2y\Big[ds_{1,3}^2+d\rho^2+\left(p^2-1\right)dz^2\Big] \notag \\
& \,\,\,\,\,\,\,\,\,\,\,\,\,\,\,  \left.+\frac{1}{4g^2\left(p^2-1\right)y}\Big[dy^2+4\left(p^2-1\right)y^2\left(d\eta^2+\cos^2\eta\,D\xi_1^2+\sin^2\eta\,D\xi_2^2\right)\Big]\right\}\,,
\end{align}
and the dilaton is
\begin{equation}
e^{\Phi_\text{D5}}\,\approx\,\frac{y}{p^5e^{-2g\rho}}\,.
\end{equation}
The metric does not get smeared by D5-brane sources. The D5-branes are
\begin{itemize}
\item extended along the $\mathbb{R}^{1,3}$, $\rho$, and $z$ directions;
\item localized at the origin of $y$, $\xi_1$, and $\xi_2$ directions;
\item not smeared along any directions. 
\end{itemize}
The $y$ and $1/y$ factors of the metric and the dilaton match the un-smeared branes reviewed in appendix \ref{appA}.

\subsection{Multi charge solution:  $A^1\ne{A}^2$: $y\in[y_1,\infty)$}

\subsubsection{$D=7$ gauged supergravity}

\begin{figure}[t] 
\begin{center}
\includegraphics[width=2.0in]{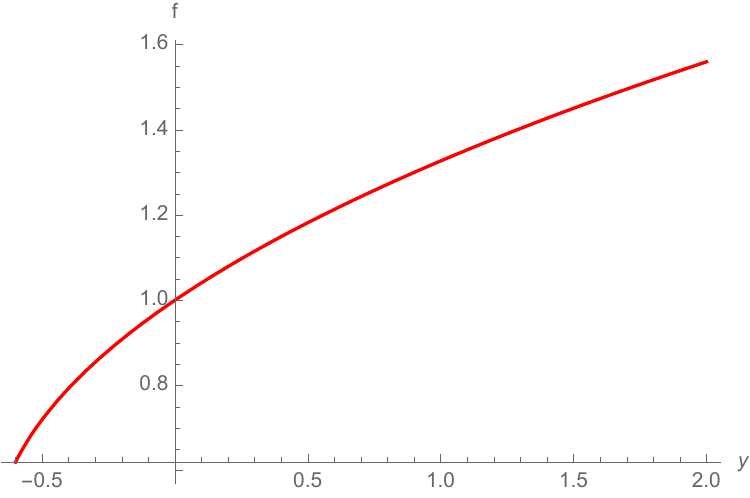} \qquad \includegraphics[width=2.0in]{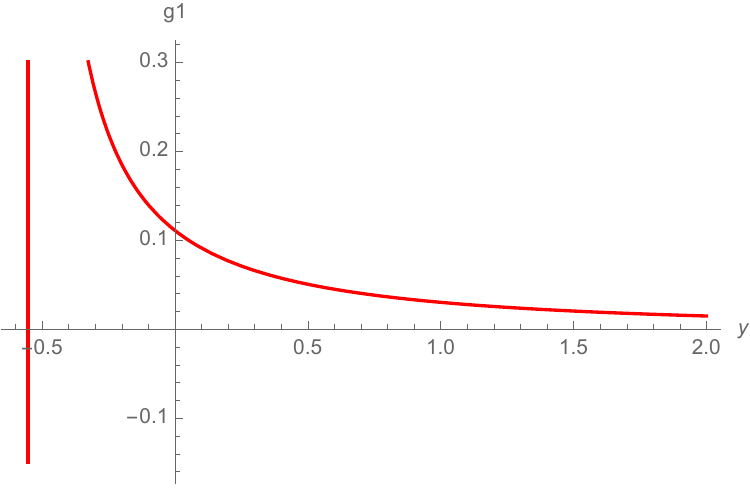} \qquad \includegraphics[width=2.0in]{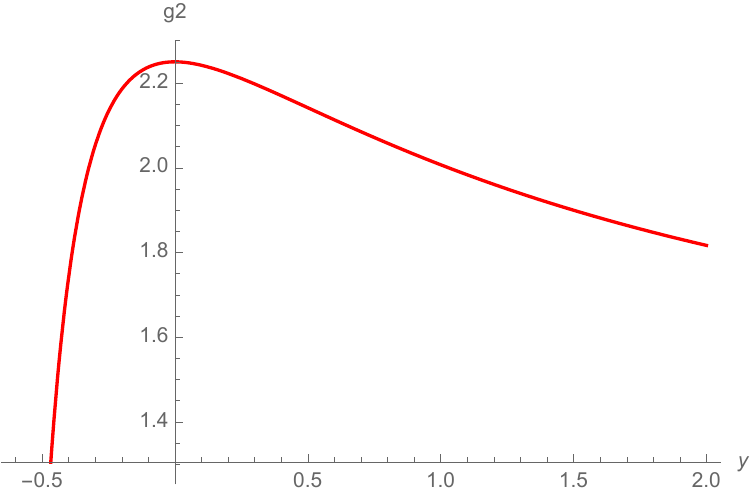}
\caption{{\it A representative solution with $g=1$, $p=1.5$, and $q=1.4$. The solution is regular in the range of $y_1=-0.553<\,y<\,\infty$.}} \label{d51three}
\end{center}
\end{figure}

In this section, we study multi charge disk solutions: $A^1\ne{A}^2$. Using the scaling symmetry, we set $q_1q_2=1$ and choose the parameters, $q_1=q$ and $q_2=q^{-1}$, for the spindle solution given in (4.26) of \cite{Boisvert:2024jrl}.{\footnote{See around (4.30) of \cite{Boisvert:2024jrl} for more on this parametrization.}} Thus two gauge fields are parametrized by $p$ and $q$. Then the solution is given by
\begin{align} \label{d5solthree}
ds_7^2\,=&\,e^{\frac{4g}{5}\rho}(y+q)^{1/5}(y+q^{-1})^{1/5}\left[ds_{1,3}^2+d\rho^2+\frac{1}{4g^2\left(y+q\right)\left(y+q^{-1}\right)h(y)}dy^2+h(y)dz^2\right]\,, \notag \\
A^1\,=&\,\frac{pq}{y+q}dz\,, \qquad A^2\,=\,\frac{pq^{-1}}{y+q^{-1}}dz\,, \notag \\
e^{\lambda_1+\frac{g}{5}\rho}\,=&\,\frac{p^{1/2}\left(y+q^{-1}\right)^{1/5}}{(y+q)^{3/10}}\,, \qquad e^{\lambda_2+\frac{g}{5}\rho}\,=\,\frac{p^{1/2}\left(y+q\right)^{1/5}}{(y+q^{-1})^{3/10}}\,,
\end{align}
where $ds_{1,3}^2$ is the metric on $\mathbb{R}^{1,3}$ and 
\begin{equation}
h(y)\,=\,\frac{p^2\left(y+q\right)\left(y+q^{-1}\right)-y^2}{\left(y+q\right)\left(y+q^{-1}\right)}\,.
\end{equation}
For $h(y)=0$, there are two roots,
\begin{equation}
y_1\,=\,\frac{p\left[p\left(1+q^2\right)-\sqrt{p^2\left(1-q^2\right)+4q^2}\right]}{2q\left(1-p^2\right)}\,, \qquad y_2\,=\,\frac{p\left[p\left(1+q^2\right)+\sqrt{p^2\left(1-q^2\right)+4q^2}\right]}{2q\left(1-p^2\right)}\,.
\end{equation}
We find disk solutions with
\begin{equation}
y_a<y<\infty\,,
\end{equation}
where $a=1,2$. We plot a representative solution with $g=1$, $p=1.5$, and $q=1.4$ in figure \ref{d51three}. The metric functions, $f(y)$, $g_1(y)$, and $g_2(y)$, are defined by
\begin{equation}
ds_7^2\,=\,e^{\frac{4g}{5}\rho}f(y)\left[ds_{1,3}^2+d\rho^2+g_1(y)dy^2+g_2(y)dz^2\right]\,.
\end{equation}

Near $y\rightarrow\infty$ the warp factor vanishes and it is a curvature singularity of the metric, 
\begin{equation}
ds_7^2\,\approx\,e^{\frac{4g}{5}\rho}y^{2/5}\left[ds_{1,3}^2+d\rho^2+\frac{1}{4g^2\left(p^2-1\right)y^2}dy^2+\left(p^2-1\right)dz^2\right]\,.
\end{equation}
Approaching $y\rightarrow{y}_a$, the metric becomes to be
\begin{equation}
ds_7^2\,\approx\,e^{\frac{4g}{5}\rho}y_1^{1/5}\left[ds_{1,3}^2+d\rho^2+\frac{d\tilde{\rho}^2+\mathcal{E}(p,q)^2\tilde{\rho}^2dz^2}{-g^2\left(y_a+q\right)\left(y_a+q^{-1}\right)h'(y_a)}\right]\,,
\end{equation}
where we have
\begin{equation}
\mathcal{E}(p,q)^2\,\equiv\,g^2\left(y_a+q\right)\left(y_a+q^{-1}\right)h'(y_a)^2\,=\,\left(\frac{gy\left[\left(q+q^{-1}\right)y+2\right]}{\left(y+q\right)^{3/2}\left(y+q^{-1}\right)^{3/2}}\right)^2\,,
\end{equation}
and we introduced a new parametrization of coordinate, $\tilde{\rho}^2\,=\,y_a-y$. Then,  the $\tilde{\rho}-z$ surface is locally an $\mathbb{R}^2/\mathbb{Z}_\ell$ orbifold if we set
\begin{equation} \label{ldz5two}
\frac{\ell\Delta{z}}{2\pi}\,=\,\frac{1}{\mathcal{E}(p,q)}\,,
\end{equation}
where $\Delta{z}$ is the period of coordinate, $z$, and $\ell\,=\,1,2,3,\ldots$. The metric spanned by $(y,z)$ has a topology of disk, $\Sigma$, with the center at $y=y_a$ and the boundary at $y=0$.

We calculate the Euler characteristic of the $y-z$ surface, $\Sigma$, 
\begin{equation}
\chi\,=\,\frac{1}{4\pi}\int_\Sigma{R}_\Sigma\text{vol}_\Sigma\,=\,\frac{1}{4\pi}2\mathcal{E}(p,q)\Delta{z}\,=\,\frac{1}{\ell}\,.
\end{equation}
This is a natural result for a disk in an $\mathbb{R}^2/\mathbb{Z}_\ell$ orbifold.

\subsubsection{Uplift to type II supergravity}

By employing the uplift formula to the NSNS sector of type IIA and IIB supergravity, $e.g.$, in \cite{Cvetic:2000dm} and in (4.8) and (4.9) of \cite{Boisvert:2024jrl}, we obtain the uplifted metric in the string frame and the dilaton, respectively,
\begin{align}
ds_\text{NS5}^2\,=&\,p^2\left[ds_{1,3}^2+d\rho^2+\frac{1}{4g^2\left(y_a+q\right)\left(y_a+q^{-1}\right)h(y)}dy^2+h(y)dz^2\right] \notag \\
& +\frac{1}{g^2}\left[d\eta^2+\frac{\left(y+q\right)\cos^2\eta\,D\xi_1^2+\left(y+q^{-1}\right)\sin^2\eta\,D\xi_2^2}{\left(y+q^{-1}\right)\cos^2\eta+\left(y+q\right)\sin^2\eta}\right]\,, \notag \\
e^{\Phi_\text{NS5}}\,=&\,\frac{p^5e^{-2g\rho}}{\left(y+q^{-1}\right)\cos^2\eta+\left(y+q\right)\sin^2\eta}\,,
\end{align}
where we have
\begin{equation}
D\xi_1\,=\,d\xi_1-gA^1\,, \qquad D\xi_2\,=\,d\xi_2-gA^2\,.
\end{equation}
We do not present the NSNS three-form flux.

When the NS5-brane disk solutions are uplifted to type IIB supergravity, D5-brane solutions can be obtained by S-dualization,
\begin{align}
ds^2_{\text{D5}}\,=&\,e^{-\Phi_{\text{NS5}}}ds^2_{\text{NS5}}\,, \notag \\
\Phi_{\text{D5}}\,=&\,-\Phi_{\text{NS5}}\,, \notag \\
*_{\text{D5}}F_{(3)}\,=&\,e^{-2\Phi_{\text{NS5}}}*_{\text{NS5}}H_{(3)}\,,
\end{align} 
where $ds^2_{\text{D5}}$, $\Phi_{\text{D5}}$, and $F_{(3)}$ are the metric, dilaton, and RR three-form flux of D5-brane solutions, respectively.

We present the metric in the string frame and the dilaton of D5-brane solutions, respectively,
\begin{align}
ds_\text{D5}^2\,=&\,\frac{\left(y+q^{-1}\right)\cos^2\eta+\left(y+q\right)\sin^2\eta}{p^5e^{-2g\rho}} \notag \\
&\times\left\{p^2\left[ds_{1,3}^2+d\rho^2+\frac{1}{4g^2\left(y_a+q\right)\left(y_a+q^{-1}\right)h(y)}dy^2+h(y)dz^2\right]\right. \notag \\
& \,\,\,\,\,\,\,\, \left.+\frac{1}{g^2}\left[d\eta^2+\frac{\left(y+q\right)\cos^2\eta\,D\xi_1^2+\left(y+q^{-1}\right)\sin^2\eta\,D\xi_2^2}{\left(y+q^{-1}\right)\cos^2\eta+\left(y+q\right)\sin^2\eta}\right]\right\}\,, \notag \\
e^{\Phi_\text{D5}}\,=&\,\frac{\left(y+q^{-1}\right)\cos^2\eta+\left(y+q\right)\sin^2\eta}{p^5e^{-2g\rho}}\,,
\end{align}
We perform the global analysis of the D5-brane solutions in the following.

The five-dimensional internal space of the uplifted metric is an $S_z^1\times{S}_{\xi_1}^1\times{S}_{\xi_2}^1$ fibration over the 2d base space, $B_2$, of $(y,\eta)$. The 2d base space is a rectangle of $(y,\eta)$ over $[y_a,\infty)\,\times\left[0,\frac{\pi}{2}\right]$ and $\xi_i\in[0,2\pi)$. However, unlike any other solutions we study in this work, there is no monopole structure in the solution. Furthermore, the geometry near the boundary, $y\rightarrow\infty$, reduces to un-smeared D5-branes, \eqref{usd5}.

\vspace{1.8cm}

\section{D4-branes wrapped on a disk}  \label{d4}

\subsection{No charge solution: $A^1=A^2=0$: $y\in[y_1,\infty)$}

\subsubsection{$D=6$ gauged supergravity}

We consider $D=6$ $SO(5)$-gauged maximal supergravity, \cite{Cowdall:1998rs}, which is obtained from the dimensional reduction of $D=7$ $SO(5)$-gauged maximal supergravity, \cite{Pernici:1984xx}, on a circle. The theory is also the reduction of type IIA supergravity on a four-sphere. In particular, the $U(1)^2$ subsector of the theory is of our interest and is obtained from the reduction of $D=7$ $U(1)^2$-gauged supergravity, \cite{Liu:1999ai}, on a circle. The field content is the metric, two $U(1)$ gauge fields, $A^1$, $A^2$, and three real scalar fields, $\lambda_1$, $\lambda_2$, $\sigma$. For details of $D=6$ $U(1)^2$-gauged supergravity, we refer section 5 of \cite{Boisvert:2024jrl}.

\begin{figure}[t] 
\begin{center}
\includegraphics[width=2.0in]{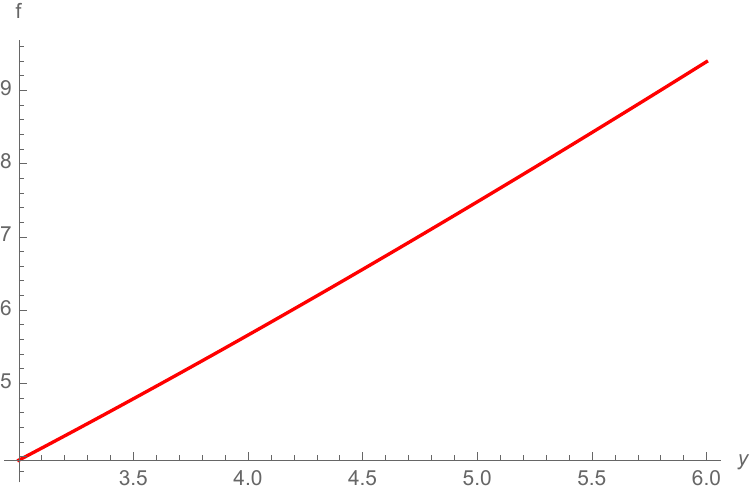} \qquad \includegraphics[width=2.0in]{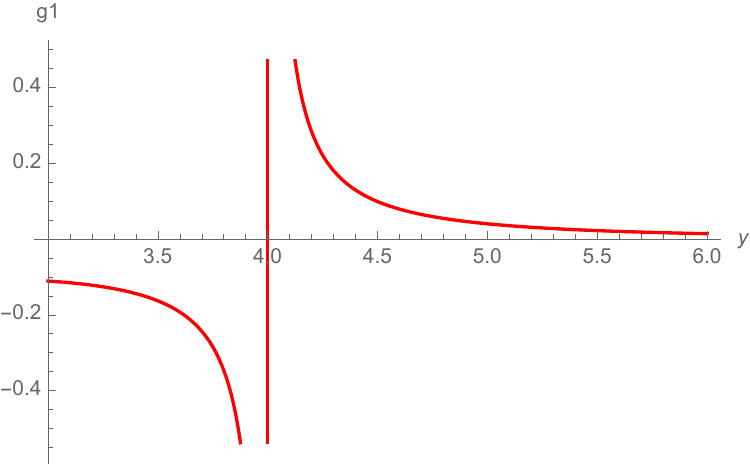} \qquad \includegraphics[width=2.0in]{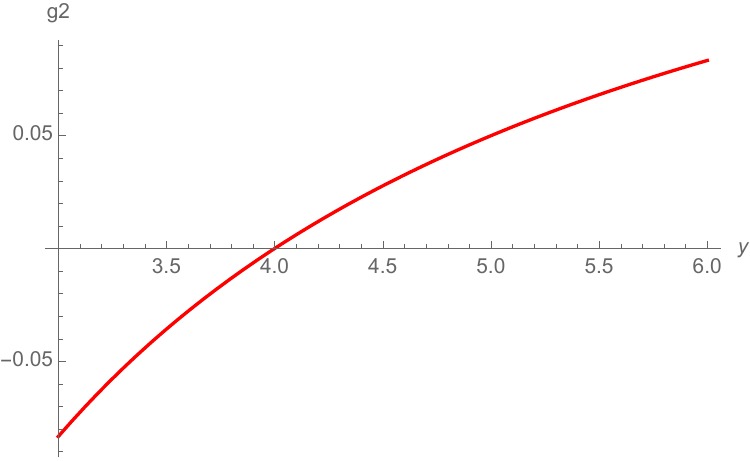}
\caption{{\it A representative solution with $g=1$ and $q=1$. The solution is regular in the range of $y_1=4<y<\infty$.}} \label{d40}
\end{center}
\end{figure}

In this section, we study no charge disk solutions, $A^1=A^2=0$. We set the parameters, $q_1=q_2=0$, for the spindle solution given in (5.20) of \cite{Boisvert:2024jrl}. Then the solution is given by
\begin{align} \label{d4sol1}
ds_6^2\,=&\,\frac{y^{5/4}}{r^{1/2}}\left[\frac{ds_{1,2}^2+dr^2}{r^2}+\frac{1}{4y^3h(y)}dy^2+h(y)dz^2\right]\,, \notag \\
e^{2\lambda_1}\,=&\,e^{2\lambda_2}\,=\,1\,, \qquad e^{8\sigma}\,=\,\frac{r}{y^{1/2}}\,, \notag \\
A^1\,=&\,A^2\,=\,0\,,
\end{align}
where $ds_{1,2}^2$ is the metric on $\mathbb{R}^{1,2}$ and 
\begin{equation}
h(y)\,=\,\frac{g^2\left(y-\frac{4}{g^2}\right)}{4y}\,.
\end{equation}
For $h(y)=0$, there is a root, $y\,=\,\frac{4}{g^2}$. We find solutions with
\begin{equation} \label{d4solr1}
y_1\,\equiv\,\frac{4}{g^2}<y<\infty\,.
\end{equation}
We plot a representative solution with $g=1$ in figure \ref{d40}. The metric functions, $f(y)$, $g_1(y)$, and $g_2(y)$, are defined by
\begin{equation}
ds_6^2\,=\,\frac{f(y)}{r^{1/2}}\left[\frac{ds_{1,2}^2+dr^2}{r^2}+g_1(y)dy^2+g_2(y)dz^2\right]\,.
\end{equation}

Near $y\rightarrow\infty$ the warp factor vanishes and it is a curvature singularity of the metric, 
\begin{equation}
ds_6^2\,\approx\,\frac{y^{5/4}}{r^{1/2}}\left[\frac{ds_{1,2}^2+dr^2}{r^2}+\frac{1}{g^2y^3}dy^2+\frac{g^2}{4}dz^2\right]\,.
\end{equation}
Approaching $y\rightarrow{y}_1$, the metric becomes to be
\begin{equation}
ds_6^2\,\approx\,\frac{y_1^{5/4}}{r^{1/2}}\left[\frac{ds_{1,2}^2+dr^2}{r^2}+\frac{d\rho^2+\mathcal{E}(q)^2\rho^2dz^2}{-y_1^3h(y_1)}\right]\,,
\end{equation}
where we have
\begin{equation}
\mathcal{E}(q)^2\,\equiv\,\left(\frac{g}{2}\right)^2\,,
\end{equation}
and we introduced a new parametrization of coordinate, $\rho^2\,=\,y_1-y$. Then,  the $\rho-z$ surface is locally an $\mathbb{R}^2/\mathbb{Z}_\ell$ orbifold if we set
\begin{equation}
\frac{\ell\Delta{z}}{2\pi}\,=\,\frac{2}{g}\,,
\end{equation}
where $\Delta{z}$ is the period of coordinate, $z$, and $\ell\,=\,1,2,3,\ldots$. The metric spanned by $(y,z)$ has a topology of disk, $\Sigma$, with the center at $y=y_1$ and the boundary at $y=0$.

We calculate the Euler characteristic of the $y-z$ surface, $\Sigma$,
\begin{equation}
\chi\,=\,\frac{1}{4\pi}\int_\Sigma{R}_\Sigma\text{vol}_\Sigma\,=\,\frac{1}{4\pi}2\frac{g}{2}\Delta{z}\,=\,\frac{1}{\ell}\,.
\end{equation}
This is a natural result for a disk in an $\mathbb{R}^2/\mathbb{Z}_\ell$ orbifold.

\subsubsection{Uplift to type IIA supergravity}

By employing the uplift formula to type IIA supergravity, \cite{Cvetic:2000ah}, which we review in appendix \ref{appB}, we obtain the uplifted metric in the string frame and the dilaton, respectively,
\begin{align}
ds_{10}^2\,=&\,\frac{1}{r}\left[y^{3/2}\left(\frac{ds_{1,2}^2+dr^2}{r^2}+\frac{1}{4y^3h(y)}dy^2+h(y)dz^2\right)\right. \notag \\
& \,\,\,\,\,\,\,\,\,\,\,\,\,\, \left.+\frac{y^{1/2}}{g^2}\Big(d\xi^2+\sin^2\xi\,d\chi_1^2+\cos^2\xi\left(d\theta^2+\cos^2\theta{d}\chi_2^2\right)\Big)\right]\,, \notag \\
e^\Phi\,=&\,\frac{y^{3/4}}{r^{3/2}}\,.
\end{align}
Explicit form of the uplift formula for flux fields is not given in \cite{Cvetic:2000ah} and it would be interesting to obtain the full uplifted solution.

We consider the singularity at $y\rightarrow\infty$ of the uplifted metric. As $y\rightarrow\infty$, the uplifted metric becomes
\begin{align}
ds_{10}^2\,\approx&\,\frac{1}{r}\left\{y^{3/2}\left[\frac{ds_{1,2}^2+dr^2}{r^2}+\frac{g^2}{4}dz^2\right]\right. \notag \\
& \,\,\,\,\,\,\,\, \left.+\frac{1}{g^2y^{3/2}}\Big[dy^2+y^2\left(d\xi^2+\sin^2\xi\,d\chi_1^2+\cos^2\xi\left(d\theta^2+\cos^2\theta{d}\chi_2^2\right)\right)\Big]\right\}\,,
\end{align}
and the dilaton is
\begin{equation}
e^\Phi\,=\,\frac{y^{3/4}}{r^{3/2}}\,.
\end{equation}
The branes are extended along the $\mathbb{R}^{1,4}$, $r$, $z$ directions and localized at the center of $y$, $\xi$, $\chi_1$, $\chi_2$, and $\theta$ directions. We observe that there is no smeared direction. The $y^{3/2}$ and $1/y^{3/2}$ factors of the metric and the dilaton match the un-smeared branes reviewed in appendix \ref{appA}.

\subsection{Single charge solution: $A^1\ne0$, $A^2=0$: $y\in[0,y_1]$}

\subsubsection{$D=6$ gauged supergravity}

In this section, we study single charge disk solutions: $A^1\ne0$, $A^2=0$. {\footnote {Disk solutions from the D4-D8-branes which asymptote to the $AdS_6$ vacuum have been studied in \cite{Suh:2021aik, Couzens:2022lvg, Suh:2024fru}.}} We set the parameters, $q\equiv{q_1}$ and $q_2=0$, for the spindle solution given in (5.20) of \cite{Boisvert:2024jrl}. Then the solution is given by
\begin{align} \label{d4sol1}
ds_6^2\,=&\,\frac{y^{3/4}(y^2+q)^{1/4}}{r^{1/2}}\left[\frac{ds_{1,2}^2+dr^2}{r^2}+\frac{1}{4y(y^2+q)h(y)}dy^2+h(y)dz^2\right]\,, \notag \\
A\,\equiv&\,A^1\,=\,\left(\frac{q}{y^2+q}-\frac{1}{4}\right)dz\,, \qquad A^2\,=\,0\,, \notag \\
e^{2\lambda_1}\,=&\,\frac{y^{6/5}}{(y^2+q)^{3/5}}\,, \qquad e^{2\lambda_2}\,=\,\frac{(y^2+q)^{2/5}}{y^{4/5}}\,, \qquad e^{8\sigma}\,=\,\frac{r}{y^{3/10}(y^2+q)^{1/10}}\,,
\end{align}
where $ds_{1,2}^2$ is the metric on $\mathbb{R}^{1,2}$ and 
\begin{equation}
h(y)\,=\,\frac{g^2\left(y^2+q\right)-4y}{4(y^2+q)}\,.
\end{equation}
The gauge choice of $-1/4$ for the gauge field, $A$, is required to observe the monopole structure in the uplifted solutions in \eqref{d4mono1}. For $h(y)=0$, there are two roots, $y\,=\,\frac{2\pm\sqrt{4-g^4q}}{g^2}$. We consider solutions with{\footnote{There is also a solution with $y_2\equiv\frac{2+\sqrt{4-g^4q}}{g^2}<y<\infty$. We briefly discuss the solution in section \ref{d4a1}.}
\begin{equation} \label{d4solr1}
q<\frac{4}{g^4}\,, \qquad 0<y<y_1\,\equiv\,\frac{2-\sqrt{4-g^4q}}{g^2}\,.
\end{equation}
We plot a representative solution with $g=1$ and $q=1$ in figure \ref{d41}. The metric functions, $f(y)$, $g_1(y)$, and $g_2(y)$, are defined by
\begin{equation}
ds_6^2\,=\,\frac{f(y)}{r^{1/2}}\left[\frac{ds_{1,2}^2+dr^2}{r^2}+g_1(y)dy^2+g_2(y)dz^2\right]\,.
\end{equation}

\begin{figure}[t] 
\begin{center}
\includegraphics[width=2.0in]{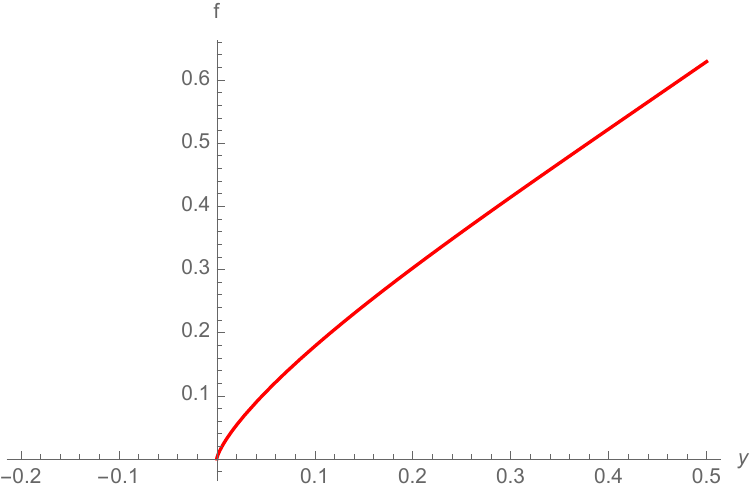} \qquad \includegraphics[width=2.0in]{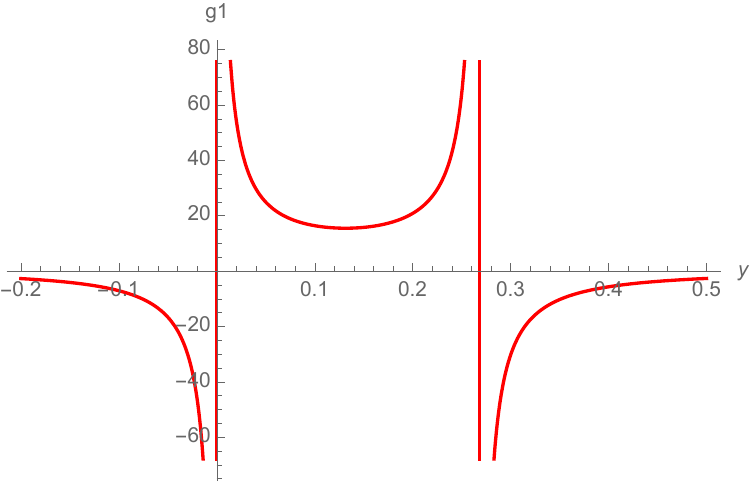} \qquad \includegraphics[width=2.0in]{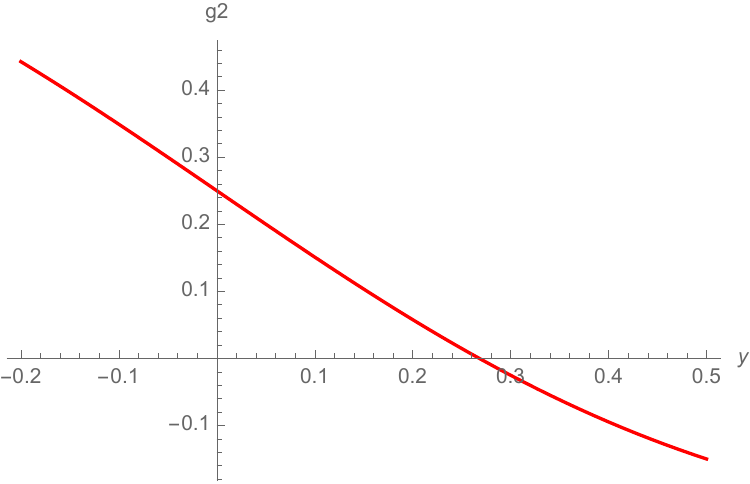}
\caption{{\it A representative solution with $g=1$ and $q=1$. The solution is regular in the range of $0<y<y_1=0.2679$.}} \label{d41}
\end{center}
\end{figure}

Near $y\rightarrow0$ the warp factor vanishes and it is a curvature singularity of the metric, 
\begin{equation}
ds_6^2\,\approx\,\frac{q^{1/4}y^{3/4}}{r^{1/2}}\left[\frac{ds_{1,2}^2+dr^2}{r^2}+\frac{1}{g^2qy}dy^2+\frac{g^2}{4}dz^2\right]\,.
\end{equation}
Approaching $y\rightarrow{y}_1$, the metric becomes to be
\begin{equation}
ds_6^2\,\approx\,\frac{y_1^{3/4}(y_1^2+q)^{1/4}}{r^{1/2}}\left[\frac{ds_{1,2}^2+dr^2}{r^2}+\frac{d\rho^2+\mathcal{E}(q)^2\rho^2dz^2}{-y_1(y_1^2+q)h(y_1)}\right]\,,
\end{equation}
where we have
\begin{equation}
\mathcal{E}(q)^2\,\equiv\,y_1(y_1^2+q)h'(y_1)^2\,=\,\frac{g\sqrt{4-g^4q}}{4}\,,
\end{equation}
and we introduced a new parametrization of coordinate, $\rho^2\,=\,y_1-y$. Then,  the $\rho-z$ surface is locally an $\mathbb{R}^2/\mathbb{Z}_\ell$ orbifold if we set
\begin{equation} \label{ldz41}
\frac{\ell\Delta{z}}{2\pi}\,=\,\frac{4}{g\sqrt{4-g^4q}}\,,
\end{equation}
where $\Delta{z}$ is the period of coordinate, $z$, and $\ell\,=\,1,2,3,\ldots$. The metric spanned by $(y,z)$ has a topology of disk, $\Sigma$, with the center at $y=y_1$ and the boundary at $y=0$.

We calculate the Euler characteristic of the $y-z$ surface, $\Sigma$,
\begin{equation}
\chi\,=\,\frac{1}{4\pi}\int_\Sigma{R}_\Sigma\text{vol}_\Sigma\,=\,\frac{1}{4\pi}\frac{g\sqrt{4-g^4q}}{2}\Delta{z}\,=\,\frac{1}{\ell}\,.
\end{equation}
This is a natural result for a disk in an $\mathbb{R}^2/\mathbb{Z}_\ell$ orbifold.

We perform the flux quantization of the $U(1)$ gauge field, $A$,
\begin{equation} \label{gf41}
p\,\equiv\,\text{hol}_{\partial\Sigma}\left(A\right)\,=\,\frac{g}{2\pi}\oint_{y=0}A\,=\,-\frac{g}{2\pi}\int_\Sigma{F}\,=\,-\frac{g}{2\pi}\frac{-y_1^2}{y_1^2+q}\Delta{z}\,.
\end{equation}
By solving \eqref{ldz41} and \eqref{gf41}, we find that $q$ and $\Delta{z}$ can be expressed in terms of $\ell$ and $p$,
\begin{equation}
q\,=\,\frac{32\pi}{g^6}p\left(g-2\pi{p}\right)\,, \qquad \Delta{z}\,=\,\frac{4\pi}{\ell\left(g-4\pi{p}\right)}\,.
\end{equation}

\subsubsection{Uplift to type IIA supergravity}

By employing the uplift formula to type IIA supergravity, \cite{Cvetic:2000ah}, which we review in appendix \ref{appB}, we obtain the uplifted metric in the string frame and the dilaton, respectively,
\begin{align}
ds_{10}^2\,=&\,\frac{\widetilde{\Delta}^{1/2}}{r}\left[\frac{ds_{1,2}^2+dr^2}{r^2}+\frac{1}{4y(y^2+q)h(y)}dy^2+h(y)dz^2\right. \notag \\
&\left.+\frac{1}{g^2y}d\xi^2+\frac{1}{g^2\widetilde{\Delta}}\Big(y^2\cos^2\xi\left(d\theta^2+\cos^2\theta{d}\chi_2^2\right)+(y^2+q)\sin^2\xi{D}\chi_1^2\Big)\right]\,, \notag \\
e^{\Phi/2}\,=&\,\frac{\widetilde{\Delta}^{1/8}}{r^{3/4}}\,,
\end{align}
where we have
\begin{equation}
\widetilde{\Delta}\,=\,y\left(y^2+q\cos^2\xi\right)\,,
\end{equation}
and
\begin{equation}
D\chi_1\,=\,d\chi_1-gA\,.
\end{equation}
Explicit form of the uplift formula for flux fields is not given in \cite{Cvetic:2000ah} and it would be interesting to obtain the full uplifted solution.

The six-dimensional internal space of the uplifted metric is an $S_z^1\times{S}_{\chi_1}^1\times{S}^2$ fibration over the 2d base space, $B_2$, of $(y,\xi)$. The two-sphere, $S^2$, is spanned by $(\theta,\chi_2)$. The 2d base space is a rectangle of $(y,\xi)$ over $[0,y_1]\,\times\left[0,\frac{\pi}{2}\right]$. See figure \ref{figd41}. We explain the geometry of the internal space by three regions of the 2d base space, $B_2$.

\begin{itemize}
\item Region I: The side of $\mathsf{P}_1\mathsf{P}_2$.
\item Region II: The sides of $\mathsf{P}_2\mathsf{P}_3$ and $\mathsf{P}_3\mathsf{P}_4$.
\item Region III: The side of $\mathsf{P}_1\mathsf{P}_4$.
\end{itemize}

\begin{figure}[t] 
\begin{center}
\includegraphics[width=4.5in]{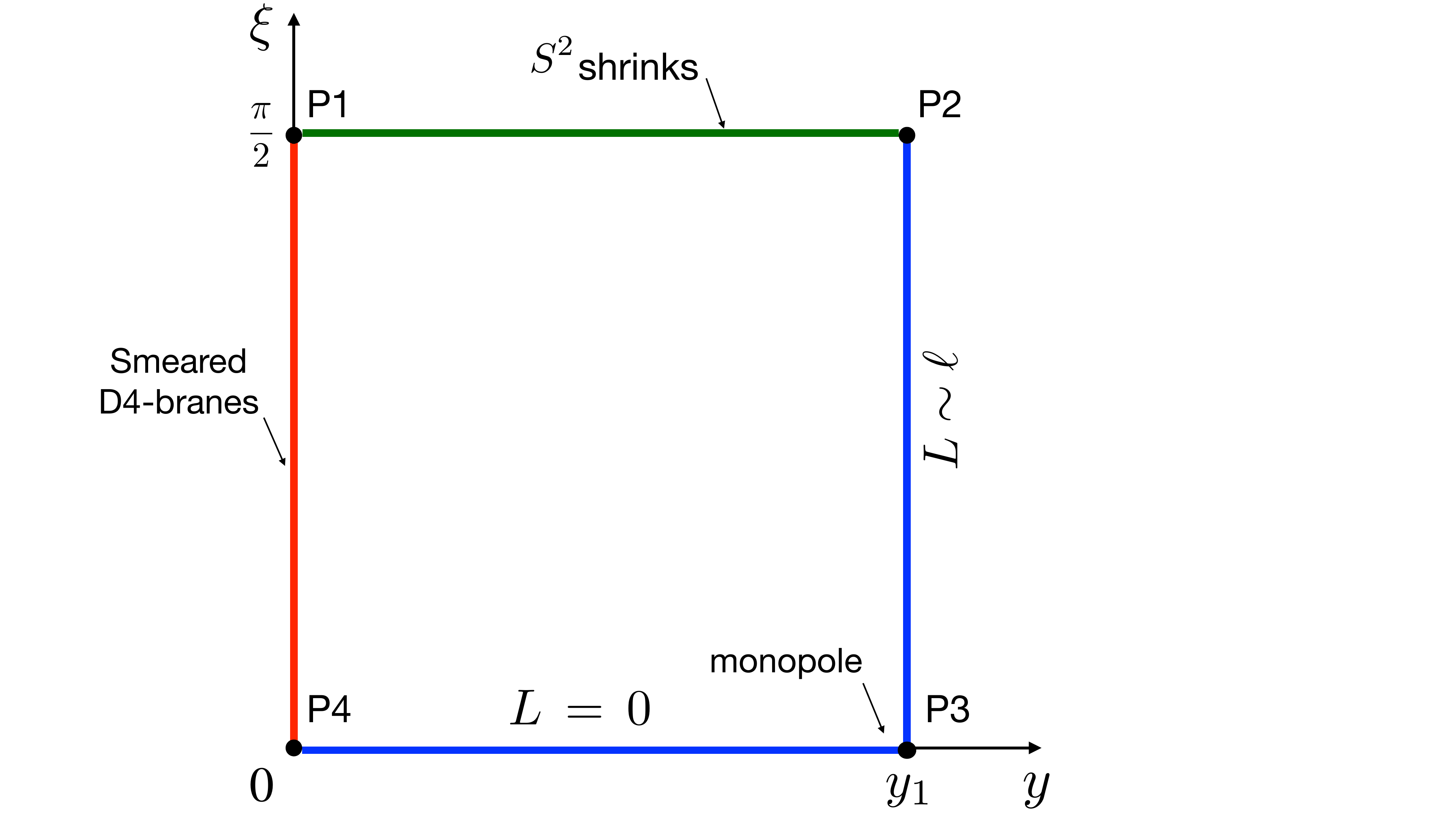}
\caption{{\it The two-dimensional base space, $B_2$, spanned by $y$ and $\xi$.}} \label{figd41}
\end{center}
\end{figure}

\noindent {\bf Region I:} On the side of $\xi\,=\,\frac{\pi}{2}$, the two-sphere, $S^2$, shrinks and the internal space caps off.

\bigskip

\noindent {\bf Region II: Monopole} We break $D\chi_1$ and complete the square of $dz$ to obtain the metric of
\begin{align} \label{d4mono1}
ds_{10}^2\,=&\,\frac{\widetilde{\Delta}^{1/2}}{r}\left[\frac{ds_{1,2}^2+dr^2}{r^2}+\frac{1}{4y(y^2+q)h(y)}dy^2\right. \notag \\
&+\frac{1}{g^2y}d\xi^2+\frac{1}{g^2\widetilde{\Delta}}y^2\cos^2\xi\left(d\theta^2+\cos^2\theta{d}\chi_2^2\right) \notag \\
&+R_z^2\left(dz+Ld\chi_1\right)^2+R_{\chi_1}^2d\chi_1^2\Big]\,.
\end{align}
The metric functions are defined to be
\begin{align}
R_z^2\,=&\,\frac{y^2\left(2g^2y-7\right)+g^2qy+q-(y^2-g^2qy+q)\cos(2\xi)}{8y(y^2+q\cos^2\xi)}\,, \notag \\
R_{\chi_1}^2\,=&\,\frac{2\left(g^2(y^2+q)-4y\right)\sin^2\xi}{g^2\left[y^2\left(2g^2y-7\right)+g^2qy+q-(y^2-g^2qy+q)\cos(2\xi)\right]}\,, \notag \\
L\,=&\,\frac{-4(y^2-q)\sin^2\xi}{g\left[y^2\left(2g^2y-7\right)+g^2qy+q-(y^2-g^2qy+q)\cos(2\xi)\right]}\,.
\end{align}

The function, $L(y,\xi)$, is piecewise constant along the sides of $y\,=\,y_1$ and $\xi\,=\,0$ of the 2d base, $B_2$,
\begin{equation}
L\left(y,0\right)\,=\,0\,, \qquad L\left(y_1,\xi\right)\,=\,\frac{4}{g\sqrt{4-g^4q}}\,=\,\frac{\ell\Delta{z}}{2\pi}\,,
\end{equation}
The jump in $L$ at the corner, $(y,\xi)\,=\,\left(y_1,0\right)$, indicates the existence of a monopole source for the $Dz$ fibration. The gauge choice of $-1/4$ for the gauge field, $A$, in \eqref{d4sol1} is required to observe monopole structure in the uplifted solutions.

\bigskip

\noindent {\bf Region III: Smeared D4-branes} We consider the singularity at $y\rightarrow0$ of the uplifted metric. As $y\rightarrow0$, the uplifted metric becomes
\begin{align}
ds_{10}^2\,\approx&\,\frac{1}{r}\left\{\left(qy\cos^2\xi\right)^{1/2}\left[\frac{ds_{1,2}^2+dr^2}{r^2}+\frac{g^2}{4}dz^2\right]\right. \notag \\
&\left.+\frac{1}{g^2\left(qy\cos^2\xi\right)^{1/2}}\Big[\cos^2\xi\left(dy^2+y^2\left(d\theta^2+\cos^2\theta{d}\chi_2^2\right)\Big)+q\left(\cos^2\xi{d}\xi^2+\sin^2\xi{D}\chi_1^2\right)\right]\right\}\,,
\end{align}
and the dilaton is
\begin{equation}
e^\Phi\,\approx\,\frac{\left(qy\cos^2\xi\right)^{1/4}}{r^{3/2}}\,.
\end{equation}
The metric implies the smeared D4-brane sources. The D4-branes are 
\begin{itemize}
\item extended along the $\mathbb{R}^{1,2}$, $r$, and $z$ directions;
\item localized at the origin of $y$, $\theta$, and $\chi_2$ directions;
\item smeared along $\xi$ and $\chi_1$ directions.
\end{itemize}
The $y^{1/2}$ and $1/y^{1/2}$ factors of the metric and the dilaton match the smeared branes reviewed in appendix \ref{appA}.

\subsubsection{Another solution: $y\in[y_2,\infty)$} \label{d4a1}

For $h(y)=0$, there are two roots, $y\,=\,\frac{2\pm\sqrt{4-g^4q}}{g^2}$. As we studied so far, there is a solution with $0<y<y_1\,\equiv\,\frac{2-\sqrt{4-g^4q}}{g^2}$ in \eqref{d4solr1}. In this section, we consider another solution with
\begin{equation}
q<\frac{4}{g^4}\,, \qquad y_2\,\equiv\,\frac{2+\sqrt{4-g^4q}}{g^2}<y<\infty\,.
\end{equation}
We plot a representative solution with $g=1$ and $q=1$ in figure \ref{d4a}.

\begin{figure}[t] 
\begin{center}
\includegraphics[width=2.0in]{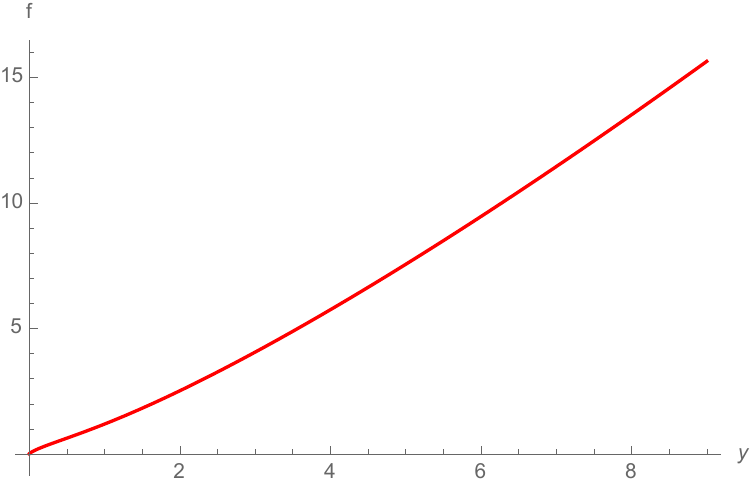} \qquad \includegraphics[width=2.0in]{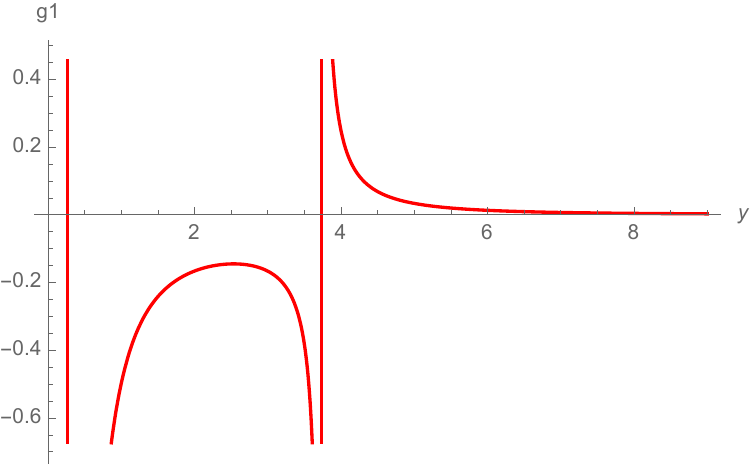} \qquad \includegraphics[width=2.0in]{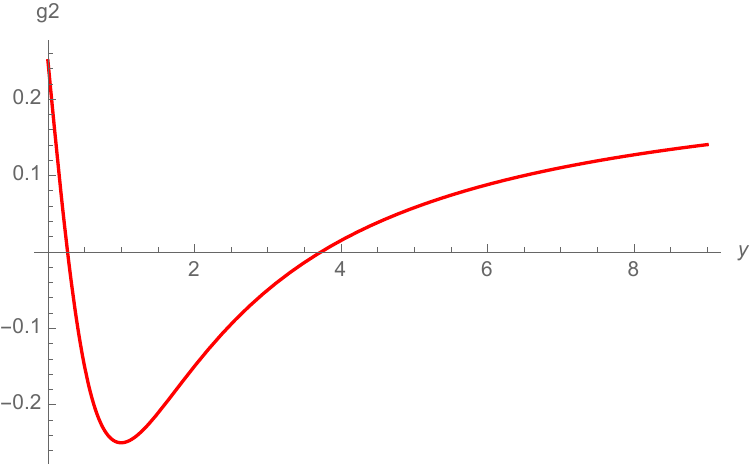}
\caption{{\it A representative solution with $g=1$ and $q=1$. The solution is regular in the range of $y_2=3.732\,<\,y\,<\,\infty$.}} \label{d4a}
\end{center}
\end{figure}

\begin{figure}[t] 
\begin{center}
\includegraphics[width=4.5in]{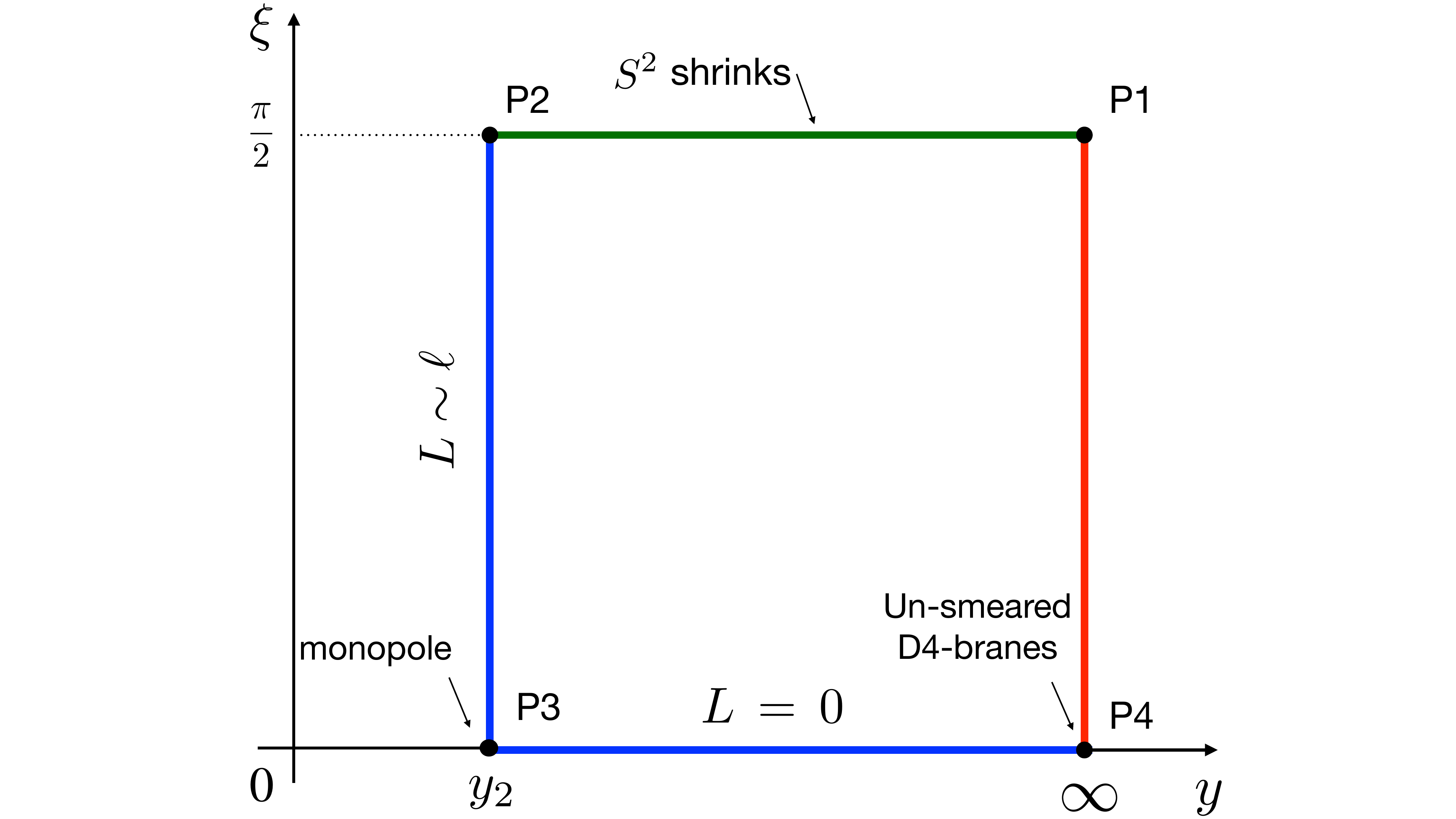}
\caption{{\it The two-dimensional base space, $B_2$, spanned by $y$ and $\xi$.}} \label{figd4a}
\end{center}
\end{figure}

Beside the structure of the solution at $y\rightarrow\infty$, which we explain below, rest of the details of the solution is identical to the solution with \eqref{d4sol1} by switching $y_1$ to $y_2$. Of course, the holographic observables, if we could calculate,  will be different for solutions.  We summarize the global structure of the uplifted metric in figure \ref{figd4a}.

\bigskip

\noindent {\bf Region III: Un-smeared D4-branes} We consider the singularity at $y\rightarrow\infty$ of the uplifted metric. As $y\rightarrow\infty$, the uplifted metric becomes
\begin{align}
ds_{10}^2\,\approx&\,\frac{1}{r}\left\{y^{3/2}\left[\frac{ds_{1,2}^2+dr^2}{r^2}+\frac{g^2}{4}dz^2\right]\right. \notag \\
&\left.+\frac{1}{g^2y^{3/2}}\Big[dy^2+y^2\Big(d\xi^2+\cos^2\xi\left(d\theta^2+\cos^2\theta{d}\chi_2^2\right)+\sin^2\xi{D}\chi_1^2\Big)\Big]\right\}\,
\end{align}
and the dilaton is
\begin{equation}
e^\Phi\,\approx\,\frac{y^{3/4}}{r^{3/2}}\,.
\end{equation}
The metric does not get smeared by D4-brane sources. The D4-branes are 
\begin{itemize}
\item extended along the $\mathbb{R}^{1,2}$, $r$, and $z$ directions;
\item localized at the origin of $y$, $\xi$, $\theta$, $\chi_1$, and $\chi_2$ directions;
\item not smeared along any directions. 
\end{itemize}
The $y^{3/2}$ and $1/y^{3/2}$ factors of the metric and the dilaton match the un-smeared branes reviewed in appendix \ref{appA}.

\subsection{Equal charge solution: $A^1=A^2\ne0$: $y\in[0,y_1]$}

\subsubsection{$D=6$ gauged supergravity}

In this section, we study equal charge disk solutions: $A^1=A^2\ne0$. We set the parameters, $q\equiv{q_1}=q_2$, for the spindle solution given in (5.20) of \cite{Boisvert:2024jrl}. Then the solution is given by
\begin{align} \label{d4sol2}
ds_6^2\,=&\,\frac{y^{1/4}(y^2+q)^{1/2}}{r^{1/2}}\left[\frac{ds_{1,2}^2+dr^2}{r^2}+\frac{y}{4(y^2+q)^2h(y)}dy^2+h(y)dz^2\right]\,, \notag \\
A\,\equiv&\,A^1\,=\,A^2\,=\,\left(\frac{q}{y^2+q}-\frac{1}{2}\right)dz\,, \notag \\
e^{2\lambda_1}\,=&\,e^{2\lambda_2}\,=\,\frac{y^{2/5}}{(y^2+q)^{1/5}}\,, \qquad e^{8\sigma}\,=\,\frac{r}{y^{1/10}(y^2+q)^{1/5}}\,,
\end{align}
where $ds_{1,2}^2$ is the metric on $\mathbb{R}^{1,2}$ and 
\begin{equation}
h(y)\,=\,\frac{g^2(y^2+q)^2-4y^3}{4(y^2+q)^2}\,.
\end{equation}
The gauge choice of $-1/2$ for the gauge field, $A$, is required to observe monopole structure in the uplifted solutions in \eqref{d4mono2}. For $h(y)=0$, there are two real and two complex roots and we denote two real roots by $y=y_1$ and $y=y_2$ where $y_1<y_2$ and they are functions of $q$ and $g$. We do not present the expressions as they are unwieldy. We consider solutions with{\footnote{There is also a solution with $y_2<y<\infty$. We briefly discuss the solution in section \ref{d4a2}.}
\begin{equation} \label{d4solr2}
0<y<y_1\,.
\end{equation}
We plot a representative solution with $g=1$ and $q=0.1$ in figure \ref{d42}. The metric functions, $f(y)$, $g_1(y)$, and $g_2(y)$, are defined by
\begin{equation}
ds_6^2\,=\,\frac{f(y)}{r^{1/2}}\left[\frac{ds_{1,2}^2+dr^2}{r^2}+g_1(y)dy^2+g_2(y)dz^2\right]\,.
\end{equation}

\begin{figure}[t] 
\begin{center}
\includegraphics[width=2.0in]{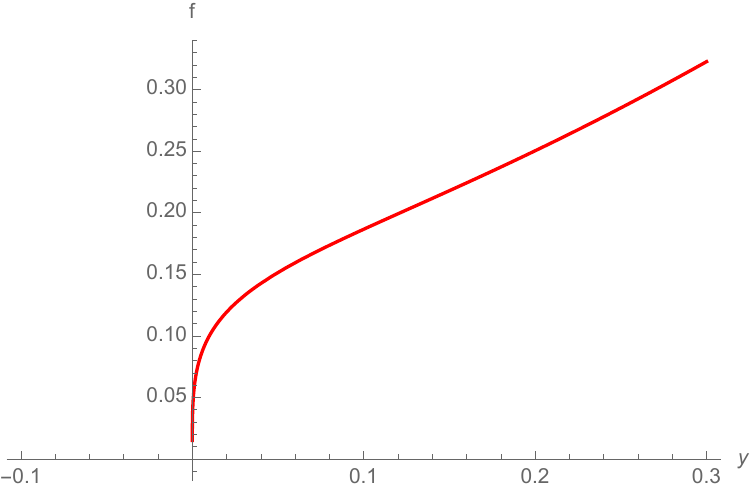} \qquad \includegraphics[width=2.0in]{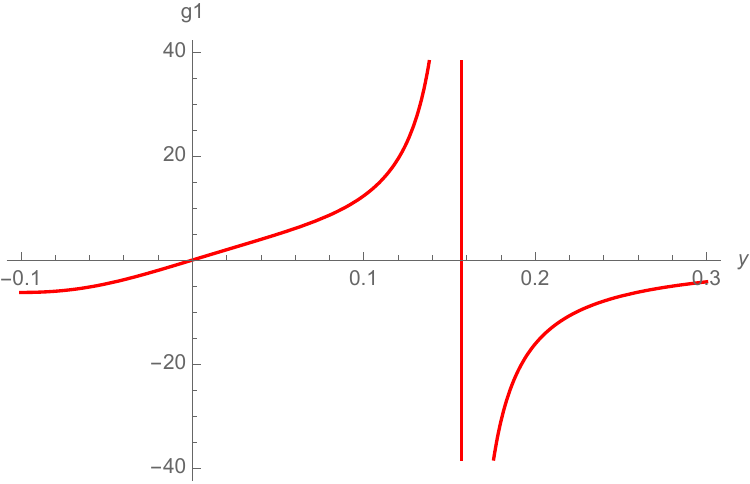} \qquad \includegraphics[width=2.0in]{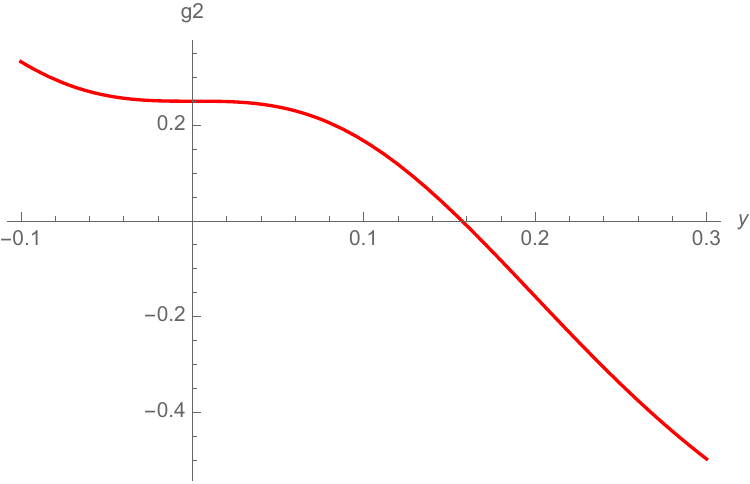}
\caption{{\it A representative solution with $g=1$ and $q=0.1$. The solution is regular in the range of $0<y<y_1=0.157$.}} \label{d42}
\end{center}
\end{figure}

Near $y\rightarrow0$ the warp factor vanishes and it is a curvature singularity of the metric, 
\begin{equation}
ds_6^2\,\approx\,\frac{q^{1/2}y^{1/4}}{r^{1/2}}\left[\frac{ds_{1,2}^2+dr^2}{r^2}+\frac{y}{g^2q^2}dy^2+\frac{g^2}{4}dz^2\right]\,.
\end{equation}
Approaching $y\rightarrow{y}_1$, the metric becomes to be
\begin{equation}
ds_6^2\,\approx\,\frac{y_1^{1/4}(y_1^2+q)^{1/2}}{r^{1/2}}\left[\frac{ds_{1,2}^2+dr^2}{r^2}+\frac{d\rho^2+\mathcal{E}(q)^2\rho^2dz^2}{-y_1^{-1}\left(y_1^2+q\right)^2h'(y_1)}\right]\,,
\end{equation}
where we have
\begin{equation}
\mathcal{E}(q)\,\equiv\,y_1^{-1}\left(y_1^2+q\right)^2h'(y_1)^2\,=\,\left(\frac{y_1^{3/2}\left(y_1^2-3q\right)}{\left(y_1^2+q\right)^2}\right)^2\,,
\end{equation}
we introduced a new parametrization of coordinate, $\rho^2\,=\,y_1-y$. Then,  the $\rho-z$ surface is locally an $\mathbb{R}^2/\mathbb{Z}_\ell$ orbifold if we set
\begin{equation} \label{ldz42}
\frac{\ell\Delta{z}}{2\pi}\,=\,\frac{1}{\mathcal{E}(q)}\,,
\end{equation}
where $\Delta{z}$ is the period of coordinate, $z$, and $\ell\,=\,1,2,3,\ldots$. The metric spanned by $(y,z)$ has a topology of disk, $\Sigma$, with the center at $y=y_1$ and the boundary at $y=0$.

We calculate the Euler characteristic of the $y-z$ surface, $\Sigma$,
\begin{equation}
\chi\,=\,\frac{1}{4\pi}\int_\Sigma{R}_\Sigma\text{vol}_\Sigma\,=\,\frac{1}{4\pi}\frac{2y_1^{3/2}\left(y_1^2-3q\right)}{\left(y_1^2+q\right)^2}\Delta{z}\,=\,\frac{1}{\ell}\,.
\end{equation}
This is a natural result for a disk in an $\mathbb{R}^2/\mathbb{Z}_\ell$ orbifold.

We perform the flux quantization of the $U(1)$ gauge field, $A$,
\begin{equation} \label{gf42}
p\,\equiv\,\text{hol}_{\partial\Sigma}\left(A\right)\,=\,\frac{g}{2\pi}\oint_{y=0}A\,=\,-\frac{g}{2\pi}\int_\Sigma{F}\,=\,-\frac{g}{2\pi}\frac{-y_1^2}{y_1^2+q}\Delta{z}\,.
\end{equation}
Although the expressions are unwieldy, by solving \eqref{ldz42} and \eqref{gf42}, one should be able to find that $q$ and $\Delta{z}$ can be expressed in terms of $\ell$ and $p$.

\subsubsection{Uplift to type IIA supergravity}

By employing the uplift formula to type IIA supergravity, \cite{Cvetic:2000ah}, which we review in appendix \ref{appB}, we obtain the uplifted metric in the string frame and the dilaton, respectively,
\begin{align}
ds_{10}^2\,=&\,\frac{\widetilde{\Delta}^{1/2}}{r}\left[\frac{ds_{1,2}^2+dr^2}{r^2}+\frac{y}{4(y^2+q)^2h(y)}dy^2+h(y)dz^2\right. \notag \\
&\left.+\frac{y}{g^2\left(y^2+q\right)}d\xi^2+\frac{y^2+q}{g^2\widetilde{\Delta}}\cos^2\xi\Big(d\theta^2+\cos^2\theta{D}\chi_1^2+\sin^2\theta{D}\chi_2^2\Big)\right]\,, \notag \\
e^{\Phi/2}\,=&\,\frac{\widetilde{\Delta}^{1/8}}{r^{3/4}}\,,
\end{align}
where we have
\begin{equation}
\widetilde{\Delta}\,=\,\frac{\left(y^2+q\right)\left(y^2+q\sin^2\xi\right)}{y}\,,
\end{equation}
and
\begin{equation}
D\chi_1\,=\,d\chi_1-gA\,, \qquad D\chi_2\,=\,d\chi_2-gA\,.
\end{equation}
Explicit form of the uplift formula for flux fields is not given in \cite{Cvetic:2000ah} and it would be interesting to obtain the full uplifted solution.

The six-dimensional internal space of the uplifted metric is an $S_z^1\times{S}^3$ fibration over the 2d base space, $B_2$, of $(y,\xi)$. The three-sphere, $S^3$, is spanned by $(\theta,\chi_1,\chi_2)$. The 2d base space is a rectangle of $(y,\xi)$ over $[0,y_1]\,\times\left[0,\frac{\pi}{2}\right]$. See figure \ref{figd42}. We explain the geometry of the internal space by three regions of the 2d base space, $B_2$.

\begin{itemize}
\item Region I: The side of $\mathsf{P}_1\mathsf{P}_2$.
\item Region II: The sides of $\mathsf{P}_2\mathsf{P}_3$ and $\mathsf{P}_3\mathsf{P}_4$.
\item Region III: The side of $\mathsf{P}_1\mathsf{P}_4$.
\end{itemize}

\begin{figure}[t] 
\begin{center}
\includegraphics[width=4.5in]{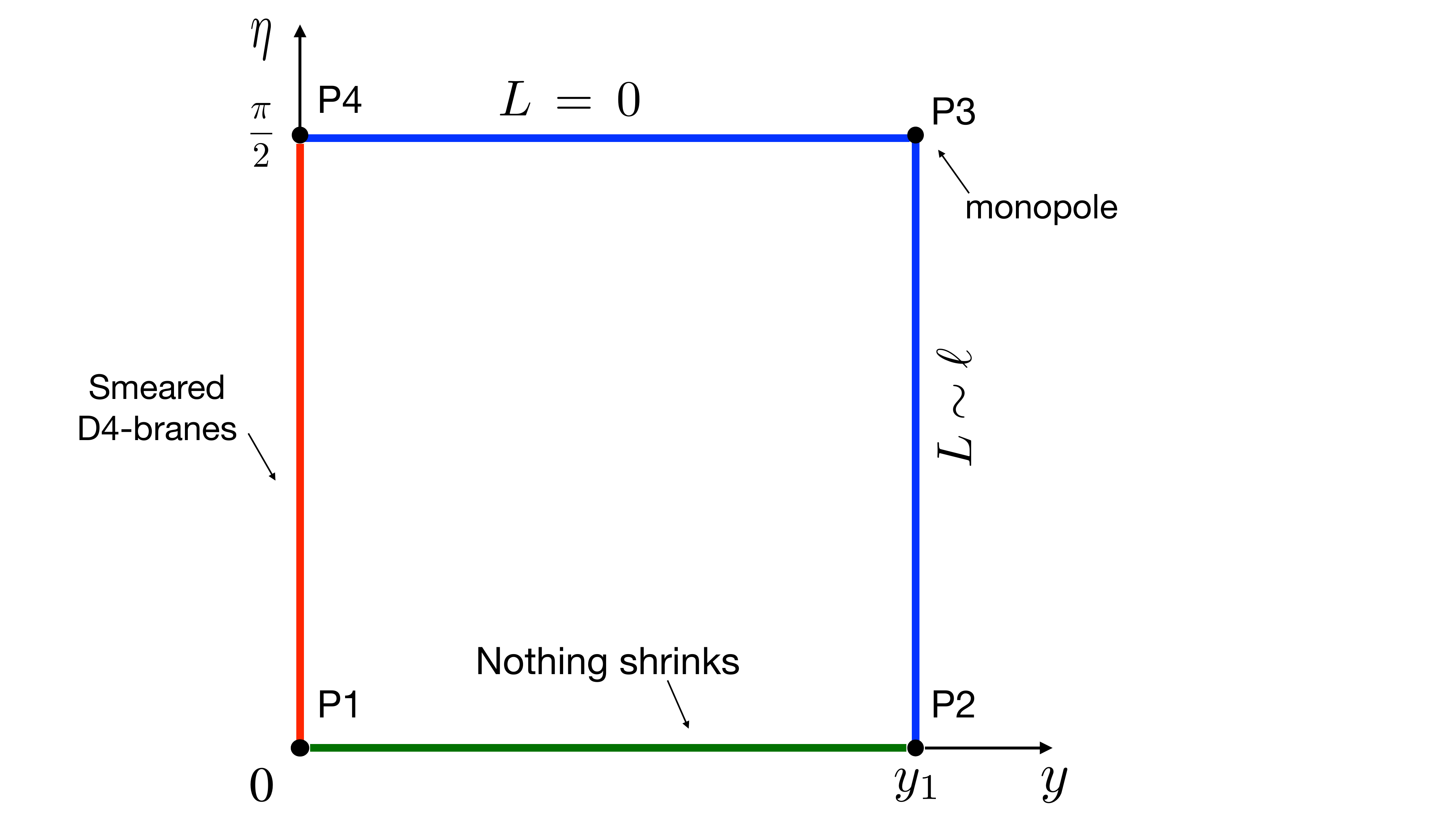}
\caption{{\it The two-dimensional base space, $B_2$, spanned by $y$ and $\xi$.}} \label{figd42}
\end{center}
\end{figure}

\noindent {\bf Region I:} On the side of $\xi\,=\,0$, unlike usual disk solutions, there is no part of the metric that shrinks.

\bigskip

\noindent {\bf Region II: Monopole} We break $D\chi_1$ and $D\chi_2$ and complete the square of $dz$ to obtain the metric of
\begin{align} \label{d4mono2}
ds_{10}^2\,=&\,\frac{\widetilde{\Delta}^{1/2}}{r}\left[\frac{ds_{1,2}^2+dr^2}{r^2}+\frac{y}{4(y^2+q)^2h(y)}dy^2\right. \notag \\
& +\frac{y}{g^2\left(y^2+q\right)}d\xi^2+\frac{y^2+q}{g^2\widetilde{\Delta}}\cos^2\xi{d}\theta^2 \notag \\
& +R_z^2\Big(dz-L\left(\cos^2\theta{d}\chi_1+\sin^2\theta{d}\chi_2\right)\Big)^2 \notag \\
& +R_{\chi_1}^2\cos^2\theta\left(d\chi_1-L_1\sin^2\theta{d}\chi_2\right)^2+R_{\chi_2}^2\sin^2\theta{d}\chi_2^2\Big]\,,
\end{align}
The metric functions are defined to be
\begin{align}
R_z^2\,=&\,\frac{X}{32\left(y^2+q\right)\left(y^2+q\sin^2\xi\right)}\,, \notag \\
R_{\chi_1}^2\,=&\,\frac{2y\cos^2\xi}{g^2\left(y^2+q\right)\left(y^2+q\sin^2\xi\right)}\frac{Y}{X}\,, \notag \\
R_{\chi_2}^2\,=&\,\frac{4y\Big(g^2\left(y^2+q\right)^2-4y^3\Big)\cos^2\xi}{g^2Y}\,, \notag \\
L\,=&\,-\frac{8y\left(y^2-3q\right)\cos^2\xi}{gX}\,, \notag \\
L_1\,=&\,\frac{y\left(y^2-3q\right)^2\cos^2\xi}{Y}\,,
\end{align}
with
\begin{align}
X\,\equiv&\,4g^2\left(2y^4+3qy^2+q^2\right)-31y^3+9qy+\Big(y^3-4g^2q\left(y^2+q\right)+9qy\Big)\cos(2\xi)\,, \notag \\
Y\,\equiv&\,4\Big(g^2\left(y^2+q\right)^2-4y^3\Big)\left(y^2+q\sin^2\xi\right)+y\left(y^2-3q\right)^2\cos^2\xi\sin^2\theta\,.
\end{align}

The function, $L(y,\xi)$, is piecewise constant along the sides of $y\,=\,y_1$ and $\xi\,=\,\pi/2$ of the 2d base, $B_2$,
\begin{equation}
L\left(y,\pi/2\right)\,=\,0\,, \qquad L\left(y_1,\xi\right)\,=\,2\frac{\left(y_1^2+q\right)^2}{y_1^{3/2}\left(y_1^2-3q\right)}\,=\,2\frac{\ell\Delta{z}}{2\pi}\,,
\end{equation}
The jump in $L$ at the corner, $(y,\xi)\,=\,\left(y_1,\pi/2\right)$, indicates the existence of a monopole source for the $Dz$ fibration. The gauge choice of $-1/2$ for the gauge field, $A$, in \eqref{d4sol2} is required to observe monopole structure in the uplifted solutions.

\bigskip

\noindent {\bf Region III: Smeared D4-branes} We consider the singularity at $y\rightarrow0$ of the uplifted metric. As $y\rightarrow0$, the uplifted metric becomes
\begin{align}
ds_{10}^2\,\approx&\,\frac{1}{r}\left\{\frac{q\sin\xi}{y^{1/2}}\left[\frac{ds_{1,2}^2+dr^2}{r^2}+\frac{g^2}{4}dz^2\right]\right. \notag \\
&\left.+\frac{y^{1/2}}{g^2q\sin\xi}\left[\sin^2\xi{d}y^2+q\Big(\sin^2\xi{d}\xi^2+\cos^2\xi\left(d\theta^2+\cos^2\theta{D}\chi_1^2+\sin^2\theta{D}\chi_2^2\right)\Big)\right]\right\}\,,
\end{align}
and the dilaton is
\begin{equation}
e^\Phi\,\approx\,\frac{\left(q\sin\xi\right)^{1/2}}{y^{1/4}r^{3/2}}\,.
\end{equation}
The metric implies the smeared D4-brane sources. The D4-branes are 
\begin{itemize}
\item extended along the $\mathbb{R}^{1,2}$, $r$, and $z$ directions;
\item localized at the origin of $y$ direction;
\item smeared along the $\xi$, $\theta$, $\chi_1$, and $\chi_2$ directions. 
\end{itemize}
The $1/y^{1/2}$ and $y^{1/2}$ factors of the metric and the dilaton match the smeared branes reviewed in appendix \ref{appA}.

\subsubsection{Another solution: $y\in[y_2,\infty)$} \label{d4a2}

We studied a solution with $0<y<y_1$ in \eqref{d4solr2}. In this section, we consider another solution with
\begin{equation}
y_2<y<\infty\,,
\end{equation}
where $y_2$ is another real root of $h(y)=0$ and is a function of $q$ and $g$. We plot a representative solution with $g=1$ and $q=0.1$ in figure \ref{d42a}.

\begin{figure}[t] 
\begin{center}
\includegraphics[width=2.0in]{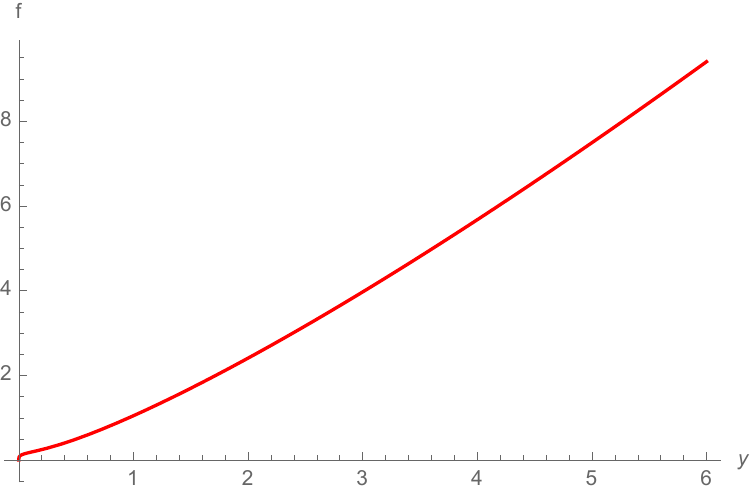} \qquad \includegraphics[width=2.0in]{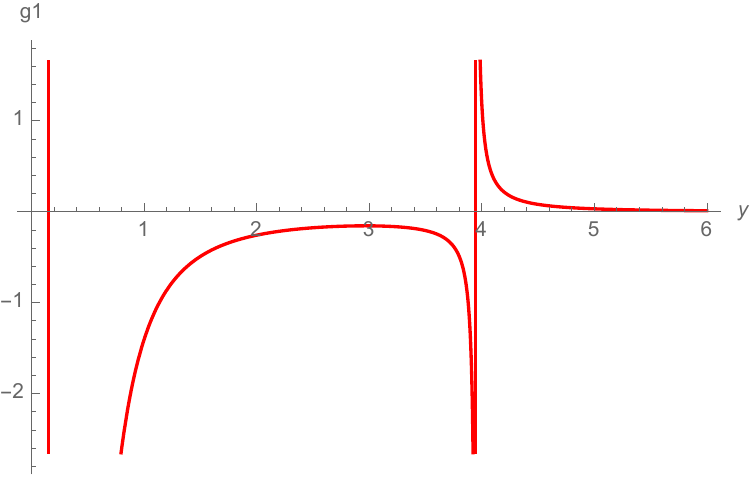} \qquad \includegraphics[width=2.0in]{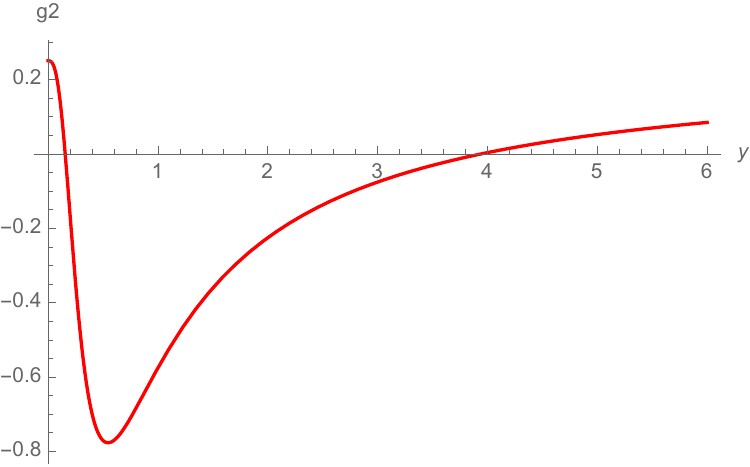}
\caption{{\it A representative solution with $g=1$ and $q=0.1$. The solution is regular in the range of $y_2=3.949<y<\infty$.}} \label{d42a}
\end{center}
\end{figure}

\begin{figure}[t] 
\begin{center}
\includegraphics[width=4.5in]{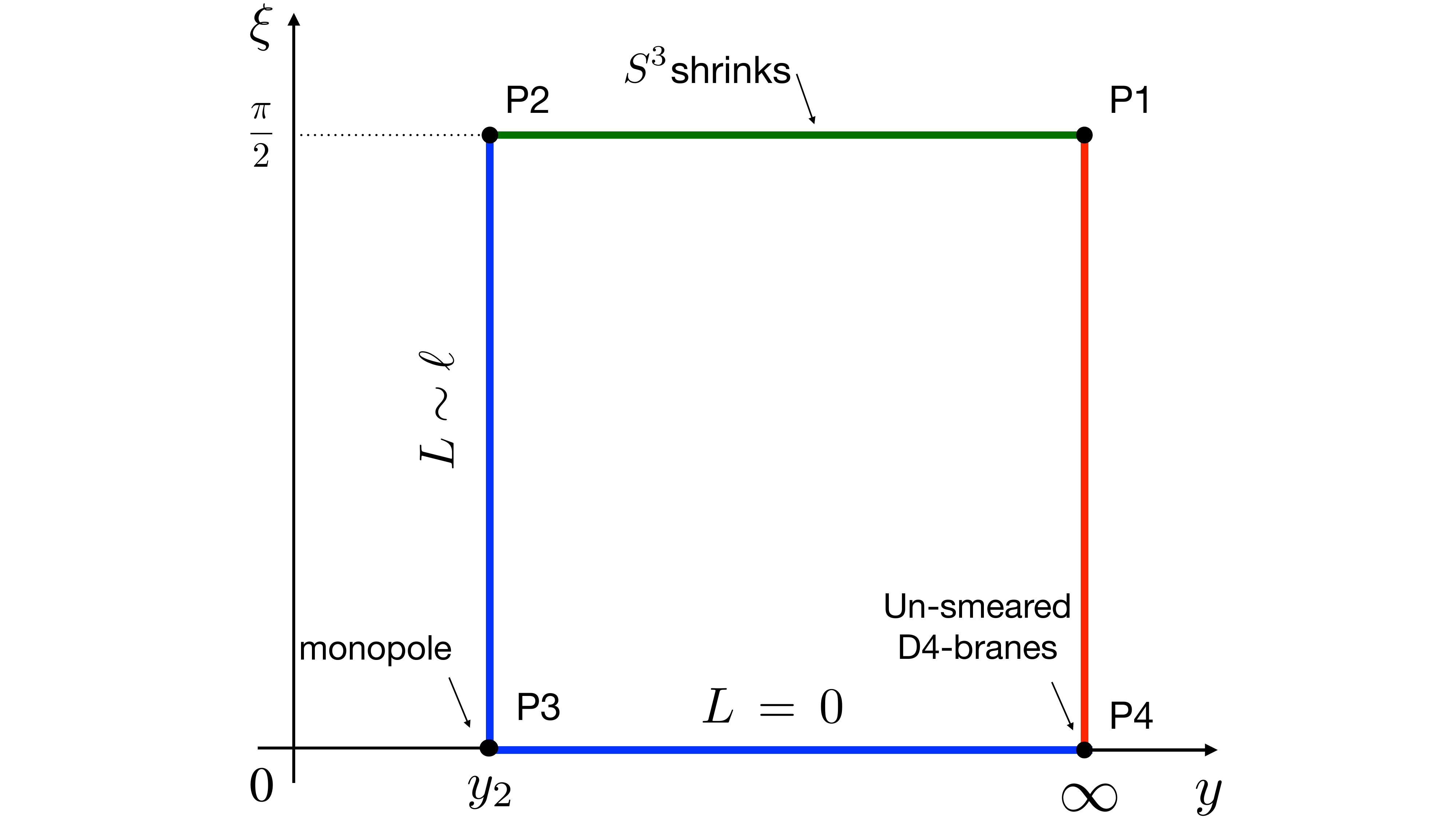}
\caption{{\it The two-dimensional base space, $B_2$, spanned by $y$ and $\xi$.}} \label{figd4aa}
\end{center}
\end{figure}

Beside the structure of the solution at $y\rightarrow\infty$, which we explain below, rest of the details of the solution is identical to the solution with \eqref{d4sol2} by switching $y_1$ to $y_2$. Of course, the holographic observables, if we could calculate,  will be different for solutions. We summarize the global structure of the uplifted metric in figure \ref{figd4aa}.

\bigskip

\noindent {\bf Region III: Un-smeared D4-branes} We consider the singularity at $y\rightarrow\infty$ of the uplifted metric. As $y\rightarrow\infty$, the uplifted metric becomes
\begin{align} \label{d4unsmeared}
ds_{10}^2\,\approx&\,\frac{1}{r}\left\{y^{3/2}\left[\frac{ds_{1,2}^2+dr^2}{r^2}+\frac{g^2}{4}dz^2\right]\right. \notag \\
&\left.+\frac{1}{g^2y^{3/2}}\left[dy^2+y^2\Big(d\xi^2+\cos^2\xi\left(d\theta^2+\cos^2\theta{D}\chi_1^2+\sin^2\theta{D}\chi_2^2\right)\Big)\right]\right\}\,,
\end{align}
and the dilaton is
\begin{equation}
e^\Phi\,\approx\,\frac{y^{3/4}}{r^{3/2}}\,.
\end{equation}
Unlike previously known disk solutions, the metric does not get smeared by D4-brane sources. The D4-branes are 
\begin{itemize}
\item extended along the $\mathbb{R}^{1,2}$, $r$, and $z$ directions;
\item localized at the origin of $y$, $\xi$, $\theta$, $\chi_1$, and $\chi_2$ directions;
\item not smeared along any directions. 
\end{itemize}
The $y^{3/2}$ and $1/y^{3/2}$ factors of the metric and the dilaton match the un-smeared branes reviewed in appendix \ref{appA}.

\subsection{Multi charge solution: $A^1\ne{A}^2$: $y\in[0,y_1]$}

\subsubsection{$D=6$ gauged supergravity}

In this section, we study multi charge disk solutions: $A^1\ne{A}^2$. We have two distinct parameters, $q_1\ne{q}_2$, for the spindle solution given in (5.20) of \cite{Boisvert:2024jrl}. Then the solution is given by
\begin{align} \label{d4sol2three}
ds_6^2\,=&\,\frac{y^{1/4}\left(y^2+q_1\right)^{1/4}\left(y^2+q_2\right)^{1/4}}{r^{1/2}}\left[\frac{ds_{1,2}^2+dr^2}{r^2}+\frac{y}{4\left(y^2+q_1\right)\left(y^2+q_2\right)h(y)}dy^2+h(y)dz^2\right]\,, \notag \\
A^1\,=&\,\left(\frac{q}{y^2+q_1}\right)dz, \qquad A^2\,=\,\left(\frac{q}{y^2+q_2}\right)dz\,, \notag \\
e^{2\lambda_1}\,=&\,\frac{y^{2/5}\left(y^2+q_2\right)^{2/5}}{\left(y^2+q_1\right)^{3/5}}\,, \quad e^{2\lambda_2}\,=\,\frac{y^{2/5}\left(y^2+q_1\right)^{2/5}}{\left(y^2+q_2\right)^{3/5}}\,, \quad e^{8\sigma}\,=\,\frac{r}{y^{1/10}\left(y^2+q_1\right)^{1/10}\left(y^2+q_2\right)^{1/10}}\,,
\end{align}
where $ds_{1,2}^2$ is the metric on $\mathbb{R}^{1,2}$ and 
\begin{equation}
h(y)\,=\,\frac{g^2\left(y^2+q_1\right)\left(y^2+q_2\right)-4y^3}{4\left(y^2+q_1\right)\left(y^2+q_2\right)}\,.
\end{equation}
For $h(y)=0$, there are two real and two complex roots and we denote two real roots by $y=y_1$ and $y=y_2$ where $y_1<y_2$ and they are functions of $q$ and $g$. We do not present the expressions as they are unwieldy. We consider solutions with{\footnote{There is also a solution with $y_2<y<\infty$. We briefly discuss the solution in section \ref{d4a3}.}
\begin{equation} \label{d4solr2three}
0<y<y_1\,.
\end{equation}
We plot a representative solution with $g=1$, $q_1=0.1$, and $q_2=0.2$ in figure \ref{d42three}. The metric functions, $f(y)$, $g_1(y)$, and $g_2(y)$, are defined by
\begin{equation}
ds_6^2\,=\,\frac{f(y)}{r^{1/2}}\left[\frac{ds_{1,2}^2+dr^2}{r^2}+g_1(y)dy^2+g_2(y)dz^2\right]\,.
\end{equation}

\begin{figure}[t] 
\begin{center}
\includegraphics[width=2.0in]{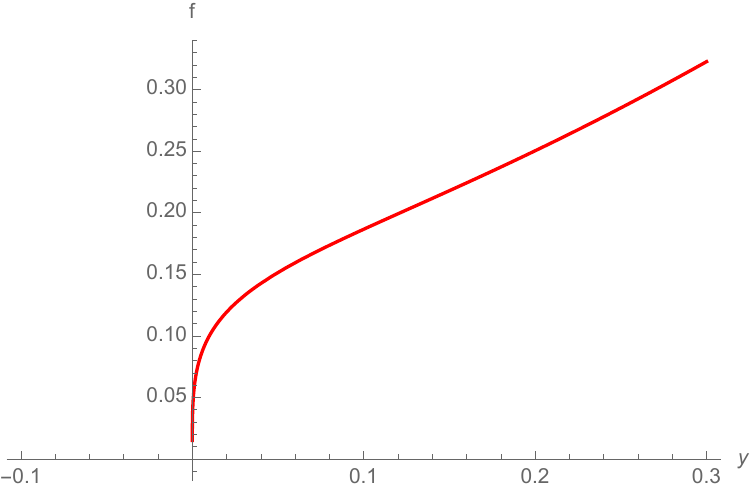} \qquad \includegraphics[width=2.0in]{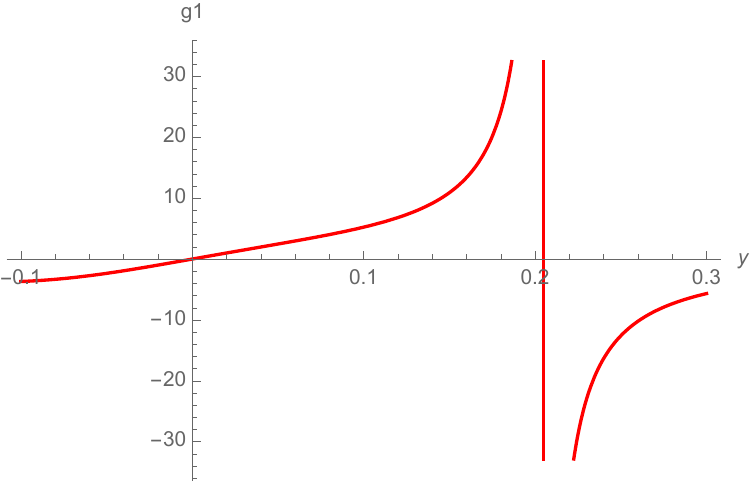} \qquad \includegraphics[width=2.0in]{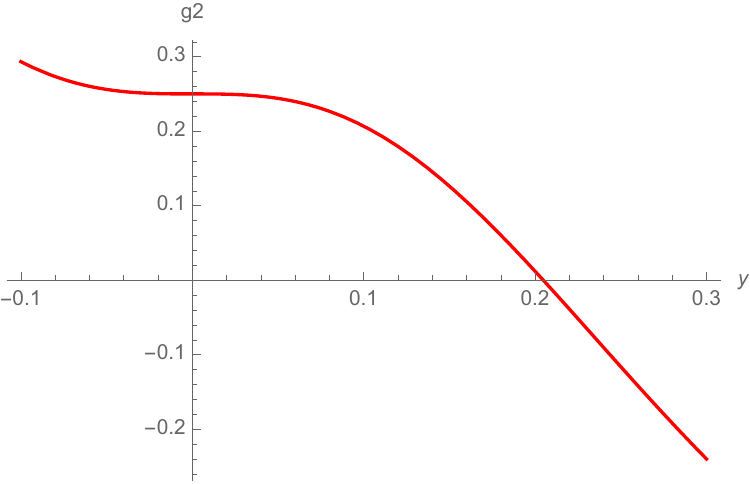}
\caption{{\it A representative solution with $g=1$, $q_1=0.1$, and $q_2=0.2$. The solution is regular in the range of $0<y<y_1=0.204$.}} \label{d42three}
\end{center}
\end{figure}

Near $y\rightarrow0$ the warp factor vanishes and it is a curvature singularity of the metric, 
\begin{equation}
ds_6^2\,\approx\,\frac{\left(q_1q_2y\right)^{1/4}}{r^{1/2}}\left[\frac{ds_{1,2}^2+dr^2}{r^2}+\frac{y}{g^2q_1q_2}dy^2+\frac{g^2}{4}dz^2\right]\,.
\end{equation}
Approaching $y\rightarrow{y}_1$, the metric becomes to be
\begin{equation}
ds_6^2\,\approx\,\frac{y_1^{1/4}\left(y_1^2+q_1\right)^{1/4}\left(y_1^2+q_2\right)^{1/4}}{r^{1/2}}\left[\frac{ds_{1,2}^2+dr^2}{r^2}+\frac{d\rho^2+\mathcal{E}(q_1,q_2)^2\rho^2dz^2}{-y_1^{-1}\left(y_1^2+q_1\right)\left(y_1^2+q_2\right)h'(y_1)}\right]\,,
\end{equation}
where we have
\begin{equation}
\mathcal{E}(q_1,q_2)\,\equiv\,y_1^{-1}\left(y_1^2+q_1\right)\left(y_1^2+q_2\right)h'(y_1)^2\,=\,\left(\frac{y_1^{3/2}\left[y_1^4-\left(q_1+q_2\right)y_1^2+3q_1q_2\right]}{\left(y_1^2+q_1\right)^{3/2}\left(y_1^2+q_2\right)^{3/2}}\right)^2\,,
\end{equation}
we introduced a new parametrization of coordinate, $\rho^2\,=\,y_1-y$. Then,  the $\rho-z$ surface is locally an $\mathbb{R}^2/\mathbb{Z}_\ell$ orbifold if we set
\begin{equation} \label{ldz42}
\frac{\ell\Delta{z}}{2\pi}\,=\,\frac{1}{\mathcal{E}(q_1,q_2)}\,,
\end{equation}
where $\Delta{z}$ is the period of coordinate, $z$, and $\ell\,=\,1,2,3,\ldots$. The metric spanned by $(y,z)$ has a topology of disk, $\Sigma$, with the center at $y=y_1$ and the boundary at $y=0$.

We calculate the Euler characteristic of the $y-z$ surface, $\Sigma$,
\begin{equation}
\chi\,=\,\frac{1}{4\pi}\int_\Sigma{R}_\Sigma\text{vol}_\Sigma\,=\,\frac{1}{4\pi}2\mathcal{E}(q_1,q_2)\Delta{z}\,=\,\frac{1}{\ell}\,.
\end{equation}
This is a natural result for a disk in an $\mathbb{R}^2/\mathbb{Z}_\ell$ orbifold.

We perform the flux quantization of the $U(1)$ gauge field, $A$,
\begin{equation} \label{gf42three}
p\,\equiv\,\text{hol}_{\partial\Sigma}\left(A\right)\,=\,\frac{g}{2\pi}\oint_{y=0}A\,=\,-\frac{g}{2\pi}\int_\Sigma{F}\,=\,-\frac{g}{2\pi}\frac{-y_1^2}{y_1^2+q}\Delta{z}\,.
\end{equation}
Although the expressions are unwieldy, by solving \eqref{ldz42} and \eqref{gf42}, one should be able to find that $q$ and $\Delta{z}$ can be expressed in terms of $\ell$ and $p$.

\subsubsection{Uplift to type IIA supergravity}

By employing the uplift formula to type IIA supergravity, \cite{Cvetic:2000ah}, which we review in appendix \ref{appB}, we obtain the uplifted metric in the string frame and the dilaton, respectively,
\begin{align}
ds_{10}^2\,&=\,\frac{\widetilde{\Delta}^{1/2}}{r}\left[\frac{ds_{1,2}^2+dr^2}{r^2}+\frac{y}{4\left(y^2+q_1\right)\left(y^2+q_2\right)h(y)}dy^2+h(y)dz^2\right] \notag \\
&+\frac{1}{g^2}\frac{\tilde{\Delta}^{-1/2}}{r}\left[\left\{y^2+\left(\frac{q_1+q_2}{2}+\frac{q_1-q_2}{2}\cos\left(2\theta\right)\right)\sin^2\xi\right\}d\xi^2\right. \notag \\
& \,\,\,\,\,\,\,\,\,\,\,\, +\cos^2\xi\left\{\left(y^2+\frac{q_1+q_2}{2}-\frac{q_1-q_2}{2}\cos\left(2\theta\right)\right)d\theta^2+\left(y^2+q_1\right)\cos^2\theta\,D\chi_1^2+\left(y^2+q_2\right)\sin^2\theta\,D\chi_2^2\right\} \notag \\
& \,\,\,\,\,\,\,\,\,\,\,\, \left.+\frac{q_1-q_2}{2}\sin(2\xi)\sin(2\theta)d\xi\,d\theta\right]\,, \notag \\
e^\Phi\,&=\,\frac{\widetilde{\Delta}^{1/8}}{r^{3/4}}\,,
\end{align}
where we have
\begin{equation}
\widetilde{\Delta}\,=\,\Big(\left(y^2+q_2\right)\cos^2\theta+\left(y^2+q_1\right)\sin^2\theta\Big)y\cos^2\xi+\frac{\left(y^2+q_1\right)\left(y^2+q_2\right)}{y}\sin^2\xi\,,
\end{equation}
and
\begin{equation}
D\chi_1\,=\,d\chi_1-gA^1\,, \qquad D\chi_2\,=\,d\chi_2-gA^2\,.
\end{equation}
Explicit form of the uplift formula for flux fields is not given in \cite{Cvetic:2000ah} and it would be interesting to obtain the full uplifted solution.

The six-dimensional internal space of the uplifted metric is an $S_z^1\times{S}^3$ fibration over the 2d base space, $B_2$, of $(y,\xi)$. The three-sphere, $S^3$, is spanned by $(\theta,\chi_1,\chi_2)$. The 2d base space is a rectangle of $(y,\xi)$ over $[0,y_1]\,\times\left[0,\frac{\pi}{2}\right]$. The internal space is more involved than the solutions we previously studied. Thus we only present the metric in the string frame and the dilaton of each solution at the boundary of the disk.

\bigskip

\noindent {\bf Smeared D4-branes} We consider the singularity at $y\rightarrow0$ of the uplifted metric. As $y\rightarrow0$, the uplifted metric becomes
\begin{align}
ds_{10}^2\,&\approx\,\frac{\left(q_1q_2\right)^{1/2}\sin\xi}{ry^{1/2}}\left[\frac{ds_{1,2}^2+dr^2}{r^2}+\frac{g^2}{4}dz^2\right] \notag \\
&+\frac{y^{1/2}}{g^2r\left(q_1q_2\right)^{1/2}\sin\xi}\left[\sin^2\xi\,dy^2+\left(\frac{q_1+q_2}{2}+\frac{q_1-q_2}{2}\cos\left(2\theta\right)\right)\sin^2\xi\,d\xi^2\right. \notag \\
& \,\,\,\,\,\,\,\,\,\,\,\, +\cos^2\xi\left\{\left(\frac{q_1+q_2}{2}-\frac{q_1-q_2}{2}\cos\left(2\theta\right)\right)d\theta^2+q_1\cos^2\theta\,D\chi_1^2+q_2\sin^2\theta\,D\chi_2^2\right\} \notag \\
& \,\,\,\,\,\,\,\,\,\,\,\, \left.+\frac{q_1-q_2}{2}\sin(2\xi)\sin(2\theta)d\xi\,d\theta\right]\,,
\end{align}
and the dilaton is
\begin{equation}
e^\Phi\,\approx\,\frac{\left(q_1q_2\right)^{1/4}\sin^{1/2}\xi}{r^{3/2}y^{1/4}}\,.
\end{equation}
The metric implies the smeared D4-brane sources. The D4-branes are 
\begin{itemize}
\item extended along the $\mathbb{R}^{1,2}$, $r$, and $z$ directions;
\item localized at the origin of $y$ direction;
\item smeared along the $\xi$, $\theta$, $\chi_1$, and $\chi_2$ directions. 
\end{itemize}
The $1/y^{1/2}$ and $y^{1/2}$ factors of the metric and the dilaton match the smeared branes reviewed in appendix \ref{appA}.

\subsubsection{Another solution: $y\in[y_2,\infty)$} \label{d4a3}

We studied a solution with $0<y<y_1$ in \eqref{d4solr2three}. In this section, we consider another solution with
\begin{equation}
y_2<y<\infty\,,
\end{equation}
where $y_2$ is another real root of $h(y)=0$ and is a function of $q$ and $g$. We plot a representative solution with $g=1$ and $q=0.1$ in figure \ref{d42athree}.

\begin{figure}[t] 
\begin{center}
\includegraphics[width=2.0in]{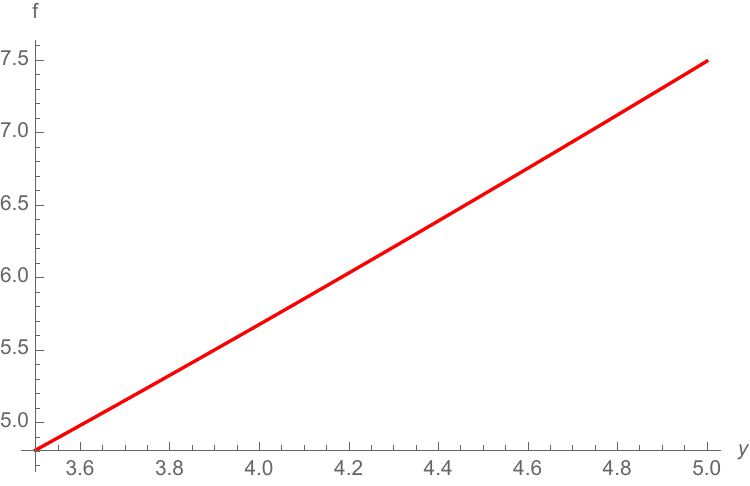} \qquad \includegraphics[width=2.0in]{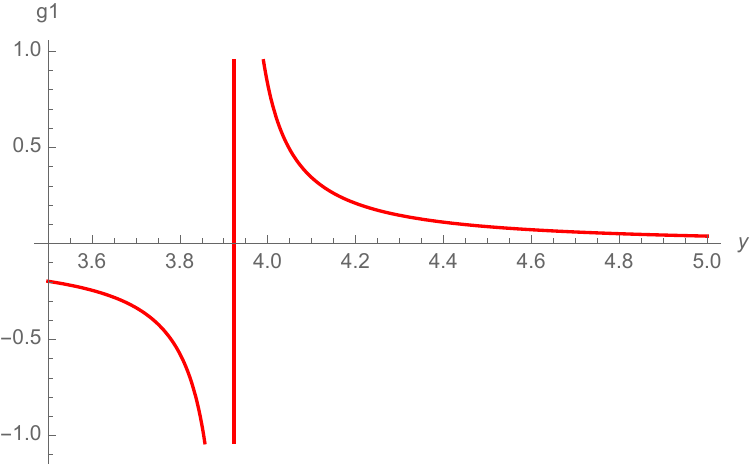} \qquad \includegraphics[width=2.0in]{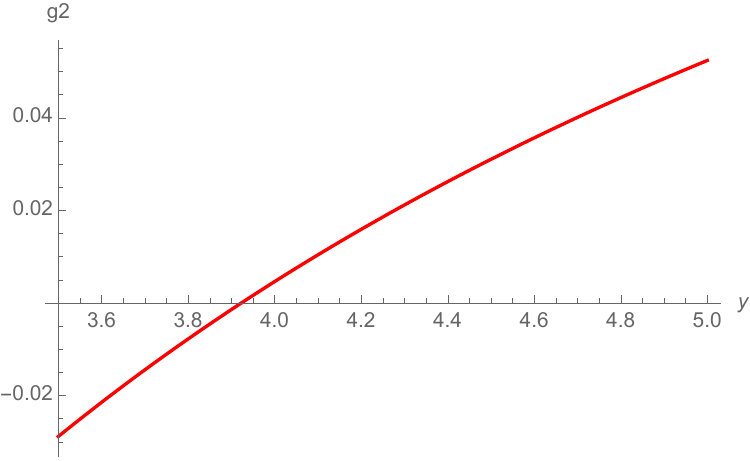}
\caption{{\it A representative solution with $g=1$ and $q=0.1$. The solution is regular in the range of $y_2=3.949<y<\infty$.}} \label{d42athree}
\end{center}
\end{figure}

Beside the structure of the solution at $y\rightarrow\infty$, which we explain below, rest of the details of the solution is identical to the solution with \eqref{d4sol2} by switching $y_1$ to $y_2$. Of course, the holographic observables, if we could calculate, will be different for solutions.

At the boundary, $y\rightarrow\infty$, the uplifted metric becomes \eqref{d4unsmeared}.

\section{D2-branes wrapped on a disk}  \label{d2}

\subsection{No charge solution: $A^1=A^2=A^3=0$: $y\in[y_1,\infty)$}

\subsubsection{$D=4$ gauged supergravity}

We consider $D=4$ $ISO(7)$-gauged maximal supergravity, \cite{Hull:1984yy} and \cite{Guarino:2015jca, Guarino:2015qaa, Guarino:2015vca}. In particular, the $U(1)^3$ subsector of the theory is of our interest. The field content is the metric, three $U(1)$ gauge fields, $A^I$, $I=1,2,3$, and four real scalar fields, $\phi_i$, $i=0,1,2,3$. For details of $D=4$ $U(1)^3$-gauged supergravity, we refer section 6 of \cite{Boisvert:2024jrl}.{\footnote{Refer v3 of \cite{Boisvert:2024jrl} on the arXiv which corrects several typographical errors in the previous version.}}

In this section, we study no charge disk solutions, $A^1=A^2=A^3=0$. We set the parameters, $q_1=q_2=q_3=0$, for the spindle solution given in (6.19) of \cite{Boisvert:2024jrl}. Then the solution is given by
\begin{align} \label{d2sol1}
ds_4^2\,=&\,\frac{y^{7/2}}{r^{1/3}}\left[\frac{-dt^2+dr^2}{r^2}+\frac{9}{4y^5h(y)}dy^2+h(y)dz^2\right]\,,  \notag \\
e^{\phi_0}\,=&\,\frac{y}{r^{2/3}}\,,\qquad e^{\phi_1}\,=\,e^{\phi_2}\,=\,e^{\phi_3}\,=\,\frac{y^{1/2}}{r^{1/3}}\,, \notag \\
A^1\,=&\,A^2\,=\,A^3\,=\,0\,,
\end{align}
and 
\begin{equation}
h(y)\,=\,\frac{9g^2\left(y^3-\frac{16}{9g^2}\right)}{16y^3}\,.
\end{equation}
For $h(y)=0$, there is a root, $y=2^{4/3}/(3g)^{2/3}$. We find solutions with
\begin{equation}
y_1\,\equiv\,\frac{2^{4/3}}{(3g)^{2/3}}<y<\infty\,.
\end{equation}
We plot a representative solution with $g=1$ in figure \ref{d210}. The metric functions, $f(y)$, $g_1(y)$, and $g_2(y)$, are defined by
\begin{equation}
ds_4^2\,=\,\frac{f(y)}{r^{1/3}}\left[\frac{-dt^2+dr^2}{r^2}+g_1(y)dy^2+g_2(y)dz^2\right]\,.
\end{equation}

\begin{figure}[t] 
\begin{center}
\includegraphics[width=2.0in]{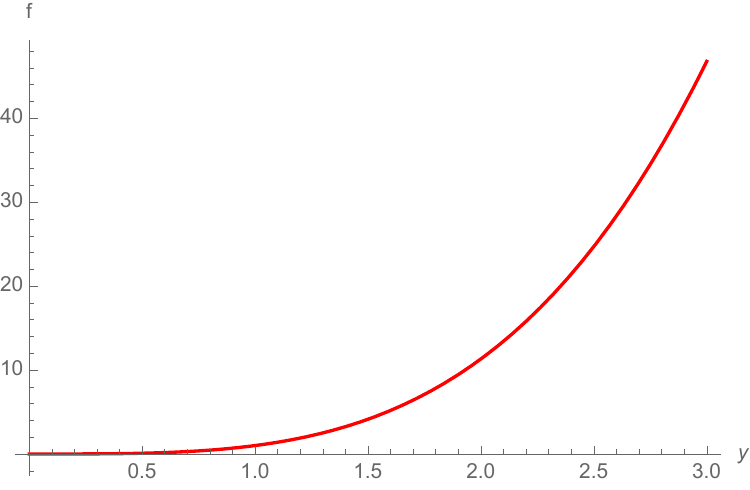} \qquad \includegraphics[width=2.0in]{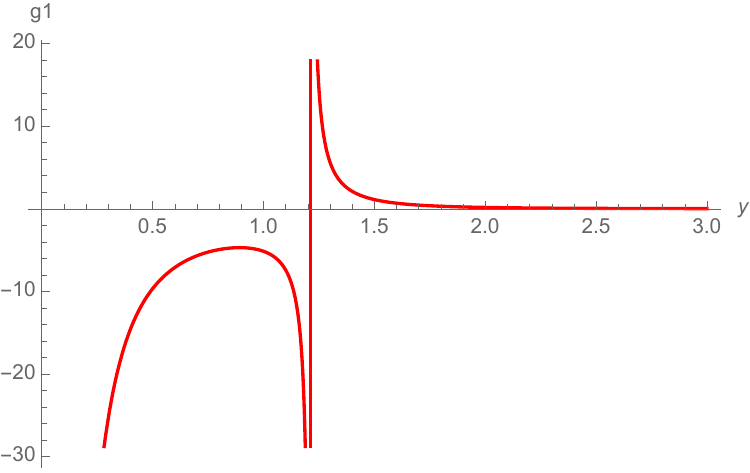} \qquad \includegraphics[width=2.0in]{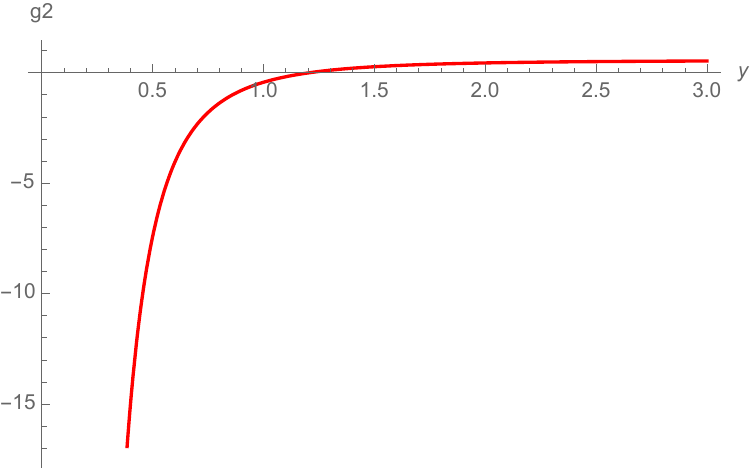}
\caption{{\it A representative solution with $g=1$. The solution is regular in the range of $y_1=2^{4/3}/3^{2/3}<y<\infty$.}} \label{d210}
\end{center}
\end{figure}

Near $y\rightarrow\infty$ the warp factor diverges and it is a curvature singularity of the metric, 
\begin{equation}
ds_4^2\,\approx\,\frac{y^{7/2}}{r^{1/3}}\left[\frac{-dt^2+dr^2}{r^2}+\frac{4}{g^2y^5}dy^2+\frac{9g^2}{16}dz^2\right]\,.
\end{equation}
Approaching $y\rightarrow{y}_1$, the metric becomes to be
\begin{equation}
ds_4^2\,\approx\,\frac{y_1^{7/2}}{r^{1/3}}\left[\frac{-dt^2+dr^2}{r^2}+\frac{d\rho^2+\mathcal{E}(q)^2\rho^2dz^2}{-y_1^5h'(y_1)/9}\right]\,,
\end{equation}
where we have
\begin{equation}
\mathcal{E}(q)^2\,\equiv\frac{1}{9}y_1^5h'(y_1)^2\,=\,\left(\frac{3g}{4}\right)^2\,,
\end{equation}
and we introduced a new parametrization of coordinate, $\rho^2\,=\,y_1-y$. Then,  the $\rho-z$ surface is locally an $\mathbb{R}^2/\mathbb{Z}_\ell$ orbifold if we set
\begin{equation} \label{ldz2}
\frac{\ell\Delta{z}}{2\pi}\,=\,\frac{4}{3g}\,,
\end{equation}
where $\Delta{z}$ is the period of coordinate, $z$, and $\ell\,=\,1,2,3,\ldots$. The metric spanned by $(y,z)$ has a topology of disk, $\Sigma$, with the center at $y=y_1$ and the boundary at $y=\infty$.

We calculate the Euler characteristic of the $y-z$ surface, $\Sigma$,
\begin{equation}
\chi\,=\,\frac{1}{4\pi}\int_\Sigma{R}_\Sigma\text{vol}_\Sigma\,=\,\frac{1}{4\pi}2\frac{3g}{4}\Delta{z}\,=\,\frac{1}{\ell}\,.
\end{equation}
This is a natural result for a disk in an $\mathbb{R}^2/\mathbb{Z}_\ell$ orbifold.

\subsubsection{Uplift to type IIA supergravity}

By employing the uplift formula to type IIA supergravity which we construct in appendix \ref{appC}, we obtain the uplifted metric in the string frame and the dilaton, respectively,
\begin{align}
&ds_{10}^2\,=r^{1/3}\left\{y^{5/2}\left[\frac{-dt^2+dr^2}{r^2}+\frac{9}{4y^5h(y)}dy^2+h(y)dz^2\right]\right. \notag \\
&\left.+\frac{1}{g^2y^{1/2}}\left[d\xi^2+\cos^2\xi\,d\varphi_1^2+\sin^2\xi\Big(d\theta^2+\cos^2\theta\left(d\psi^2+\cos^2\psi\,d\varphi_2^2+\sin^2\psi\,d\varphi_3^2\right)\Big)\right]\right\}\,, \notag \\
&e^{\Phi}\,=\,\frac{r^{5/6}}{y^{5/4}}\,.
\end{align}

We consider the singularity at $y\rightarrow\infty$ of the uplifted metric. As $y\rightarrow\infty$, the uplifted metric becomes
\begin{align}
&ds_{10}^2\,\approx\,r^{1/3}\left\{y^{5/2}\left[\frac{-dt^2+dr^2}{r^2}+\frac{9g^2}{16}dz^2\right]\right. \notag \\
&\left.+\frac{1}{g^2y^{5/2}}\Big[4dy^2+y^2\left(d\xi^2+\cos^2\xi\,d\varphi_1^2+\sin^2\xi\Big(d\theta^2+\cos^2\theta\left(d\psi^2+\cos^2\psi\,d\varphi_2^2+\sin^2\psi\,d\varphi_3^2\right)\Big)\right)\Big]\right\}\,,
\end{align}
and the dilaton is
\begin{equation}
e^\Phi\,=\,\frac{r^{5/6}}{y^{5/4}}\,.
\end{equation}
The branes are extended along the $\mathbb{R}^{1,4}$, $r$, $z$ directions and localized at the center of $y$, $\xi$, $\theta$, $\psi$, $\varphi_1$, $\varphi_2$, and $\varphi_3$ directions. We observe that there is no smeared direction. The $y^{5/2}$ and $1/y^{5/2}$ factors of the metric and the dilaton match the un-smeared branes reviewed in appendix \ref{appA}.

\subsection{Single charge solution: $A^1\ne0$, $A^2=A^3=0$: $y\in[y_1,\infty)$}

\subsubsection{$D=4$ gauged supergravity}

In this section, we study single charge disk solutions: $A^1\ne0$, $A^2=A^3=0$. We turn off two gauge fields, $A^2=A^3=0$, by setting $q_2=q_3=0$ and $q\equiv{q_1}$ for the spindle solution given in (6.19) of \cite{Boisvert:2024jrl}. Then the solution is given by
\begin{align} \label{d2sol1}
ds_4^2\,=&\,\frac{y^{5/2}\left(y^2+q\right)}{r^{1/3}}\left[\frac{-dt^2+dr^2}{r^2}+\frac{9}{4y^3\left(y^2+q\right)h(y)}dy^2+h(y)dz^2\right]\,,  \notag \\
e^{\phi_0}\,=&\,\frac{y}{r^{2/3}}\,,\qquad e^{\phi_1}\,=\,\frac{y^3}{y^2+q}\frac{1}{r^{1/3}}\,, \qquad e^{\phi_2}\,=\,e^{\phi_3}\,=\,\frac{y^2+q}{y}\frac{1}{r^{1/3}}\,, \notag \\
A\,\equiv&\,A^1\,=\,\left(\frac{q}{y^2+q}-\frac{3}{2}\right)dz\,, \qquad A^2\,=\,A^3\,=\,0\,,
\end{align}
and 
\begin{equation}
h(y)\,=\,\frac{9g^2\left[y^4\left(y^2+q\right)-\frac{16}{9g^2}y^3\right]}{16y^4\left(y^2+q\right)}\,.
\end{equation}
The gauge choice of $-3/2$ for the gauge field, $A$, is required to observe monopole structure in the uplifted solutions in \eqref{d2mono1}. For $h(y)=0$, there are one real and two complex roots and we denote the real root by $y=y_1$ and it is a function of $q$ and $g$. We do not present the expression as it is unwieldy. We consider solutions with
\begin{equation}
y_1<y<\infty\,.
\end{equation}
We plot a representative solution with $g=1$ and $q=0.5$ in figure \ref{d21one}. The metric functions, $f(y)$, $g_1(y)$, and $g_2(y)$, are defined by
\begin{equation}
ds_4^2\,=\,\frac{f(y)}{r^{1/3}}\left[\frac{-dt^2+dr^2}{r^2}+g_1(y)dy^2+g_2(y)dz^2\right]\,.
\end{equation}

\begin{figure}[t] 
\begin{center}
\includegraphics[width=2.0in]{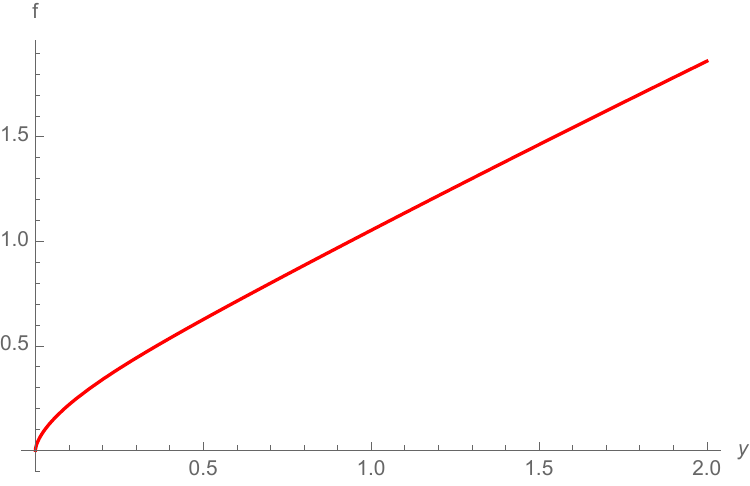} \qquad \includegraphics[width=2.0in]{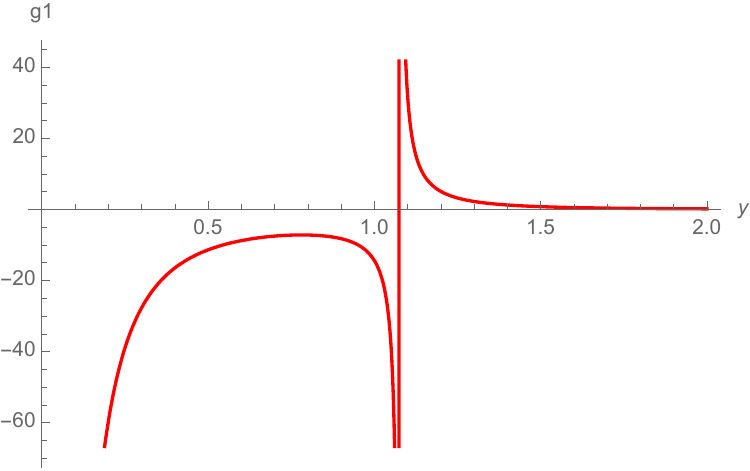} \qquad \includegraphics[width=2.0in]{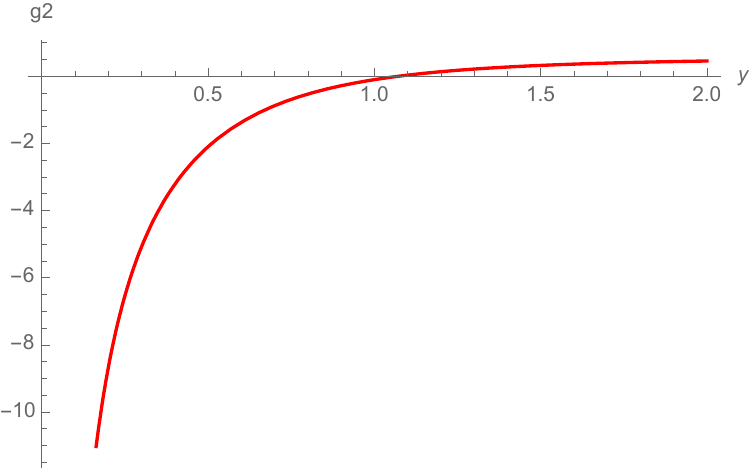}
\caption{{\it A representative solution with $g=1$, $q=0.5$. The solution is regular in the range of $y_1=1.0749<y<\infty$.}} \label{d21one}
\end{center}
\end{figure}

Near $y\rightarrow\infty$ the warp factor diverges and it is a curvature singularity of the metric, 
\begin{equation}
ds_4^2\,\approx\,\frac{y^{7/2}}{r^{1/3}}\left[\frac{-dt^2+dr^2}{r^2}+\frac{4}{g^2y^5}dy^2+\frac{9g^2}{16}dz^2\right]\,.
\end{equation}
Approaching $y\rightarrow{y}_1$, the metric becomes to be
\begin{equation}
ds_4^2\,\approx\,\frac{y_1^{5/2}\left(y_1^2+q\right)^{1/2}}{r^{1/3}}\left[\frac{-dt^2+dr^2}{r^2}+\frac{d\rho^2+\mathcal{E}(q)^2\rho^2dz^2}{-y_1^3\left(y_1^2+q\right)h'(y_1)/9}\right]\,,
\end{equation}
where we have
\begin{equation}
\mathcal{E}(q)^2\,\equiv\frac{1}{9}y_1^3\left(y_1^2+q\right)h'(y_1)^2\,=\,\left(\frac{3y_1^2+q}{3\sqrt{y_1}\left(y_1^2+q\right)^{3/2}}\right)^2\,,
\end{equation}
and we introduced a new parametrization of coordinate, $\rho^2\,=\,y_1-y$. Then,  the $\rho-z$ surface is locally an $\mathbb{R}^2/\mathbb{Z}_\ell$ orbifold if we set
\begin{equation} \label{ldz2}
\frac{\ell\Delta{z}}{2\pi}\,=\,\frac{1}{\mathcal{E}(q)}\,,
\end{equation}
where $\Delta{z}$ is the period of coordinate, $z$, and $\ell\,=\,1,2,3,\ldots$. The metric spanned by $(y,z)$ has a topology of disk, $\Sigma$, with the center at $y=y_1$ and the boundary at $y=\infty$.

We calculate the Euler characteristic of the $y-z$ surface, $\Sigma$,
\begin{equation}
\chi\,=\,\frac{1}{4\pi}\int_\Sigma{R}_\Sigma\text{vol}_\Sigma\,=\,\frac{1}{4\pi}2\mathcal{E}(q)\Delta{z}\,=\,\frac{1}{\ell}\,.
\end{equation}
This is a natural result for a disk in an $\mathbb{R}^2/\mathbb{Z}_\ell$ orbifold.

We perform the flux quantization of the $U(1)$ gauge fields, $A$,
\begin{equation} \label{gf2}
p\,\equiv\,\text{hol}_{\partial\Sigma}\left(A\right)\,=\,\frac{g}{2\pi}\oint_{y=\infty}A\,=\,-\frac{g}{2\pi}\int_\Sigma{F}\,=\,-\frac{g}{2\pi}\frac{q}{y_1^2+q}\Delta{z}\,.
\end{equation}
Although the expressions are unwieldy, by solving \eqref{ldz2} and \eqref{gf2}, one should be able to find that $q$ and $\Delta{z}$ can be expressed in terms of $\ell$ and $p$.

\subsubsection{Uplift to type IIA supergravity}

By employing the uplift formula to type IIA supergravity which we construct in appendix \ref{appC}, we obtain the uplifted metric in the string frame and the dilaton, respectively,
\begin{align} \label{d210one}
&ds_{10}^2\,=r^{1/3}y^{3/2}\widetilde{\Delta}^{1/2}\left[\frac{-dt^2+dr^2}{r^2}+\frac{9}{4y^3\left(y^2+q\right)h(y)}dy^2+h(y)dz^2\right. \notag \\
&\left.+\frac{1}{g^2y^3\widetilde{\Delta}}\Big\{\widetilde{\Delta}d\xi^2+\left(y^2+q\right)\cos^2\xi\,D\varphi_1^2+y^2\sin^2\xi\Big(d\theta^2+\cos^2\theta\left(d\psi^2+\cos^2\psi\,d\varphi_2^2+\sin^2\psi\,d\varphi_3^2\right)\Big)\Big\}\right]\,, \notag \\
&e^{4\Phi}\,=\,\frac{r^{10/3}}{y^3\widetilde{\Delta}}\,,
\end{align}
where we have
\begin{equation}
\widetilde{\Delta}\,=\,y^2+q\sin^2\xi\,,
\end{equation}
and
\begin{equation}
D\varphi_1\,=\,d\varphi_1-gA\,.
\end{equation}

The eight-dimensional internal space of the uplifted metric is an $S_z^1\times{S}_{\varphi_1}^1\times{S}^4$ fibration over the 2d base space, $B_2$, of $(y,\xi)$. The four-sphere, $S^4$, is spanned by $(\theta,\psi, \varphi_1, \varphi_2)$. The 2d base space is a rectangle of $(y,\xi)$ over $[y_1,\infty)\,\times\left[0,\frac{\pi}{2}\right]$. See figure \ref{figd21}. We explain the geometry of the internal space by three regions of the 2d base space, $B_2$.

\begin{itemize}
\item Region I: The side of $\mathsf{P}_1\mathsf{P}_2$.
\item Region II: The sides of $\mathsf{P}_2\mathsf{P}_3$ and $\mathsf{P}_3\mathsf{P}_4$.
\item Region III: The side of $\mathsf{P}_1\mathsf{P}_4$.
\end{itemize}

\begin{figure}[t] 
\begin{center}
\includegraphics[width=4.5in]{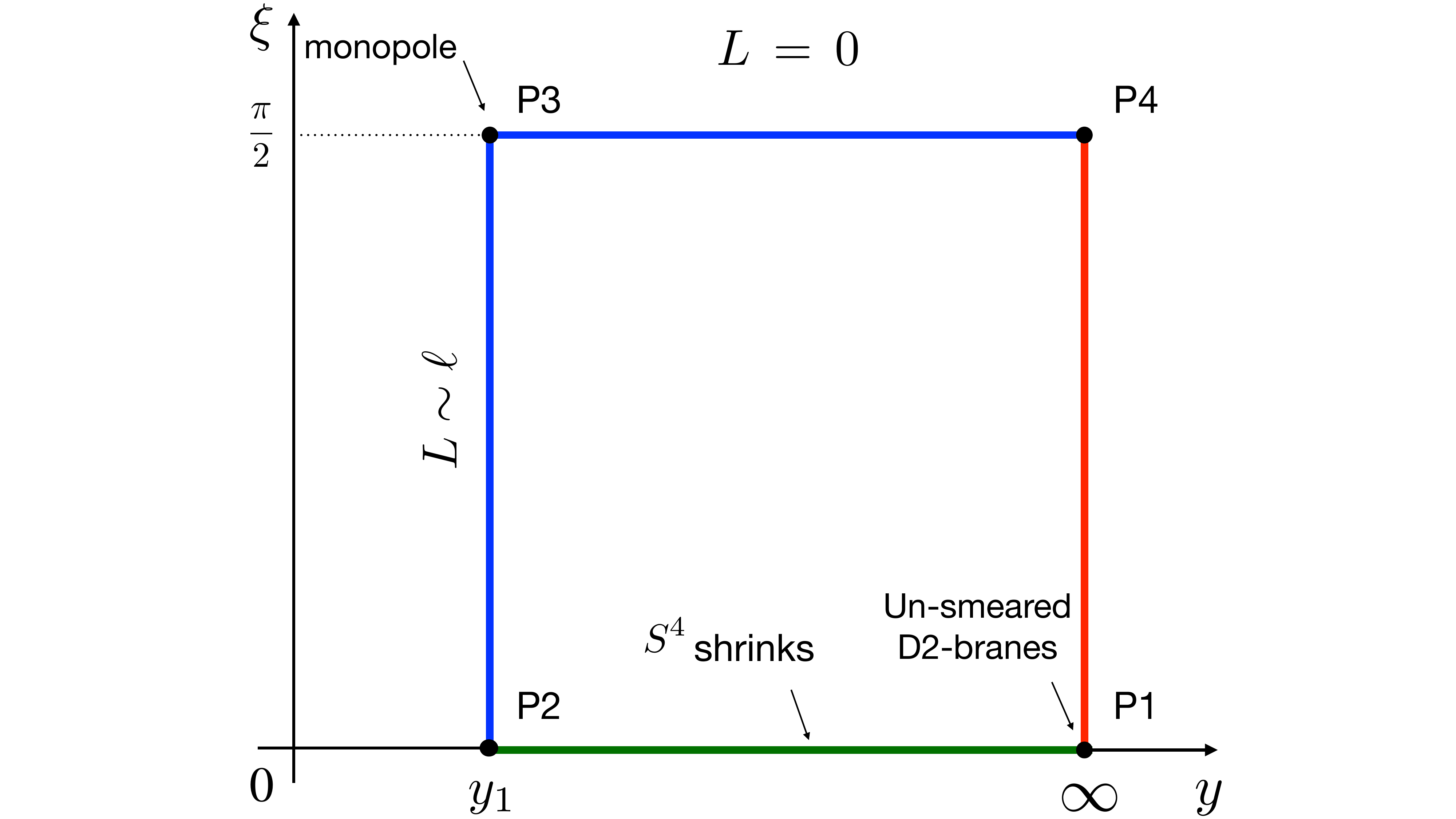}
\caption{{\it The two-dimensional base space, $B_2$, spanned by $y$ and $\xi$.}} \label{figd21}
\end{center}
\end{figure}

\noindent {\bf Region I:} On the side of $\xi\,=\,0$, the four-sphere, $S^4$, shrinks and the internal space caps off.

\bigskip

\noindent {\bf Region II: Monopole} We break $D\varphi_1$ and complete the square of $dz$ to obtain the metric of
\begin{align} \label{d2mono1}
ds_{10}^2\,&=\,r^{1/3}y^{3/2}\widetilde{\Delta}^{1/2}\left[\frac{-dt^2+dr^2}{r^2}+\frac{9}{4y^3\left(y^2+q\right)h(y)}dy^2\right. \notag \\
&+\frac{1}{g^2y^3}\Big\{d\xi^2+\widetilde{\Delta}^{-1}y^2\sin^2\xi\Big(d\theta^2+\cos^2\theta\left(d\psi^2+\cos^2\psi\,d\varphi_2^2+\sin^2\psi\,d\varphi_3^2\right)\Big)\Big\} \notag \\
&+R_z^2\left(dz-Ld\varphi_1\right)^2+R_{\chi_1}^2d\varphi_1^2\Big]\,.
\end{align}
The metric functions are defined to be
\begin{align}
R_z^2\,=&\,\frac{18g^2y^5+4y^2+q\left(9g^2y^3+4\right)+\left[36y^2-q\left(9g^2y^3-4\right)\right]\cos\left(2\xi\right)}{32y^3(y^2+q\sin^2\xi)}\,, \notag \\
R_{\varphi_1}^2\,=&\,\frac{2\left(9g^2y\left(y^2+q\right)-16\right)\cos^2\xi}{g^2y\left\{18g^2y^5+4y^2+q\left(9g^2y^3+4\right)+\left[36y^2-q\left(9g^2y^3-4\right)\right]\cos\left(2\xi\right)\right\}}\,, \notag \\
L\,=&\,\frac{16\left(3y^2+q\right)\cos^2\xi}{g\left\{18g^2y^5+4y^2+q\left(9g^2y^3+4\right)+\left[36y^2-q\left(9g^2y^3-4\right)\right]\cos\left(2\xi\right)\right\}}\,.
\end{align}

The function, $L(y,\xi)$, is piecewise constant along the sides of $y\,=\,y_1$ and $\xi\,=\,\pi/2$ of the 2d base, $B_2$,
\begin{equation}
L\left(y,\pi/2\right)\,=\,0\,, \qquad L\left(y_1,\xi\right)\,=\,\frac{4}{\mathcal{E}(q)}\,=\,4\frac{\ell\Delta{z}}{2\pi}\,,
\end{equation}
The jump in $L$ at the corner, $(y,\xi)\,=\,\left(y_1,\pi/2\right)$, indicates the existence of a monopole source for the $Dz$ fibration. The gauge choice of $-3/2$ for the gauge field, $A$, in \eqref{d2sol1} is required to observe monopole structure in the uplifted solutions.

\bigskip

\noindent {\bf Region III: Un-smeared D2-branes} We consider the singularity at $y\rightarrow\infty$ of the uplifted metric. As $y\rightarrow\infty$, the uplifted metric becomes
\begin{equation}
ds_{10}^2\,\approx\,r^{1/3}\left[y^{5/2}\left(\frac{-dt^2+dr^2}{r^2}+\frac{9g^2}{16}dz^2\right)+\frac{1}{g^2}\frac{1}{y^{5/2}}\left(4dy^2+y^2ds_{S^6}^2\right)\right]\,,
\end{equation}
and the dilaton is
\begin{equation}
e^\Phi\,\approx\,\frac{r^{5/6}}{y^{5/4}}\,.
\end{equation}
The metric does not get smeared by D2-brane sources. The D2-branes are 
\begin{itemize}
\item extended along the $t$, $r$, and $z$ directions;
\item localized at the origin of $y$ and $S^6$ directions;
\item not smeared along any directions. 
\end{itemize}
The $y^{5/2}$ and $1/y^{5/2}$ factors of the metric and the dilaton match the un-smeared branes reviewed in appendix \ref{appA}.

\subsection{Equal charge solution: $A^1=A^2\ne0$, $A^3=0$: $y\in[0,y_1]$}

\subsubsection{$D=4$ gauged supergravity}

In this section, we study equal charge disk solutions: $A^1=A^2\ne0$, $A^3=0$. We turn off one of the  gauge fields, $A^3=0$, by setting $q_3=0$ and $q\equiv{q_1}=q_2$ for the spindle solution given in (6.19) of \cite{Boisvert:2024jrl}. Then the solution is given by
\begin{align} \label{d2sol1twotwo}
ds_4^2\,=&\,\frac{y^{3/2}\left(y^2+q\right)}{r^{1/3}}\left[\frac{-dt^2+dr^2}{r^2}+\frac{9}{4y\left(y^2+q\right)^2h(y)}dy^2+h(y)dz^2\right]\,,  \notag \\
e^{\phi_0}\,=&\,\frac{y}{r^{2/3}}\,,\qquad e^{\phi_1}\,=\,e^{\phi_2}\,=\,\frac{y^{1/2}}{r^{1/3}}\,, \qquad e^{\phi_3}\,=\,\frac{y^2+q}{y^{3/2}}\frac{1}{r^{1/3}}\,, \notag \\
A\,\equiv&\,A^1\,=A^2\,=\,\left(\frac{q}{y^2+q}-\frac{3}{4}\right)dz\,, \qquad A^3\,=\,0\,,
\end{align}
and 
\begin{equation}
h(y)\,=\,\frac{9g^2\left[y^2\left(y^2+q\right)^2-\frac{16}{9g^2}y^3\right]}{16y^2\left(y^2+q\right)^2}\,.
\end{equation}
The gauge choice of $-3/4$ for the gauge field, $A$, is required to observe monopole structure in the uplifted solutions in \eqref{d2mono1}. For $h(y)=0$, there are two real and two complex roots and we denote two real roots by $y=y_1$ and $y=y_2$ where $y_1<y_2$ and they are functions of $q$ and $g$. We do not present the expressions as they are unwieldy. We consider solutions with{\footnote{There is also a solution with $y_2<y<\infty$. We briefly discuss the solution in section \ref{d2a}.}
\begin{equation}
0<y<y_1\,.
\end{equation}
We plot a representative solution with $g=1$ and $q=0.5$ in figure \ref{d21twotwo}. The metric functions, $f(y)$, $g_1(y)$, and $g_2(y)$, are defined by
\begin{equation}
ds_4^2\,=\,\frac{f(y)}{r^{1/3}}\left[\frac{-dt^2+dr^2}{r^2}+g_1(y)dy^2+g_2(y)dz^2\right]\,.
\end{equation}

\begin{figure}[t] 
\begin{center}
\includegraphics[width=2.0in]{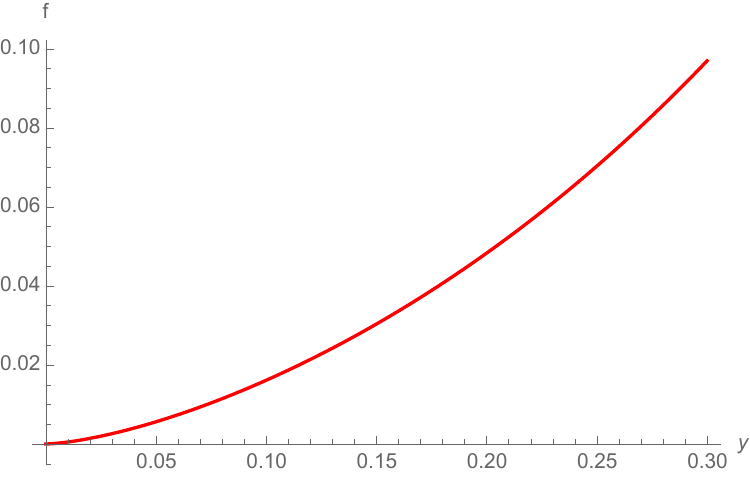} \qquad \includegraphics[width=2.0in]{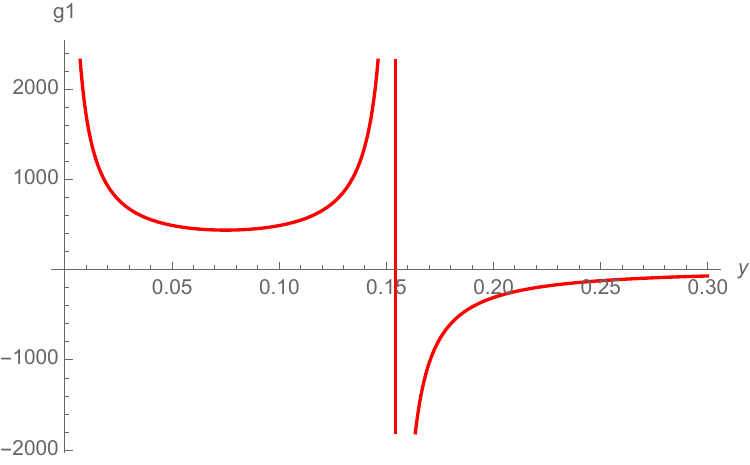} \qquad \includegraphics[width=2.0in]{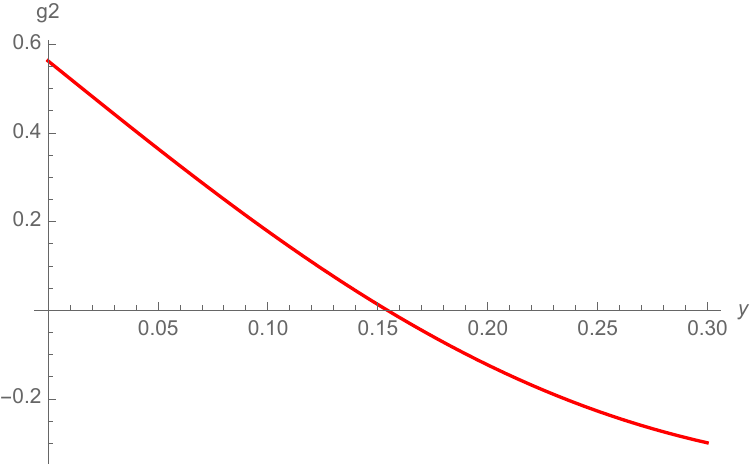}
\caption{{\it A representative solution with $g=1$, $q=0.5$. The solution is regular in the range of $0<y<y_1=0.154$.}} \label{d21twotwo}
\end{center}
\end{figure}

Near $y\rightarrow0$ the warp factor diverges and it is a curvature singularity of the metric, 
\begin{equation}
ds_4^2\,\approx\,\frac{q^{1/2}y^{3/4}}{r^{1/3}}\left[\frac{-dt^2+dr^2}{r^2}+\frac{4}{g^2q^2y}dy^2+\frac{9g^2}{16}dz^2\right]\,.
\end{equation}
Approaching $y\rightarrow{y}_1$, the metric becomes to be
\begin{equation}
ds_4^2\,\approx\,\frac{y_1^{3/2}\left(y_1^2+q\right)}{r^{1/3}}\left[\frac{-dt^2+dr^2}{r^2}+\frac{d\rho^2+\mathcal{E}(q)^2\rho^2dz^2}{-y_1\left(y_1^2+q\right)^2h'(y_1)/9}\right]\,,
\end{equation}
where we have
\begin{equation}
\mathcal{E}(q)^2\,\equiv\frac{1}{9}y_1\left(y_1^2+q\right)^2h'(y_1)^2\,=\,\left(\frac{\sqrt{y_1}\left(3y_1^2-q\right)}{3\left(y_1^2+q\right)^2}\right)^2\,,
\end{equation}
and we introduced a new parametrization of coordinate, $\rho^2\,=\,y_1-y$. Then,  the $\rho-z$ surface is locally an $\mathbb{R}^2/\mathbb{Z}_\ell$ orbifold if we set
\begin{equation} \label{ldz2twotwo}
\frac{\ell\Delta{z}}{2\pi}\,=\,\frac{1}{\mathcal{E}(q)}\,,
\end{equation}
where $\Delta{z}$ is the period of coordinate, $z$, and $\ell\,=\,1,2,3,\ldots$. The metric spanned by $(y,z)$ has a topology of disk, $\Sigma$, with the center at $y=y_1$ and the boundary at $y=\infty$.

We calculate the Euler characteristic of the $y-z$ surface, $\Sigma$,
\begin{equation}
\chi\,=\,\frac{1}{4\pi}\int_\Sigma{R}_\Sigma\text{vol}_\Sigma\,=\,\frac{1}{4\pi}2\mathcal{E}(q)\Delta{z}\,=\,\frac{1}{\ell}\,.
\end{equation}
This is a natural result for a disk in an $\mathbb{R}^2/\mathbb{Z}_\ell$ orbifold.

We perform the flux quantization of the $U(1)$ gauge fields, $A$,
\begin{equation} \label{gf2twotwo}
p\,\equiv\,\text{hol}_{\partial\Sigma}\left(A\right)\,=\,\frac{g}{2\pi}\oint_{y=0}A\,=\,-\frac{g}{2\pi}\int_\Sigma{F}\,=\,-\frac{g}{2\pi}\frac{q}{y_1^2+q}\Delta{z}\,.
\end{equation}
Although the expressions are unwieldy, by solving \eqref{ldz2twotwo} and \eqref{gf2twotwo}, one should be able to find that $q$ and $\Delta{z}$ can be expressed in terms of $\ell$ and $p$.

\subsubsection{Uplift to type IIA supergravity}

By employing the uplift formula to type IIA supergravity which we construct in appendix \ref{appC}, we obtain the uplifted metric in the string frame and the dilaton, respectively,
\begin{align} \label{d210two}
ds_{10}^2\,=&\,r^{1/3}\widetilde{\Delta}^{1/2}\left[\frac{-dt^2+dr^2}{r^2}+\frac{9}{4y\left(y^2+q\right)^2h(y)}dy^2+h(y)dz^2\right. \notag \\
&+\frac{1}{g^2\widetilde{\Delta}}\Big(\left(y^2+q\sin^2\xi\right)d\xi^2+\left(y^2+q\right)\cos^2\xi\left(d\theta^2+\cos^2\theta\,D\varphi_1^2+\sin^2\theta\,D\varphi_2^2\right) \notag \\
&\left.+y^2\sin^2\xi\left(d\psi^2+\cos^2\psi\,d\varphi_3^3\right)\Big)\right]\,, \\
e^{4\Phi}\,=&\,\frac{r^{10/3}}{\widetilde{\Delta}}\,,
\end{align}
where we have
\begin{equation}
\widetilde{\Delta}\,=\,y\left(y^2+q\right)\left(y^2+q\sin^2\xi\right)\,,
\end{equation}
and
\begin{equation}
D\varphi_1\,=\,d\varphi_1-gA\,, \qquad D\varphi_2\,=\,d\varphi_2-gA\,.
\end{equation}

The eight-dimensional internal space of the uplifted metric is an $S_z^1\times{S}^3\times{S}^2$ fibration over the 2d base space, $B_2$, of $(y,\xi)$. The two-sphere, $S^2$, is spanned by $(\psi, \varphi_3)$. The 2d base space is a rectangle of $(y,\xi)$ over $[0,y_1]\,\times\left[0,\frac{\pi}{2}\right]$. See figure \ref{figd21twotwo}. The internal space is more involved than the solutions we previously studied. Thus we only present the metric and the dilaton at the boundary of the disk.

\begin{itemize}
\item Region I: The side of $\mathsf{P}_1\mathsf{P}_2$.
\item Region II: The sides of $\mathsf{P}_2\mathsf{P}_3$ and $\mathsf{P}_3\mathsf{P}_4$.
\item Region III: The side of $\mathsf{P}_1\mathsf{P}_4$.
\end{itemize}

\begin{figure}[t] 
\begin{center}
\includegraphics[width=4.5in]{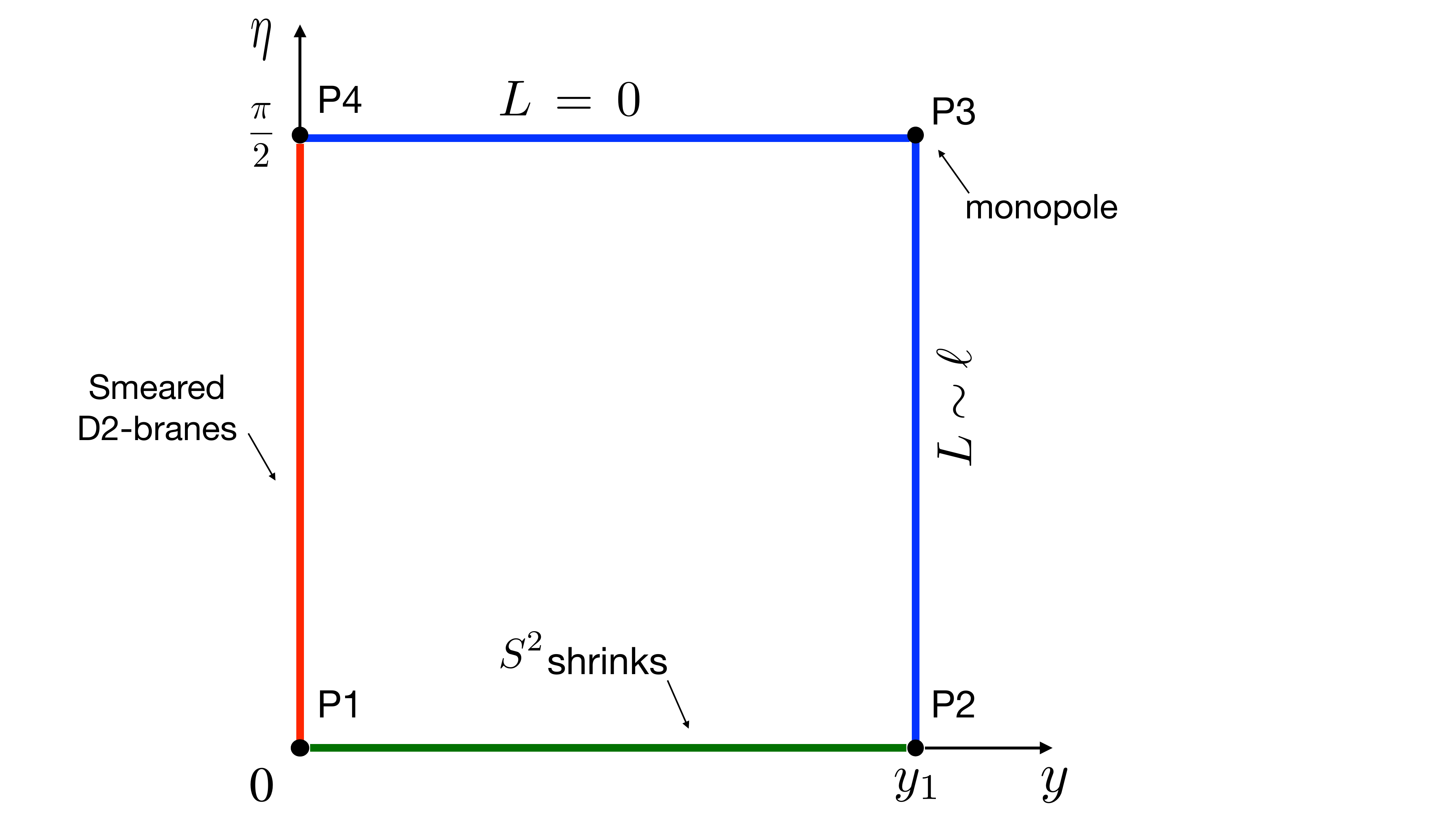}
\caption{{\it The two-dimensional base space, $B_2$, spanned by $y$ and $\xi$.}} \label{figd21twotwo}
\end{center}
\end{figure}

\noindent {\bf Region I:} On the side of $\xi\,=\,0$, the two-sphere, $S^2$, shrinks and the internal space caps off.

\bigskip

\noindent {\bf Region II: Monopole} We break $D\varphi_1$ and $D\varphi_2$ and complete the square of $dz$ to obtain the metric of
\begin{align} \label{d2mono1}
ds_{10}^2\,=&\,r^{1/3}\widetilde{\Delta}^{1/2}\left[\frac{-dt^2+dr^2}{r^2}+\frac{9}{4y\left(y^2+q\right)^2h(y)}dy^2\right. \notag \\
&+\frac{1}{g^2\widetilde{\Delta}}\Big(\left(y^2+q\sin^2\xi\right)d\xi^2+\left(y^2+q\right)\cos^2\xi\,d\theta^2 \notag \\
&+y^2\sin^2\xi\left(d\psi^2+\cos^2\psi\,d\varphi_3^3\right)\Big) \notag \\
&\left.+R_z^2\Big(dz+L\left(\cos^2\theta\,d\varphi_1+\sin^2\theta\,d\varphi_2\right)\Big)^2+R_{\varphi_1}^2\cos^2\theta\left(d\varphi_1-L_1\sin^2\theta\,d\varphi_2\right)^2+R_{\varphi_2}^2\sin^2\theta\,d\varphi_2^2\right]\,.
\end{align}
The metric functions are defined to be
\begin{align}
L\,=&\,\frac{-4\left(3y^2-q\right)\left(y^2+q\right)\cos^2\xi}{\left(3y^2-q\right)^2\cos^2\xi+g^2y\left(9g^2\left(y^2+q\right)^2-16y\right)\left(y^2+q\sin^2\xi\right)}\,, \notag \\
R_z^2\,=&\,\frac{\left(3y^2-q\right)^2\cos^2\xi+g^2y\left(9g^2\left(y^2+q\right)^2-16y\right)\left(y^2+q\sin^2\xi\right)}{16g^2y\left(y^2+q\right)^2\left(y^2+q\sin^2\xi\right)}\,, \notag \\
R^2_{\varphi_1}\,=&\,\frac{g^2y\left(9g^2\left(y^2+q\right)^2-16y\right)\left(y^2+q\sin^2\xi\right)\cos^2\xi+\left(3y^2-q\right)^2\cos^4\xi\sin^2\theta}{g^2y\left[g^2y\left(9g^2\left(y^2+q\right)^2-16y\right)\left(y^2+q\sin^2\xi\right)+\left(3y^2-q\right)^2\cos^2\xi\right]\left(y^2+q\sin^2\xi\right)}\,, \notag \\
R^2_{\varphi_2}\,=&\,\frac{\left(9g^2\left(y^2+q\right)^2-16y\right)\cos^2\xi}{g^2y\left(9g^2\left(y^2+q\right)^2-16y\right)\left(y^2+q\sin^2\xi\right)+\left(3y^2-q\right)^2\cos^2\xi\sin^2\theta}\,, \notag \\
L_1\,=&\,\frac{\left(3y^2-q\right)^2\cos^2\xi}{g^2y\left(9g^2\left(y^2+q\right)^2-16y\right)\left(y^2+q\sin^2\xi\right)+\left(3y^2-q\right)^2\cos^2\xi\sin^2\theta}\,.
\end{align}

The function, $L(y,\xi)$, is piecewise constant along the sides of $y\,=\,y_1$ and $\xi\,=\,\pi/2$ of the 2d base, $B_2$,
\begin{equation}
L\left(y,\pi/2\right)\,=\,0\,, \qquad L\left(y_1,\xi\right)\,=\,\frac{1}{\mathcal{E}(q)}\,=\,\frac{\ell\Delta{z}}{2\pi}\,,
\end{equation}
The jump in $L$ at the corner, $(y,\xi)\,=\,\left(y_1,\pi/2\right)$, indicates the existence of a monopole source for the $Dz$ fibration. The gauge choice of $-3/4$ for the gauge field, $A$, in \eqref{d2sol1twotwo} is required to observe monopole structure in the uplifted solutions.

\bigskip

\noindent {\bf Region III: Smeared D2-branes} We consider the singularity at $y\rightarrow0$ of the uplifted metric. As $y\rightarrow0$, the uplifted metric becomes
\begin{align}
&ds_{10}^2\,\approx\,r^{1/3}\left[\left(q^2y\sin^2\xi\right)^{1/2}\left(\frac{-dt^2+dr^2}{r^2}+\frac{9g^2}{16}dz^2\right)\right. \notag \\
&\left.+\frac{1}{g^2}\frac{\sin^2\xi}{\left(q^2y\sin^2\xi\right)^{1/2}}\Big(4dy^2+y^2\left(d\psi^2+\cos^2\psi\,d\varphi_3^2\right)+qd\xi^2+q\cot^2\xi\left(d\theta^2+\cos^2\theta\,D\varphi_1^2+\sin^2\theta\,D\varphi_2^2\right)\Big)\right]\,,
\end{align}
and the dilaton is
\begin{equation}
e^\Phi\,\approx\,\frac{r^{5/6}}{\left(q^2y\sin^2\xi\right)^{1/4}}\,.
\end{equation}
The metric implies the smeared D2-brane sources. The D2-branes are 
\begin{itemize}
\item extended along the $t$, $r$, and $z$ directions;
\item localized at the origin of $y$, $\psi$, and $\varphi_3$ directions;
\item smeared along $\xi$, $\theta$, $\varphi_1$, and $\varphi_2$ directions. 
\end{itemize}
The $y^{1/2}$ and $1/y^{1/2}$ factors of the metric and the dilaton match the smeared branes reviewed in appendix \ref{appA}.

\subsubsection{Another solution: $y\in[y_2,\infty)$} \label{d2a}

We studied a solution with $0<y<y_1$ in \eqref{d2sol1twotwo}. In this section, we consider another solution with
\begin{equation}
y_2<y<\infty\,,
\end{equation}
where $y_2$ is another real root of $h(y)=0$ and is a function of $q$ and $g$. We plot a representative solution with $g=1$ and $q=0.5$ in figure \ref{d22atwotwo}.

\begin{figure}[t] 
\begin{center}
\includegraphics[width=2.0in]{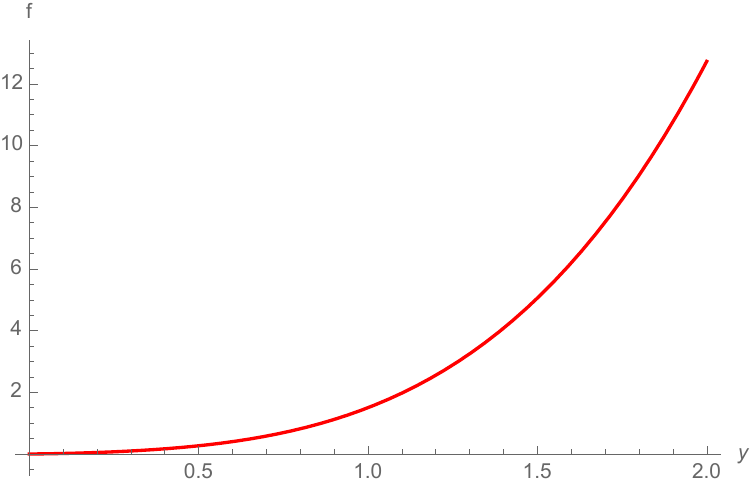} \qquad \includegraphics[width=2.0in]{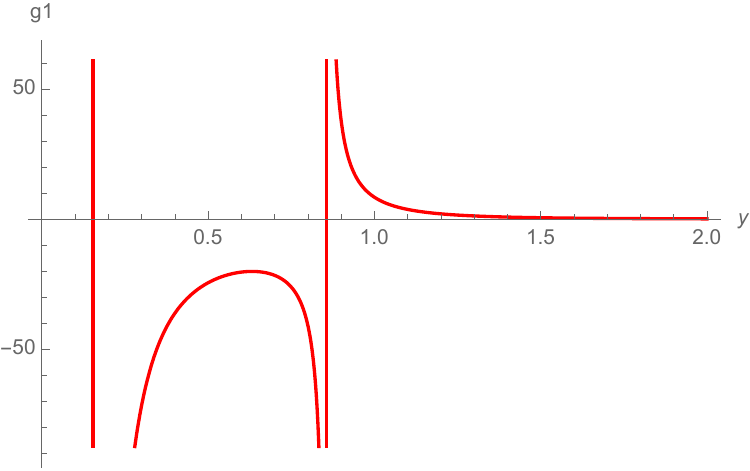} \qquad \includegraphics[width=2.0in]{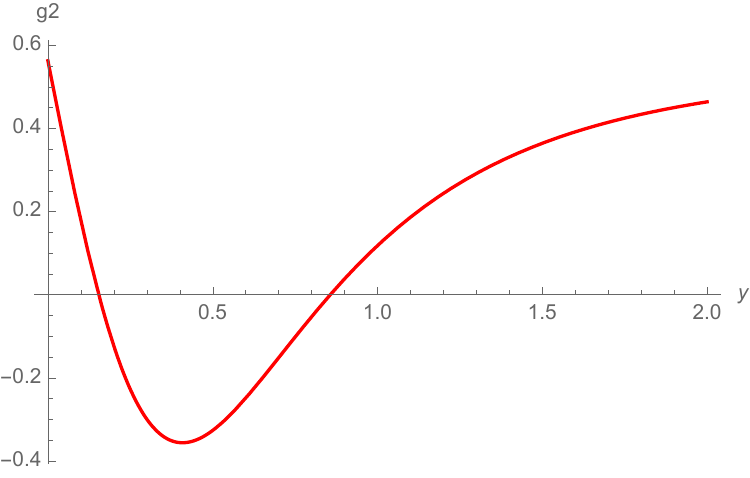}
\caption{{\it A representative solution with $g=1$ and $q=0.5$. The solution is regular in the range of $y_2=0.856<y<\infty$.}} \label{d22atwotwo}
\end{center}
\end{figure}

\begin{figure}[t] 
\begin{center}
\includegraphics[width=4.5in]{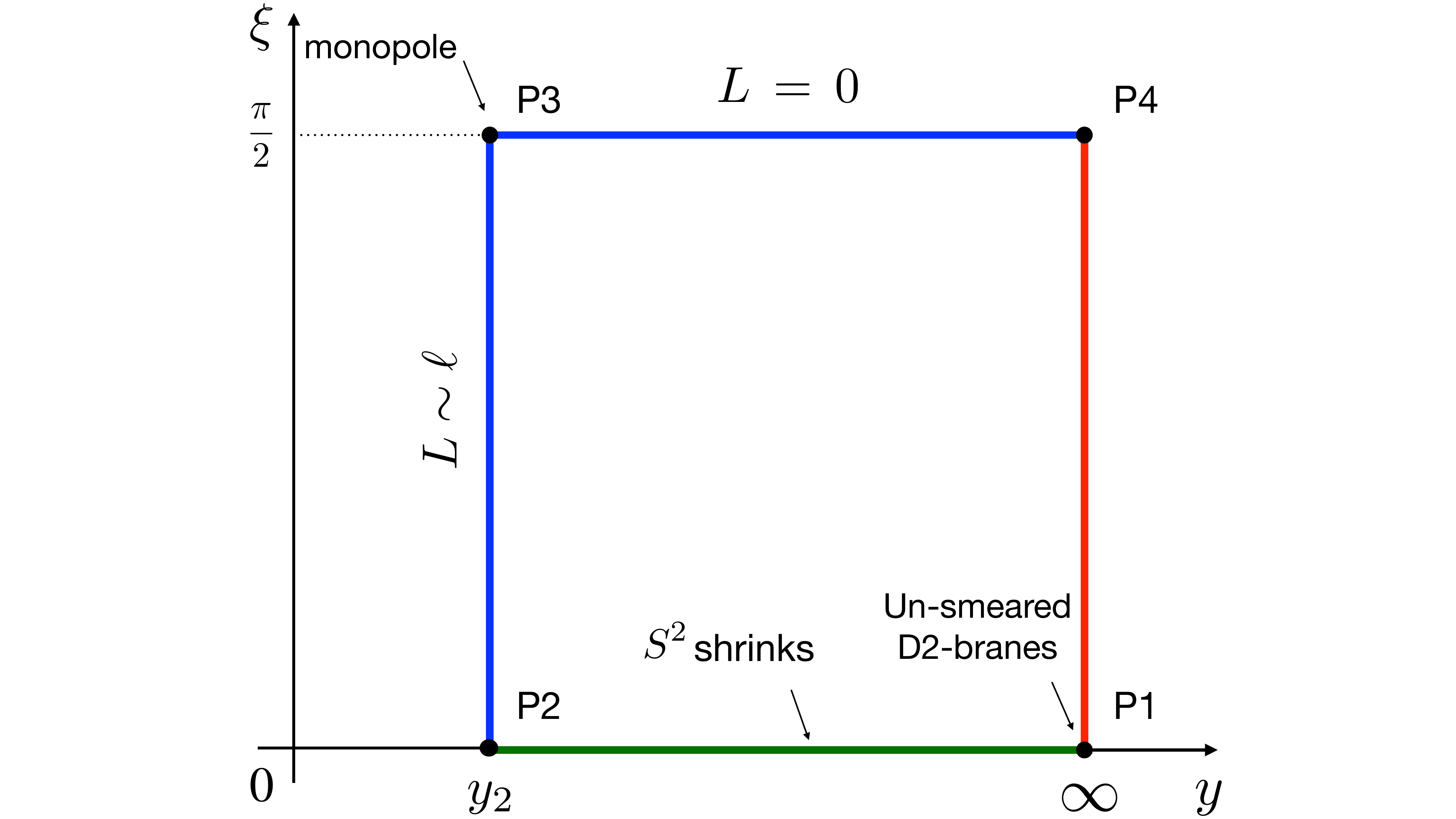}
\caption{{\it The two-dimensional base space, $B_2$, spanned by $y$ and $\xi$.}} \label{figd2aa}
\end{center}
\end{figure}

Beside the structure of the solution at $y\rightarrow\infty$, which we explain below, rest of the details of the solution is identical to the solution with \eqref{d2sol1twotwo} by switching $y_1$ to $y_2$. Of course, the holographic observables, if we could calculate,  will be different for solutions. We summarize the global structure of the uplifted metric in figure \ref{figd2aa}.

\bigskip

\noindent {\bf Region III: Un-smeared D2-branes} We consider the singularity at $y\rightarrow\infty$ of the uplifted metric. As $y\rightarrow\infty$, the uplifted metric becomes
\begin{align} \label{d2unsmeared}
&ds_{10}^2\,\approx\,r^{1/3}\left\{y^{5/2}\left[\frac{-dt^2+dr^2}{r^2}+\frac{9g^2}{16}dz^2\right]\right. \notag \\
&\left.+\frac{1}{g^2y^{5/2}}\left[4dy^2+y^2\Big(d\xi^2+\cos^2\xi\left(d\theta^2+\cos^2\theta{D}\varphi_1^2+\sin^2\theta{D}\varphi_2^2\right)+\sin^2\xi\left(d\psi^2+\cos^2\psi\,d\varphi_3^2\right)\Big)\right]\right\}\,,
\end{align}
and the dilaton is
\begin{equation}
e^\Phi\,\approx\,\frac{y^{5/6}}{r^{5/4}}\,.
\end{equation}
The metric does not get smeared by D2-brane sources. The D2-branes are 
\begin{itemize}
\item extended along the $t$, $r$, and $z$ directions;
\item localized at the origin of $y$, $S^6$ directions;
\item not smeared along any directions. 
\end{itemize}
The $y^{5/2}$ and $1/y^{5/2}$ factors of the metric and the dilaton match the un-smeared branes reviewed in appendix \ref{appA}.

\vspace{0.8cm}

\subsection{Multi charge solution: $A_1\ne{A}_2$, $A_3=0$: $y\in[0,y_1]$, $y\in[y_2,\infty)$}

\subsubsection{$D=4$ gauged supergravity}

In this section, we study multi charge disk solutions: $A_1\ne{A}_2$, $A_3=0$. We turn off one of the gauge fields, $A^3=0$, by setting $q_3=0$ for the spindle solution given in (6.19) of \cite{Boisvert:2024jrl}. Then the solution is given by
\begin{align}
ds_4^2\,=&\,\frac{y^{3/2}\left(y^2+q_1\right)^{1/2}\left(y^2+q_2\right)^{1/2}}{r^{1/3}}\left[\frac{-dt^2+dr^2}{r^2}+\frac{9}{4y\left(y^2+q_1\right)\left(y^2+q_2\right)h(y)}dy^2+h(y)dz^2\right]\,,  \notag \\
e^{\phi_0}\,=&\,\frac{y}{r^{2/3}}\,, \notag \\
\qquad e^{\phi_1}\,=&\,\frac{1}{r^{1/3}}\sqrt{y\frac{y^2+q_2}{y^2+q_1}}\,, \qquad e^{\phi_2}\,=\,\frac{1}{r^{1/3}}\sqrt{y\frac{y^2+q_1}{y^2+q_2}}\,, \qquad e^{\phi_3}\,=\,\frac{1}{r^{1/3}}\sqrt{\frac{\left(y^2+q_1\right)\left(y^2+q_2\right)}{y^3}}\,, \notag \\
A^1\,=&\,\frac{q_1}{y^2+q_1}dz\,, \qquad A^2\,=\,\frac{q_2}{y^2+q_2}dz\,, \qquad A^3\,=\,0\,,
\end{align}
and 
\begin{equation}
h(y)\,=\,\frac{9g^2\left[y^2\left(y^2+q_1\right)\left(y^2+q_2\right)-\frac{16}{9g^2}y^3\right]}{16y^2\left(y^2+q_1\right)\left(y^2+q_2\right)}\,.
\end{equation}
For $h(y)=0$, there are two real and two complex roots and we denote two real roots by $y=y_1$ and $y=y_2$ where $y_1<y_2$ and they are functions of $q_1$, $q_2$, and $g$. We do not present the expressions as they are unwieldy. In this case, there are two classes of solutions with
\begin{equation}
0<y<y_1\,,
\end{equation}
and
\begin{equation}
y_2<y<\infty\,.
\end{equation}
We plot a representative solution with $g=4/3$, $q_1=3/10$, $q_2=7/10$ in figure \ref{d21}. The metric functions, $f(y)$, $g_1(y)$, and $g_2(y)$, are defined by
\begin{equation}
ds_4^2\,=\,\frac{f(y)}{r^{1/3}}\left[\frac{-dt^2+dr^2}{r^2}+g_1(y)dy^2+g_2(y)dz^2\right]\,.
\end{equation}

\begin{figure}[t] 
\begin{center}
\includegraphics[width=2.0in]{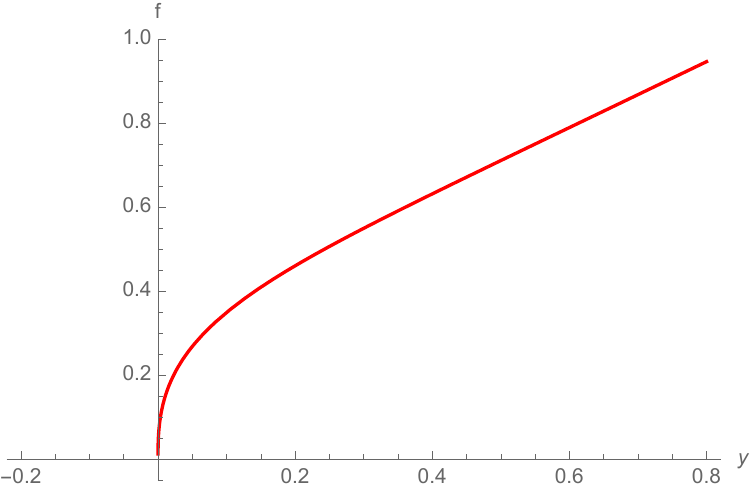} \qquad \includegraphics[width=2.0in]{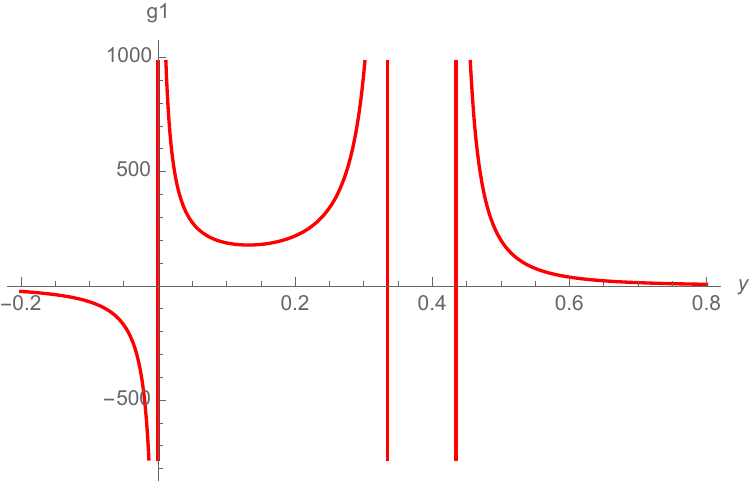} \qquad \includegraphics[width=2.0in]{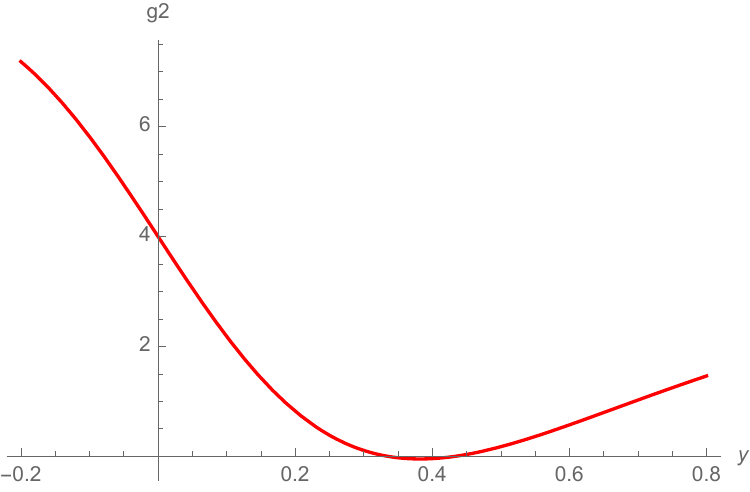}
\caption{{\it Representative solutions with $g=4/3$, $q_1=3/10$, $q_2=7/10$. A solution is regular in the range of $0\,<\,y\,<\,y_1=0.334$ and another in $y_2=0.434<y<\infty$.}} \label{d21}
\end{center}
\end{figure}

Near $y\rightarrow0$ the warp factor vanishes and it is a curvature singularity of the metric, 
\begin{equation}
ds_4^2\,\approx\,\frac{q_1^{1/2}q_2^{1/2}y^{3/2}}{r^{1/3}}\left[\frac{-dt^2+dr^2}{r^2}+\frac{4}{g^2q_1q_2y}dy^2+\frac{9g^2}{16}dz^2\right]\,.
\end{equation}
Near $y\rightarrow\infty$ the warp factor diverges and it is a curvature singularity of the metric, 
\begin{equation}
ds_4^2\,\approx\,\frac{y^{7/2}}{r^{1/3}}\left[\frac{-dt^2+dr^2}{r^2}+\frac{4}{g^2y^5}dy^2+\frac{9g^2}{16}dz^2\right]\,.
\end{equation}
Approaching $y\rightarrow{y}_a$, $a=1,2$, the metric becomes to be
\begin{equation}
ds_4^2\,\approx\,\frac{y_a^{7/2}}{r^{1/3}}\left[\frac{-dt^2+dr^2}{r^2}+\frac{d\rho^2+\mathcal{E}(q_1,q_2)^2\rho^2dz^2}{-y_a\left(y_a^2+q_1\right)\left(y_a^2+q_2\right)h'(y_a)/9}\right]\,,
\end{equation}
where we have
\begin{equation}
\mathcal{E}(q_1,q_2)^2\,\equiv\frac{1}{9}y_a\left(y_a^2+q_1\right)\left(y_a^2+q_2\right)h'(y_a)^2\,=\,\frac{y_a\left(3y_a^4+(q_1+q_2)y_a^2-q_1q_2\right)^2}{9\left(y_a^2+q_1\right)^3\left(y_a^2+q_2\right)^3}\,,
\end{equation}
and we introduced a new parametrization of coordinate, $\rho^2\,=\,y_a-y$. Then,  the $\rho-z$ surface is locally an $\mathbb{R}^2/\mathbb{Z}_\ell$ orbifold if we set
\begin{equation} \label{ldz22}
\frac{\ell\Delta{z}}{2\pi}\,=\,\frac{1}{\mathcal{E}(q_1,q_2)}\,,
\end{equation}
where $\Delta{z}$ is the period of coordinate, $z$, and $\ell\,=\,1,2,3,\ldots$. The metric spanned by $(y,z)$ has a topology of disk, $\Sigma$, with the center at $y=y_a$ and the boundary at $y=0$ or $y=\infty$.

We calculate the Euler characteristic of the $y-z$ surface, $\Sigma$,
\begin{equation}
\chi\,=\,\frac{1}{4\pi}\int_\Sigma{R}_\Sigma\text{vol}_\Sigma\,=\,\frac{1}{4\pi}2\mathcal{E}(q_1,q_2)\Delta{z}\,=\,\frac{1}{\ell}\,.
\end{equation}
This is a natural result for a disk in an $\mathbb{R}^2/\mathbb{Z}_\ell$ orbifold.

\subsubsection{Uplift to type IIA supergravity}

By employing the uplift formula to type IIA supergravity which we construct in appendix \ref{appC}, we obtain the uplifted metric in the string frame and the dilaton. When the solutions are uplifted to type IIA supergravity by employing the uplift formula presented in appendix \ref{appC}, the internal space is more involved than the solutions we previously studied. Thus we only present the metric in the string frame and the dilaton of each solution at the boundary of the disk.

For the solution with $0<y<y_1$, as $y\rightarrow0$, the uplifted metric becomes
\begin{align}
ds_{10}^2\,\approx\,r^{1/3}&\left[\left(q_1q_2y\sin^2\xi\right)^{1/2}\left(\frac{-dt^2+dr^2}{r^2}+\frac{9g^2}{16}dz^2\right)\right. \notag \\
&\left.+\frac{1}{g^2\left(q_1q_2y\sin^2\xi\right)^{1/2}}\right(\sin^2\xi\Big(4\,dy^2+y^2\left(d\psi^2+\cos^2\psi\,d\varphi_3^2\right)+\left(q_1\cos^2\theta+q_2\sin^2\theta\right)d\xi^2\Big) \notag \\
&+\cos^2\xi\Big(\left(q_2\cos^2\theta+q_1\sin^2\theta\right)d\theta^2+q_1\cos^2\theta\,D\varphi_1^2+q_2\sin^2\theta\,D\varphi_2^2\Big) \notag \\
&\left.\left.+\frac{1}{2}\left(q_1-q_2\right)\sin\left(2\xi\right)\sin\left(2\theta\right)d\xi\,d\theta\right)\right]\,,
\end{align}
and the dilaton is
\begin{equation}
e^\Phi\,\approx\,\frac{r^{5/6}}{\left(q_1q_2y\sin^2\xi\right)^{1/4}}\,.
\end{equation}
The metric implies the smeared D2-brane sources. The D2-branes are 
\begin{itemize}
\item extended along the $t$, $r$, and $z$ directions;
\item localized at the origin of $y$, $\psi$, and $\varphi_3$ directions;
\item smeared along $\xi$, $\theta$, $\varphi_1$, and $\varphi_2$ directions. 
\end{itemize}
The $y^{1/2}$ and $1/y^{1/2}$ factors of the metric and the dilaton match the smeared branes reviewed in appendix \ref{appA}.

For the solution with $<y_2<y<\infty$, as $y\rightarrow\infty$, the uplifted metric becomes \eqref{d2unsmeared}.

\bigskip
\bigskip
\leftline{\bf Acknowledgements}
\noindent We are grateful to Pietro Ferrero and Adolfo Guarino for helpful communications. This work was supported by the Kumoh National Institute of Technology.

\appendix
\section{Smeared branes} \label{appA}
\renewcommand{\theequation}{A.\arabic{equation}}
\setcounter{equation}{0} 

The metric and the dilaton of D$p$-branes smeared over $s$ directions are given by $e.g.$, (3.38) of \cite{Lozano:2019emq},
\begin{align}
ds_{10}^2\,\approx&\,r^{\frac{7-p-s}{2}}ds_{1,p}^2+r^{-\frac{7-p-s}{2}}\left(dr^2+r^2ds_{8-p-s}^2+ds_s^2\right)\,, \notag \\
e^\Phi\,\approx&\,r^{\frac{(p-3)(7-p-s)}{4}}\,.
\end{align}
If $s=0$, it reduces to the asymptotics of usual un-smeared D$p$-branes.

\section{Uplift formula for $D=6$ gauged supergravity} \label{appB}
\renewcommand{\theequation}{B.\arabic{equation}}
\setcounter{equation}{0} 

For the metric and the dilaton of type IIA supergravity, \cite{Giani:1984wc, Campbell:1984zc, Huq:1983im}, the uplift formula of $D=6$ gauged supergravity is given in (67) of \cite{Cvetic:2000ah}. In $D=6$ gauged supergravity, $ds_6^2$ and $A^{ij}$ are the metric and the gauge fields, respectively,  and $\sigma$ and $T^{ij}$ parametrize the scalar fields where $i,j$ are the $SO(5)$ indices. There are also axions parametrized by $\chi^{ij}$. In the absence of axions, $\chi_{ij}=0$, the uplift formula for the metric in the Einstein frame and the dilaton is given by
\begin{align}
ds_{10}^2\,=&\,\Omega^{1/8}\left[\Delta^{1/3}e^{4\sigma}ds_6^2+\frac{1}{g^2}\Delta^{-2/3}T^{-1}_{ij}D\mu^iD\mu^j\right]\,, \notag \\
e^{\frac{4}{3}\Phi}\,=&\,\Omega\,,
\end{align}
where we have
\begin{equation}
\Omega\,\equiv\,\Delta^{1/3}e^{-16\sigma}\,, \qquad \Delta\,\equiv\,T_{ij}\mu^i\mu^j\,, \qquad D\mu^i\,\equiv\,d\mu^i+gA^{ij}\mu^j\,.
\end{equation}
The uplift of fluxes are not given in \cite{Cvetic:2000ah}.

For the $U(1)^2$-invariant truncation of $D=6$ gauged supergravity, we have two gauge fields, $A^1\equiv{A}^{12}$ and $A^2\equiv{A}^{34}$, and the scalar fields are parametrized by $(\lambda_1, \lambda_2,\sigma)$ with
\begin{equation}
T_{ij}\,=\,\text{diag}\left(e^{2\lambda_1},e^{2\lambda_1},e^{2\lambda_2},e^{2\lambda_2},e^{-4\lambda_1-4\lambda_2}\right)\,.
\end{equation}
See section 5 of \cite{Boisvert:2024jrl} for details of $D=6$ $U(1)^2$-gauged supergravity. In order to uplift the solutions to type IIA supergravity, we employ the parametrization of four-sphere by the constrained coordinates, $e.g.$, in appendix A of \cite{Cheung:2022ilc},
\begin{equation}
\mu^1+i\mu^2\,=\,\cos\xi\cos\theta{e}^{i\chi_1}\,, \qquad \mu^3+i\mu^4\,=\,\cos\xi\sin\theta{e}^{i\chi_2}\,, \qquad \mu^5\,=\,\sin\xi\,,
\end{equation}
with $-\pi/2\le\xi\le\pi/2$, $0\le\theta\le\pi/2$, $0\le\chi_1,\chi_2\le2\pi$. With the parametrization, the metric on four-sphere is
\begin{equation}
T^{-1}_{ij}D\mu^iD\mu^j\,=\,e^{4\lambda_1+4\lambda_2}dw_0^2+e^{-2\lambda_1}\left(dw_1^2+w_1^2d\chi_1^2\right)+e^{-2\lambda_2}\left(dw_2^2+w_2^2d\chi_2^2\right)\,,
\end{equation}
and
\begin{equation}
\Delta\,=\,e^{-4\lambda_1-4\lambda_2}w_0^2+e^{2\lambda_1}w_1^2+e^{2\lambda_2}w_2^2\,,
\end{equation}
where we introduced
\begin{equation}
w_0\,=\,\sin\xi\,, \qquad w_1\,=\,\cos\xi\cos\theta\,, \qquad w_2\,=\,\cos\xi\sin\theta\,,
\end{equation}
satisfying $w_0^2+w_1^2+w_2^2\,=\,1$.

\section{Uplift formula for $D=4$ gauged supergravity} \label{appC}
\renewcommand{\theequation}{C.\arabic{equation}}
\setcounter{equation}{0} 

We consider $D=4$ $ISO(7)$-gauged maximal supergravity, \cite{Hull:1984yy} and \cite{Guarino:2015jca, Guarino:2015qaa, Guarino:2015vca}. In particular, the $U(1)^3$ subsector of the theory is of our interest. The field content is the metric, three $U(1)$ gauge fields, $A^I$, $I=1,2,3$, and four real scalar fields, $\phi_i$, $i=0,1,2,3$. For details of $D=4$ $U(1)^3$-gauged supergravity, we refer section 6 of \cite{Boisvert:2024jrl}. We introduce a parametrization for scalar fields, 
\begin{equation}
X_0\,=\,e^{\frac{1}{2}\left(\phi_1+\phi_2+\phi_3\right)-\phi_0}\,, \quad X_1\,=\,e^{\frac{1}{2}\left(\phi_1-\phi_2-\phi_3\right)}\,, \quad X_2\,=\,e^{\frac{1}{2}\left(-\phi_1+\phi_2-\phi_3\right)}\,, \quad X_3,=\,e^{\frac{1}{2}\left(-\phi_1-\phi_2+\phi_3\right)}\,.
\end{equation}

We construct the uplift formula to type IIA supergravity for the metric in the Einstein frame and the dilaton,
\begin{align}
ds_{10}^2\,=&\,\left(X_0X_1^2X_2^2X_3^2\right)^{1/8}\Delta^{5/8}\left[ds_4^2+g^{-2}\Delta^{-1}\Big(X_0^{-1}d\mu_0^2+X_i^{-1}\left(d\mu_i^2+\mu_i^2D\varphi_i^2\right)\Big)\right]\,, \notag \\
e^{4\Phi}\,=&\,\left(X_0X_1^2X_2^2X_3^2\right)^3\Delta^{-1}\,,
\end{align}
where we introduce
\begin{equation}
\Delta\,\equiv\,\sum^3_{i=0}X_i\mu_i^2\,, \qquad D\varphi_I\,\equiv\,d\varphi_I-gA\,.
\end{equation}
We have chosen the constrained coordinates for the solution in \eqref{d210one},
\begin{equation}
\mu_0\,=\,\sin\xi\sin\theta\,, \qquad \mu_1\,=\,\cos\xi\,, \qquad \mu_2\,=\,\sin\xi\cos\theta\cos\psi\,, \qquad \mu_3\,=\,\sin\xi\cos\theta\sin\psi\,,
\end{equation}
and, for the solution in \eqref{d210two},
\begin{equation}
\mu_0\,=\,\sin\xi\sin\psi\,, \qquad \mu_1\,=\,\cos\xi\cos\theta\,, \qquad \mu_2\,=\,\cos\xi\sin\theta\,, \qquad \mu_3\,=\,\sin\xi\cos\psi\,.
\end{equation}
We have not obtained the uplift formula for the three-form potential and leave this for a future work.

Employing the uplift formula, one can uplift the truncation given in (3.7) of \cite{Guarino:2015qaa} to the solution in (2.6) of \cite{Varela:2015uca}, if we turn off the axionic scalar fields, $\chi=a=\zeta=\tilde{\zeta}=0$, and identify the scalar fields by
\begin{align}
\phi_1=\phi_2=\phi_3\,=&\,-\varphi^\text{GV}\,, \notag \\
 \phi_0\,=&\,-2\phi^{GV}\,,
\end{align}
where $\varphi^\text{GV}$ and $\phi^\text{GV}$ are the scalar fields in \cite{Guarino:2015qaa} and \cite{Varela:2015uca}.

\section{No charge solutions from M5-branes} \label{appD}
\renewcommand{\theequation}{D.\arabic{equation}}
\setcounter{equation}{0} 

\subsection{$D=7$ gauged supergravity}

We consider $D=7$ $SU(2)$-gauged maximal supergravity, \cite{Pernici:1984xx}, which is obtained from the reduction of eleven-dimensional supergravity on a four-sphere. In particular, a $U(1)^2$ subsector of the theory, \cite{Liu:1999ai}, is of our interest. The field content is the metric, two $U(1)$ gauge fields, $A^1$, $A^2$, and two real scalar fields, $X^1$, $X^2$.

In this section, we consider solution without a gauge field by setting $A^1=A^2=0$ or, equivalently, $q_1=q_2=0$ in the the spindle solution of \cite{Ferrero:2021wvk}. Then the solution is given by
\begin{align}
ds_7^2\,=&\,y\left[ds_{AdS_5}^2+\frac{1}{4y^3h(y)}dy^2+h(y)dz^2\right]\,, \notag \\ 
X_1\,=&\,X_2\,=\,1\, \notag \\
A_1\,=&\,A_2\,=\,0\,,
\end{align}
where we have
\begin{equation}
h(y)\,=\,\frac{y-4}{4y}\,.
\end{equation}
For $h(y)=0$, there is a single root, $y_1\,\equiv\,4$. We find solutions with
\begin{equation}
y_1\,\equiv\,4<y<\infty\,,
\end{equation}

Approaching $y\rightarrow{y}_1$, the metric becomes to be
\begin{equation}
ds_7^2\,\approx\,y_1\left[ds_{AdS_5}^2+\frac{d\rho^2+\mathcal{E}^2\rho^2dz^2}{-y_1^2h'(y_1)}\right]\,,
\end{equation}
where we have
\begin{equation}
\mathcal{E}^2\,\equiv\,y_1^3h'(y_1)^2\,=\,\left(\frac{1}{y_1^{1/2}}\right)^2\,,
\end{equation}
and we introduced a new parametrization of coordinate, $\rho^2\,=\,y_1-y$. Then,  the $\rho-z$ surface is locally an $\mathbb{R}^2/\mathbb{Z}_\ell$ orbifold if we set
\begin{equation}
\frac{\ell\Delta{z}}{2\pi}\,=\,y_1^{1/2}\,,
\end{equation}
where $\Delta{z}$ is the period of coordinate, $z$, and $\ell\,=\,1,2,3,\ldots$. The metric spanned by $(y,z)$ has a topology of disk, $\Sigma$, with the center at $y=y_1$ and the boundary at $y=0$.

We calculate the Euler characteristic of the $y-z$ surface, $\Sigma$, 
\begin{equation}
\chi\,=\,\frac{1}{4\pi}\int_\Sigma{R}_\Sigma\text{vol}_\Sigma\,=\,\frac{1}{4\pi}\frac{2}{y_1^{1/2}}\Delta{z}\,=\,\frac{1}{\ell}\,.
\end{equation}
This is a natural result for a disk in an $\mathbb{R}^2/\mathbb{Z}_\ell$ orbifold.

\subsection{Uplift to eleven-dimensional supergravity}

By employing the uplift formula to eleven-dimensional supergravity, $e.g.$, in \cite{Liu:1999ai}, we obtain the metric and the four-form flux,
\begin{align}
L^{-2}ds_{11}^2\,=&\,y\left[ds_{AdS_5}^2+\frac{1}{y^2\left(y-4\right)}dy^2+\frac{y-4}{4y}dz^2\right]+ds_{S^4}^2\,, \notag \\
L^{-6}*_{11}G_{(4)}\,=&\,-3\text{vol}_7\,,
\end{align}
where $L$ is the radius of $AdS_7$ and $\text{vol}_7$ is the volume form of seven-dimensional metric.

We consider the singularity at $y\rightarrow\infty$ of the uplifted metric. As $y\rightarrow\infty$, the uplifted metric becomes
\begin{equation}
L^{-2}ds_{11}^2\,\approx\,y\left[ds_{AdS_5}^2+\frac{1}{4}dz^2\right]+\frac{1}{y^2}\left[dy^2+y^2ds_{S^4}^2\right]\,,
\end{equation}
The branes are extended along the $AdS_5$ and $z$ directions and localized at the center of $y$ and $S^4$ directions. We observe that there is no smeared direction. The $y$ and $1/y^{2}$ factors of the metric match the un-smeared M5-branes.

We calculated the holographic central charge of dual 4d SCFTs. For the metric of
\begin{equation}
ds_{11}^2\,=\,L^2e^{2\lambda}\left[ds_{AdS_5}^2+ds_{M_6}^2\right]\,,
\end{equation}
the holographic central charge is given by, \cite{Ferrero:2021wvk},
\begin{equation}
a\,=\,\frac{1}{2^7\pi^6}\left(\frac{L}{\ell_P}\right)^9\int_{M_6}d^6x\sqrt{g_{M_6}}e^{9\lambda}\,,
\end{equation}
where $\ell_P$ is the eleven-dimensional Planck length. We find, \cite{Ferrero:2021wvk},
\begin{equation}
a\,=\,\frac{L^9}{2^7\pi^6\ell_P^9}\text{vol}_{S^4}\int_{y_1}^\infty\frac{y}{2}dydz\,,
\end{equation}
where
\begin{equation}
\text{vol}_{S^4}\,=\,\frac{8\pi^2}{3}\,.
\end{equation}
Thus the integral diverges.

No charge solutions from D3-, D4-D8-, and M2-branes are all analogous to the case of M5-branes.

\bibliographystyle{JHEP}
\bibliography{20250515}

\end{document}